\documentclass[reqno,11pt]{amsart}
\usepackage[linktocpage]{hyperref}
\usepackage[utf8]{inputenc}
\usepackage{graphicx}
\usepackage{slashed}
\usepackage{amscd}
\usepackage{amssymb}
\usepackage{enumitem}
\usepackage[mathscr]{eucal}
\usepackage{pstricks}
\usepackage{pst-all}

\numberwithin{equation}{section}
\allowdisplaybreaks[1]

\title[A Positive Mass Theorem for Static Causal Fermion Systems]
{A Positive Mass Theorem for \\ Static Causal Fermion Systems}

\author[F.\ Finster]{Felix Finster}

\author[A.\ Platzer]{Andreas Platzer \\ \\ January 2020}
\address{Fakult\"at f\"ur Mathematik \\ Universit\"at Regensburg \\ D-93040 Regensburg \\ Germany}
\email{finster@ur.de, andy.platzer@web.de}

\newtheorem{Def}{Definition}[section]
\newtheorem{Thm}[Def]{Theorem}
\newtheorem{Prp}[Def]{Proposition}
\newtheorem{Lemma}[Def]{Lemma}
\newtheorem{Remark}[Def]{Remark}

\newcommand{\Thanks}{\vspace*{.5em} \noindent \thanks}
\newcommand{\beq}{\begin{equation}}
\newcommand{\eeq}{\end{equation}}
\newcommand{\Proof}{\begin{proof}}
\newcommand{\QED}{\end{proof} \noindent}
\newcommand{\QEDrem}{\ \hfill $\Diamond$}

\newcommand{\la}{\langle}
\newcommand{\ra}{\rangle}

\newcommand{\Sl}{\mbox{$\prec \!\!$ \nolinebreak}}
\newcommand{\Sr}{\mbox{\nolinebreak $\succ$}}

\newcommand{\C}{\mathbb{C}}
\newcommand{\R}{\mathbb{R}}
\newcommand{\1}{\mbox{\rm 1 \hspace{-1.05 em} 1}}

\newcommand{\N}{\mathbb{N}}

\renewcommand{\Tr}{\text{\rm{Tr}}}
\DeclareMathOperator{\tr}{tr}
\renewcommand{\O}{{\mathscr{O}}}
\renewcommand{\L}{{\mathcal{L}}}
\newcommand{\Sact}{{\mathcal{S}}}
\newcommand{\T}{{\mathcal{T}}}
\newcommand{\scrt}{{\mathfrak{t}}}
\newcommand{\M}{{\scrM}}
\newcommand{\Mass}{{\mathfrak{M}}}

\newcommand{\U}{\text{\rm{U}}}

\newcommand{\Cisc}{C^\infty_{\text{\rm{sc}}}}

\newcommand{\Dir}{{\mathcal{D}}}
\DeclareMathOperator{\supp}{supp}
\renewcommand{\H}{\mathscr{H}}
\newcommand{\Lin}{\text{\rm{L}}}
\newcommand{\F}{{\mathscr{F}}}
\newcommand{\G}{{\mathscr{G}}}

\newcommand{\D}{\mathscr{D}}

\newcommand{\scrM}{\mycal M}
\newcommand{\scrN}{\mycal N}
\newcommand{\itemD}{\item[{\raisebox{0.125em}{\tiny $\blacktriangleright$}}]}
\newcommand{\scrU}{{\mathscr{U}}}

\newcommand{\J}{\mathfrak{J}}
\newcommand{\s}{\mathfrak{s}}
\newcommand{\Jdiff}{\mathfrak{J}^\text{\rm{\tiny{diff}}}}
\newcommand{\Jtest}{\mathfrak{J}^\text{\rm{\tiny{test}}}}

\newcommand{\Jin}{\mathfrak{J}^\text{\rm{\tiny{in}}}}
\newcommand{\Jlin}{\mathfrak{J}^\text{\rm{\tiny{lin}}}}

\newcommand{\Gdiff}{\Gamma^\text{\rm{\tiny{diff}}}}
\newcommand{\Gtest}{\Gamma^\text{\rm{\tiny{test}}}}
\newcommand{\Ctest}{C^\text{\rm{\tiny{test}}}}
\renewcommand{\u}{\mathfrak{u}}
\renewcommand{\v}{\mathfrak{v}}

\newcommand{\bitem}{\begin{itemize}[leftmargin=2.5em]}
\newcommand{\eitem}{\end{itemize}}
\renewcommand{\div}{{\rm{div}}\,}
\newcommand{\scrA}{{\mathscr{A}}}

\newcommand{\x}{\mathbf{x}}
\newcommand{\y}{\mathbf{y}}
\newcommand{\z}{\mathbf{z}}

\DeclareFontFamily{OT1}{rsfso}{}
\DeclareFontShape{OT1}{rsfso}{m}{n}{ <-7> rsfso5 <7-10> rsfso7 <10-> rsfso10}{}
\DeclareMathAlphabet{\mycal}{OT1}{rsfso}{m}{n}

\begin{document}
\maketitle

\begin{abstract}
Asymptotically flat static causal fermion systems are introduced.
Their total mass is defined as a limit of surface layer integrals which compare the measures
describing the asymptotically flat spacetime and a vacuum spacetime near spatial infinity.
Our definition does not involve any regularity assumptions; it even applies to singular
or generalized ``quantum'' spacetimes. A positive mass theorem is proven.
Our methods and results explain why and how the causal action principle incorporates the nonlinear effects
of gravity for static systems.
\end{abstract}

\tableofcontents

\section{Introduction and Outline of Results} \label{secintro}
In Newtonian physics, the total energy is obtained simply by
adding the energies of all particles and fields of the system, including the energy
of the gravitational field.
In general relativity, however, the situation is much more difficult due to the nonlinearity
of the gravitational interaction. Nevertheless, as shown by Arnowitt, Deser and Misner~\cite{adm},
it is possible to define total energy and momentum of an isolated gravitational system,
expressed in terms of the asymptotics of the metric tensor near infinity.
Since in the present paper we restrict attention throughout to the {\em{static}} situation,
we also recall the definition by Arnowitt, Deser and Misner (ADM) only in this setting.
Let~$\scrM$ be a Lorentzian manifold which is the topological product~$\scrM = \R \times \scrN$.
We denote the spacetime points by~$x=(t, \x)$
with~$t \in \R$ and~$\x \in \scrN$. Next, we assume that the Lorentzian metric is time independent
(in other words, we assume that~$\partial_t$ is a Killing field), that the induced metric~$g$ on~$\scrN$
is Riemannian, and that the second fundamental form vanishes on~$\scrN$
(this is sometimes referred to as the time-symmetric case).
Finally, we assume that~$(\scrN, g)$ is {\em{asymptotically flat}}.
Stated for simplicity in three spatial dimensions and with one asymptotic end, this means
that there is a compact set~$K \subset \scrN$ such that~$\scrN \setminus K$ is diffeomorphic to the
region~$\R^3 \setminus \overline{B_R(0)}$ outside a closed ball of radius $R$. 
In the chart defined by this diffeomorphism, the metric should be of the form
\beq \label{metricdecay}
g_{\alpha \beta}(\x) = \delta_{\alpha \beta} + a_{\alpha \beta}(\x) \:,\qquad \x \in \R^3 \setminus \overline{B_R(0)} \:,
\eeq
where $a_{\alpha \beta}$ decays at infinity as
\[ a_{\alpha \beta} = \O(1/|\x|) \:,\quad
\partial_\gamma a_{\alpha \beta} = \O(1/|\x|^2) \quad \text{and} \quad
\partial_{\gamma \delta} a_{\alpha \beta} = \O(1/|\x|^3) \:. \]
Under these assumptions, the total energy is also referred to as
the {\em{total mass}} or {\em{ADM mass}}. It is defined by
\beq \label{adm}
\Mass_\text{ADM} = \frac{1}{16 \pi} \lim_{R \rightarrow \infty} \sum_{\alpha, \beta=1}^3 \int_{S_R}
(\partial_\beta g_{\alpha \beta} - \partial_\alpha g_{\beta \beta}) \:\nu^\alpha\: d\Omega \:,
\eeq
where~$d\Omega$ is the area form on the coordinate sphere~$S_R$, and $\nu$ is
the normal vector to~$S_R$ (both defined in the coordinate chart).

In Newtonian physics, the fact that the energies of all particles and all energy densities of fields
are positive implies that the total energy is also positive.
Again, in general relativity the connection is much more involved. It is made precise by the
{\em{positive mass theorem}} proved by Schoen and Yau~\cite{schoen+yau}.
It states that if a suitable {\em{local energy condition}} is fulfilled, which in the static case reduces
to the condition that the scalar curvature of~$g$ is non-negative,
\[ 
\text{scal} \geq 0 \:, \]
then the total mass
non-negative. Moreover, if the total mass vanishes, then~$(\scrN, g)$ is flat.
The positive mass theorem makes a profound statement on the nature of the gravitational interaction.
We remark that it can be generalized to higher dimensions; see for example the recent papers~\cite{herzlich,
schoen+yau3} and the references therein.

The theory of {\em{causal fermion systems}} is a recent approach to fundamental physics
where spacetime is no longer modelled by a Lorentzian manifold but may instead have a
nontrivial, possibly discrete structure on a microscopic length scale (which can be thought of
as the Planck scale). In the setting of causal fermion systems, the physical equations are formulated
via a variational principle, the {\em{causal action principle}}.
In~\cite[Chapter~4]{cfs} it is shown that in a specific limiting case, the so-called
{\em{continuum limit}}, the Euler-Lagrange (EL) equations of the causal action principle
give rise to the Einstein equations,
up to possible higher order corrections in curvature (which scale in powers of~$(\delta^2\:
\text{Riem})$, where~$\delta$ is the Planck length and~$\text{Riem}$ is the
curvature tensor). In this limiting case, spacetime goes over to a Lorentzian manifold,
whereas the gravitational coupling constant~$G \sim \delta^2$ is determined by
the length scale~$\delta$ of the microscopic spacetime structure.

The derivation of the Einstein equations in~\cite[Chapter~4]{cfs} has two disadvantages.
First, it is rather technical, because it relies on the detailed form of the regularized light-cone expansion of the
kernel of the fermionic projector. Consequently, the derivation does not give a good intuitive
understanding of the underlying mechanisms.
Second and more importantly, the Einstein equations are recovered
only in the continuum limit, but the methods do not give any insight into the geometric meaning
of the EL equations for more general ``quantum'' spacetimes.

In view of theses disadvantages, it is an important task to study the nature of the gravitational
interaction as described by the causal action principle without referring to limiting cases,
but instead by analyzing directly the corresponding EL equations.
One step in this direction is the recent paper~\cite{jacobson}, where the connection between
area change and matter flux is worked out for two-dimensional surfaces propagating
in a null Killing direction. In the present paper we go a step in a different, somewhat complementary
direction which aims at understanding gravity for static systems directly from the causal action principle.
We succeed in giving a general definition of the total mass for
static causal fermion systems. Moreover, a positive mass theorem is proved which states that if a
suitable local energy condition is fulfilled, then the total mass is positive.
We also explain how the ADM mass is recovered as a limiting case.

More precisely, our results are stated as follows.
In the theory of causal fermion systems, a physical system (consisting of spacetime
and all structures therein) is described by a Borel measure~$\rho$
on a set of linear operator~$\F \subset \Lin(\H)$ on a Hilbert space~$\H$
(for details see the preliminaries in Section~\ref{seccfs}).
Spacetime~$M$ is defined as the support of this measure,
\[ M := \supp \rho \:. \]
In the {\em{static}} situation to be considered here, there is a global time coordinate~$t \in \R$, i.e.
\[ M = \R \times N \;\ni\; x = (t, \x) \:. \]
Moreover, the system should be time independent. This is made precise by a one-parameter group~$(\scrU_t)_{t \in \R}$
of unitary transformations which leaves the measure~$\rho$ invariant (for details see Definition~\ref{defstatic}).
In particular, the measure~$\rho$ can be written as
\beq \label{rhodecomp}
d\rho = dt\, d\mu
\eeq
where~$\mu$ is a Borel measure on~$N$. The dynamics of a causal fermion system is
described by a variational principle, the {\em{causal action principle}} (see Section~\ref{seccfs}).
In the {\em{static}} case, it reduces to minimizing the action~$\Sact$ given by
\[ \Sact (\mu) = \int_\G d\mu(\x) \int_\G d\mu(\y)\: \L_\kappa(\x,\y) \]
under variations of the measure~$\mu$, leaving the total volume fixed.
Here in~$\G := \F /\R$ we divided out the action of the group~$(\scrU_t)_{t \in \R}$, and the Lagrangian~$\L_\kappa$ is
of the form
\[ \L_\kappa = \L + \kappa\, \T \qquad \text{with given functions} \qquad \L, \T \::\: \G \times \G \rightarrow \R^+_0 \:, \]
and~$\kappa>0$ is a Lagrange parameter (for details see Sections~\ref{seccfs} and~\ref{secSstatic}).
The functions~$\L$ and~$\T$ are symmetric, i.e.\
\[ 
\L(\x,\y) = \L(\y,\x) \quad \text{and} \quad \T(\x,\y) = \T(\y,\x) \qquad \text{for all~$\x, \y \in \G$}\:. \]
The causal action principle in the static case is a specific example of
a {\em{causal variational principle}} (for the general context see Section~\ref{seccvp}).

A minimizer~$\mu$ of the above variational principle satisfies the
\beq \label{ELintro}
\text{\em{Euler-Lagrange (EL) equations}} \qquad \ell_\kappa|_N \equiv \inf_\G \ell_\kappa = 0 \:,
\eeq
where the function~$\ell_\kappa$ is defined by
\beq \label{ellkappa}
\ell_\kappa : \G \rightarrow \R^+_0 \:,\qquad \ell_\kappa(\x) = \ell(\x) + \kappa\, \scrt(\x)
\eeq
with
\begin{align}
\ell(\x) &:= \int_N \L(\x, \y)\: d\mu(\y) - \s \label{ellintro} \\
\scrt(\x) &:= \int_N \T(\x, \y)\: d\mu(\y) \:, \label{scrtintro}
\end{align}
and~$\s$ is a positive parameter which for convenience is chosen such that the infimum in~\eqref{ELintro} is zero.

Using the above notions, the total mass can be introduced as follows.
Let~$\mu$ and~$\tilde{\mu}$ be two measures which are jointly static
(meaning that they are both static with respect to the same one-parameter group~$(\scrU_t)_{t \in \R}$;
for details see Definition~\ref{defstatic}) and are both minimizers or critical points
of the static causal action principle for the same values of the parameters~$\kappa$ and~$\s$
(for details on the definition of critical measures see the paragraph after~\eqref{ELtest} in~Section~\ref{seccvp}).
We define the functions~$\tilde{\ell}$ and~$\tilde{\scrt}$
by adding tildes to~$\mu$, $\ell$ and~$\scrt$ in~\eqref{ellintro} and~\eqref{scrtintro}.
In order to compare the measures~$\mu$ and~$\tilde{\mu}$, we relate them by the Lagrangian.
To this end, we introduce the functions
\beq \label{ndef}
\left\{ \begin{array}{rl} 
n \,:\, N \rightarrow \R^+_0 \cup \{\infty\} \:,\qquad
n(\x) &\!\!\!\!= \displaystyle \int_{\tilde{N}} \L_\kappa(\x,\y)\: d\tilde{\mu}(\y) \\[1em]
\tilde{n} \,:\, \tilde{N} \rightarrow \R^+_0 \cup \{\infty\} \:,\qquad
\tilde{n}(\x) &\!\!\!\!= \displaystyle \int_N \L_\kappa(\x,\y)\: d\mu(\y)
\end{array} \right.
\eeq
and define the {\em{correlation measures}}~$\nu$ and~$\tilde{\nu}$ by
\beq \label{nuintro}
d\nu(\x) = n(\x)\: d\mu(\x) \qquad \text{and} \qquad d\tilde{\nu}(\x) = \tilde{n}(\x)\: d\tilde{\mu}(\x) \:.
\eeq

\begin{Def} \label{defasyclose}
The measures~$\tilde{\mu}$ and~$\mu$ are {\bf{asymptotically close}} if
they are both $\sigma$-finite with infinite total volume,
\beq \label{nuinf}
\tilde{\mu}(\tilde{N}) = \mu(N) = \infty \:,
\eeq
but
\[ \int_N \big| n(x) - \s \big| \: d\mu(x) < \infty \qquad \text{and} \qquad
\int_{\tilde{N}} \big| \tilde{n}(x) - \s \big| \: d\tilde{\mu}(x) < \infty 
 \:. \]
\end{Def}

We begin with the most general definition of the total mass.
\begin{Def} \label{defmassgen}
Assume that~$\mu$ and~$\tilde{\mu}$ are asymptotically close.
Then the {\bf{total mass}}~$\Mass$ of~$\tilde{\mu}$ relative to~$\mu$ is defined by
\begin{align}
&\Mass(\tilde{\mu}, \mu) := \lim_{\Omega \nearrow N} \;\lim_{\tilde{\Omega} \nearrow \tilde{N}} 
\bigg( -\s \Big( \tilde{\mu}(\tilde{\Omega}) - \mu(\Omega) \Big) \notag \\
& \qquad
+ \int_{\tilde{\Omega}} \!d\tilde{\mu}(\x) \int_{N\setminus \Omega} \!\!\!\!\!\!\!d\mu(\y)\: \L_\kappa(\x,\y)
- \int_{\Omega} \!d\mu(\x) \int_{\tilde{N} \setminus \tilde{\Omega}}  \!\!\!\!d\tilde{\mu}(\y)\: \L_\kappa(\x,\y) \bigg) \:,
\label{massgen}
\end{align}
where the notation~$\Omega \nearrow N$ means that we take an exhaustion of~$N$ by 
sets of finite $\mu$-measure.
\end{Def} \noindent
Here the limits exist and are independent of the choice of the exhaustions (see Proposition~\ref{prpindepend}).

This definition is extremely general because it does not involve any smoothness or continuity assumptions.
It even applies to singular or discrete measures. It is not necessary (and would not even be possible
in this generality) to specify the dimension of spacetime. Clearly, in order to compare 
this notion of total mass with the ADM mass, we need
to specialize the setting. We now explain step by step how this can be done.
Along the way, we will also explain the structure of the formula~\eqref{massgen}.

The first step is to assume that~$\mu$ or~$\tilde{\mu}$ is a continuous measure in the sense
that it is non-atomic (for details see Definition~\ref{defnonatomic}).
In this case, in~\eqref{massgen} one may restrict attention to sets~$\Omega$ and~$\tilde{\Omega}$ with the
same measure, making it possible to write the total mass as
\begin{align}
&\Mass(\tilde{\mu}, \mu) = \lim_{\Omega_n \nearrow N, \; \tilde{\Omega}_n \nearrow \tilde{N}
\text{ with } \mu(\Omega_n) = \tilde{\mu}(\tilde{\Omega}_n) < \infty} \notag \\
&\quad \times \bigg(
\int_{\tilde{\Omega}_n} \!\!d\tilde{\mu}(\x) \int_{N\setminus \Omega_n} \!\!\!\!\!\!\!\!d\mu(\y)\: \L_\kappa(\x,\y)
- \int_{\Omega_n} \!\!d\mu(\x) \int_{\tilde{N} \setminus \tilde{\Omega}_n}  \!\!\!\!\!\!\!d\tilde{\mu}(\y)\: \L_\kappa(\x,\y) \bigg) \:,
\label{massintro}
\end{align}
where~$(\Omega_n)_{n \in \N}$ and~$(\tilde{\Omega}_n)_{n \in \N}$ are exhaustions
of~$N$ and~$\tilde{N}$, respectively
(for details see Proposition~\ref{prpnonatomic}).

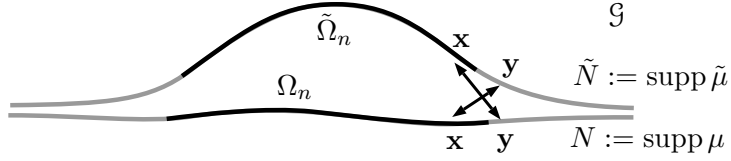
\begin{figure}
%
\psscalebox{1.0 1.0} 
{
\begin{pspicture}(-2.3,-1.045158)(14.810092,1.045158)
\definecolor{colour0}{rgb}{0.6,0.6,0.6}
\rput[bl](7.510092,-0.13922058){\normalsize{$\tilde{N}:= \supp \tilde{\mu}$}}
\psbezier[linecolor=colour0, linewidth=0.06](0.025092011,-0.33725628)(1.3550938,-0.3220137)(1.8907287,-0.2754297)(2.415092,0.12238656180246096)(2.9394553,0.5202028)(3.5928288,0.99649286)(4.270092,1.0057794)(4.9473553,1.0150659)(5.42014,0.74156153)(5.890092,0.37292227)(6.3600445,0.0042830044)(6.88906,-0.33742145)(8.315092,-0.37422058)
\psbezier[linecolor=colour0, linewidth=0.06](0.0,-0.4718888)(0.9616981,-0.4688655)(1.8381602,-0.579791)(2.4229932,-0.48749685024003725)(3.007826,-0.39520273)(3.6470058,-0.39106253)(4.160086,-0.43922058)(4.6731663,-0.48737863)(5.3516655,-0.5919167)(5.8220882,-0.5892501)(6.292511,-0.5865835)(6.88234,-0.49192157)(8.310092,-0.49922058)
\rput[bl](7.465092,-0.9692206){\normalsize{$N:=\supp \mu$}}
\rput[bl](7.970092,0.7157794){\normalsize{$\G$}}
\psbezier[linecolor=black, linewidth=0.06](2.310092,0.050779417)(3.0729995,0.60274875)(3.5402741,1.006699)(4.365092,1.0057794189453124)(5.18991,1.0048599)(5.619156,0.5635629)(6.220092,0.13077942)
\psbezier[linecolor=black, linewidth=0.06](2.105092,-0.5192206)(2.5967734,-0.4562537)(3.4201167,-0.35720322)(4.055092,-0.4242205810546875)(4.6900673,-0.49123794)(5.7186456,-0.63545233)(6.385092,-0.5642206)
\rput[bl](4.095092,0.47077942){\normalsize{$\tilde{\Omega}_n$}}
\rput[bl](3.570092,-0.25422058){\normalsize{$\Omega_n$}}
\psline[linecolor=black, linewidth=0.04, arrowsize=0.05291667cm 2.0,arrowlength=1.4,arrowinset=0.0]{<->}(5.930092,0.23077942)(6.545092,-0.5292206)
\psline[linecolor=black, linewidth=0.04, arrowsize=0.05291667cm 2.0,arrowlength=1.4,arrowinset=0.0]{<->}(5.900092,-0.49422058)(6.535092,-0.07422058)
\rput[bl](6.495092,-0.9592206){\normalsize{$\y$}}
\rput[bl](5.820092,-0.9142206){\normalsize{$\x$}}
\rput[bl](5.9050922,0.44077942){\normalsize{$\x$}}
\rput[bl](6.565092,0.03577942){\normalsize{$\y$}}
\end{pspicture}
}
\caption{The surface layer integral defining the total mass.}
\label{figosi}
\end{figure}

Let us briefly explain the structure of the above formulas for the total mass.
The double integrals in~\eqref{massgen} and~\eqref{massintro}
are so-called {\em{surface layer integrals}}, which generalize surface integrals to the
setting of causal variational principles (see~\cite[Section~2.3]{noether}, \cite[Section~7.1]{intro} or Sections~\ref{secosi}
and~\ref{secosinonlin} in the preliminaries). The main point is that of the two
arguments~$\x$ and~$\y$ of the Lagrangian~$\L_\kappa$, one lies in the interior region
($\Omega_n$ or~$\tilde{\Omega}_n$), whereas the other lies in the exterior region of the
other spacetime ($\tilde{N} \setminus \tilde{\Omega}_n$ and~$N \setminus \Omega_n$, respectively),
as is indicated by the arrows in Figure~\ref{figosi}.
Since~$\L_\kappa$ typically decays if its arguments are far apart,
the main contribution to the integrals is obtained when both~$\x$ and~$\y$ are
close to the boundaries~$\partial \Omega_n \subset N$ or~$\partial \tilde{\Omega}_n \subset \tilde{N}$.
Therefore, each of the two double integrals~\eqref{massgen} and~\eqref{massintro}
can be thought of as an integral over a ``thin strip'' around the boundaries of~$\partial \Omega_n \subset N$
or~$\partial \tilde{\Omega}_n \subset \tilde{N}$. Moreover, it is important that in~\eqref{massgen}
and~\eqref{massintro} we take the
difference of these double integrals. This is needed for getting a connection to the 
general conservation law as first derived in~\cite[Section~4 and Appendix~A]{fockbosonic}.

Using this intuitive picture, it is evident that the expressions~\eqref{massgen} and~\eqref{massintro}
depend only on the geometry (as encoded in the measures~$\mu$ and~$\tilde{\mu}$) near infinity,
but not on the geometry in any compact subset.
This statement will be made mathematically precise in Theorem~\ref{thmindepend} in Section~\ref{secind}.
However, it is important to observe that the {\em{inner volume}} does come into play,
as one sees from the term~$\tilde{\mu}(\tilde{\Omega}) - \mu(\Omega)$ in~\eqref{massgen}
and the constraint~$\tilde{\mu}(\tilde{\Omega}_n)=\mu(\Omega_n)$ in~\eqref{massintro}.
Therefore, our definition of the total mass can be understood as a limit of surface layer integrals
to be evaluated on the boundaries of large subsets~$\Omega \subset N$
and~$\tilde{\Omega} \subset \tilde{N}$ which have the same volume.

At first sight, the fact that the inner volume comes into play seems to be a major difference to the ADM mass~\eqref{adm}, which is computable purely from the geometric data near infinity. 
However, as will be explained later in this introduction (after the statement of Theorem~\ref{thmcorrespond}),
for causal fermion systems constructed in a Lorentzian spacetime (so-called static Dirac system; see
Section~\ref{seclorentz}), the inner volume drops out of the computation,
giving formulas for~$\Mass$ purely in terms of the geometry near infinity.
This can be understood non-technically as follows:
The main contribution to the surface layer integral in~\eqref{massgen} is~$\s$ times the
difference of the volumes of~$\tilde{\Omega}$ and~$\Omega$.
If this contribution vanishes, the next-to-leading contribution comes into play.
This next-to-leading order contribution gives the total mass.
For static Dirac systems, it can be computed purely from the geometric data at infinity, without
referring to the volumes of~$\tilde{\Omega}$ or~$\Omega$.
This argument will be made precise in Proposition~\ref{prpnoinner}
and Appendix~\ref{appscalingmass}.

In order to avoid confusion, we point out that the ``volume'' considered here
refers to the spatial volume measure~$\mu$ obtained by rewriting the spacetime volume
measure~$\rho$ in the form~\eqref{rhodecomp}
(and similarly for~$\tilde{\mu}$). In examples of spacetimes described by static Lorentzian manifolds
(like the Schwarzschild geometry), this measure does {\em{not}} agree with the volume element
of the induced Riemannian metric on~$N$, but it coincides instead with the volume form of spacetime contracted
with the Killing field. This will be explained in more detail in Section~\ref{secex}.

We next outline our positivity results for the mass.
In order to prove our results, we need to impose stronger assumptions on the asymptotics at infinity,
We first state these assumptions and explain the afterward.
\begin{Def} \label{defae}
The measure~$\mu$ has {\bf{one asymptotic end of dimension~$k \in \N$}} if there is a relatively compact
open set~$I \subset N$ and a diffeomorphism
\beq \label{Phidef}
\Phi \::\: N \setminus I \rightarrow \R^k \setminus B_R
\eeq
(where~$B_R$ is the open ball of radius~$R>0$) with the property that the push-forward of~$\mu$
is of the form
\[ d(\Phi_* \mu)(\x) = h(\x)\: d^k\x \]
with a smooth function~$h \in C^\infty(\R^k \setminus B_R, \R^+)$ 
with~$\lim_{\x \rightarrow \infty} h(\x) =1$.

A sequence~$(\x_n)_{n \in \N}$ in~$N$ {\bf{tends to infinity}}, $\x_n \rightarrow \infty$,
if almost all elements of the sequence are in~$N \setminus I$, and if the
image sequence~$\Phi(\x_n)$ tends to infinity in~$\R^k$.
A function~$g$ on~$N$ converges at infinity if the sequence~$f(\x_n)$ converges
for every~$\x_n$ which tends to infinity. The limit is denoted by~$\lim_{N \ni \x \rightarrow \infty} g(x)$.
\end{Def} \noindent
For simplicity, we shall restrict attention to one asymptotic end. But
all our methods and results could be extended in a straightforward way
to several asymptotic ends.

\begin{Def} \label{defvacuum}The measure~$\mu$ is a {\bf{vacuum measure}} of dimension~$k \in \N$
if it satisfies the following conditions:
\bitem
\item[\rm{(i)}] It has one asymptotic end of dimension~$k$ (see Definition~\ref{defae}).
\item[\rm{(ii)}] The function~$\ell$ is constant on~$N$,
\beq \label{ellvac}
\ell(\x) =: \ell_\infty \qquad \text{for all~$\x \in N$}\:.
\eeq
\eitem 
\end{Def}

\begin{Def} \label{defasyflat}
The measure~$\tilde{\mu}$ is {\bf{asymptotically flat}} of dimension~$k \in \N$ if it satisfies the following conditions:
\bitem
\item[\rm{(i)}] It is asymptotically close to a vacuum measure~$\mu$ of dimension~$k \in \N$
(see Definitions~\ref{defasyclose} and~\ref{defvacuum}).
\item[\rm{(ii)}] It has one asymptotic end of dimension~$k$ (see Definition~\ref{defae}).
The following limit exists,
\[ \lim_{\tilde{N} \ni \x \rightarrow \infty} \tilde{\ell}(\x) =: \tilde{\ell}_\infty \in \R \:. \]
The function~$\tilde{\ell} - \tilde{\ell}_\infty$ is integrable, i.e.\
\[ \int_{\tilde{N}} \big| \tilde{\ell}(\x) - \tilde{\ell}_\infty \big|\: d\tilde{\mu}(\x) < \infty \:.  \]
\item[\rm{(iii)}] There is a mapping~$F : N \rightarrow \tilde{N}$ such that
\[ \tilde{\mu}  = F_* (\mu) \]
(where $F_* \mu$ is the push-forward measure defined by $(F_* \mu)(\tilde{\Omega})
= \mu(F^{-1}(\tilde{\Omega}))$).
When restricted to~$N \setminus I$ (with~$I$ as in Definition~\ref{defvacuum}),
this mapping is a diffeomorphism to its image. Moreover, it 
tends to the identity at infinity in the sense that the surface layer integral
in~\eqref{massintro} can be linearized to obtain
\begin{align}
\Mass(\tilde{\mu}, \mu) &= \lim_{\Omega \nearrow N}
\int_\Omega \!d\mu(\x) \int_{N\setminus \Omega} \!\!\!\!\!\!\!d\mu(\y)\: \Big( \L_\kappa\big( F(\x),\y \big)
- \L_\kappa\big(\x, F(\y) \big) \Big) \label{osilin1} \\
&= \lim_{\Omega \nearrow N}
\int_\Omega \!d\mu(\x) \int_{N\setminus \Omega} \!\!\!\!\!\!\!d\mu(\y)\: \big( D_{1,w} - D_{2,w} \big) \L_\kappa(\x,\y) \Big)
\label{osilin2}
\end{align}
for a suitable vector field~$w \in \Gamma(N, T\G)$ on~$\G$ along~$N$.
\eitem
\end{Def} \noindent
We remark for clarity that~\eqref{osilin1} follows immediately from~\eqref{massintro} using the
definition of the push-forward measure (together with the fact that the limits in~\eqref{massintro} exist).
The directional derivatives in~\eqref{osilin2} act on the first and second argument of the Lagrangian, respectively.
We also note that, setting~$\tilde{I} = F(I)$, the mapping~$\tilde{\Phi} := \Phi \circ F^{-1}|_{\tilde{N} \setminus \tilde{I}}$
is a diffeomorphism from~$\tilde{N} \setminus \tilde{I}$ to~$\R^k \setminus B_R$
with the property that it changes the measure~$\tilde{\mu}$ according to
\[ d(\tilde{\Phi}_* \tilde{\mu})(\x) = \tilde{h}(\x)\: d^k\x \qquad \text{with}
\qquad \lim_{\x \rightarrow \infty} \tilde{h}(\x) =1 \:. \]

The assumptions so far can be regarded as the analogs of the decay conditions
for the metric in~\eqref{metricdecay} in the setting of asymptotically flat Riemannian manifolds.
We next introduce another important assumption, which means in words that
both measures~$\mu$ and~$\tilde{\mu}$ should be extendable to families of critical
measures~$(\mu_\tau)_\tau$ and~$(\tilde{\mu}_\tau)_\tau$ for a variable value of the
parameter~$\kappa=\kappa(\tau)$.

\begin{Def} \label{defkappaextend}
The measure~$\mu$ is {\bf{$\kappa$-extendable}} if the following conditions hold:
\bitem
\item[\rm{(i)}] There is a family of measures~$(\mu_\tau)_{\tau \in (-1,1)}$ of the form
\[ \mu_\tau = (F_\tau)_* \mu \:, \]
each of which satisfies the EL equations~\eqref{ELintro} with a parameter~$\kappa(\tau)$ and
\[ F_0 = \text{\rm{id}}_N \qquad\text{and} \qquad \kappa'(0) \neq 0 \:. \]
\item[\rm{(ii)}] For every~$x \in N$, the curve~$F_\tau(x)$ is differentiable at~$\tau=0$,
giving rise to a vector field
\beq \label{dFdef}
v := \frac{d}{d \tau} F_\tau \Big|_{\tau=0} \;\in\; \Gamma(N, T\G)\:.
\eeq
\eitem
\end{Def} \noindent
For a convenient normalization, we always choose the parametrization such that
\beq \label{lognorm}
\frac{d}{d\tau} \log \kappa(\tau) \Big|_{\tau=0} = -1 \:.
\eeq
We remark that all our results hold as well for families of critical measures if
one merely replaces the EL equations~\eqref{ELintro} in~(i) by the weak EL equations
(see~\eqref{ELtest} in~Section~\ref{seccvp} with~$\ell$ carrying an additional subscript~$\kappa$).

\begin{Def} \label{defkappascale}
Let~$\tilde{\mu}$ be asymptotically flat with respect to the
vacuum measure~$\mu$. Then~$\tilde{\mu}$ is said to be {\bf{$\kappa$-scalable}}
if both~$\tilde{\mu}$ and~$\mu$ are $\kappa$-extendable
and if, for suitable choices of the mappings~$F_\tau$ and~$\tilde{F}_\tau$
in Definition~\ref{defkappaextend},
the vector field~$w$ in~\eqref{osilin2} is related to the vector fields~$v$ and~$\tilde{v}$
in~\eqref{dFdef} by
\beq \label{wdef}
w = g\, \big(\tilde{v} - v \big)
\eeq
with a constant~$g \in \R$, referred to as the {\bf{gravitational coupling constant}}.
\end{Def}

We can now state our main result.
\begin{Def} \label{deflec}
The asymptotically flat measure~$\tilde{\mu}$ satisfies the {\bf{local energy condition}} if
\beq \label{ellmin}
\tilde{\ell}(\x) \geq \tilde{\ell}_\infty \qquad \text{for all~$\x \in \tilde{N}$}\:.
\eeq
\end{Def}

\begin{Thm} {\bf{(Positive mass theorem)}} \label{thmpmt}
Assume that~$\tilde{\mu}$ is asymptotically flat (see Definition~\ref{defasyflat})
and $\kappa$-scalable (see Definition~\ref{defkappascale}).
Then the total mass can be written as
\beq \label{Mid}
\Mass(\tilde{\mu}, \mu) = g \int_{\tilde{N}} \big( \tilde{\ell} - \tilde{\ell}_\infty \big)\: d\tilde{\mu}  \:.
\eeq
If~$\tilde{\mu}$ satisfies the local energy condition and the gravitational coupling constant~$g$ is positive,
then the total mass is non-negative,
\[ \Mass(\tilde{\mu}, \mu) \geq 0 \:. \]
Moreover, if the total mass vanishes, then~$\tilde{\mu}$ is a vacuum measure (see Definition~\ref{defvacuum}).
\end{Thm}
For the proof of this theorem, we consider a linear integral equation, referred to as the
{\em{equations of linearized gravity}} (see~\eqref{inhom} in Section~\ref{seclininhom}). These equations
are obtained by linearizing the EL equations of the causal action principle in the parameter~$\kappa$.
Working with linear equations is inspired by the spinor proof of the positive energy theorem
(see~\cite{witten, parker+taubes}). The reason why these spinorial methods extend to the
setting of causal fermion systems can be understood from the fact that in interesting examples
(see the static Dirac systems described below), the causal fermion system is built up of solutions of the
Dirac equation. The integral formula~\eqref{Mid} for the total mass can be regarded as the analog
of the formula
\[ \Mass_\text{ADM} = c(k) \int_{\scrN} \bigg( |\nabla \psi|^2 + \frac{\text{scal}}{4}\: |\psi|^2 \bigg) \, d\mu_\scrN \:, \]
where~$\psi$ is the Witten spinor.

Clearly, the above theorem leaves the following questions open:
\bitem
\item[\bf{(a)}] How is~$\Mass(\tilde{\mu}, \mu)$ related to the ADM mass in~\eqref{adm}?
\item[\bf{(b)}] Why is the measure~$\tilde{\mu}$ $\kappa$-scalable (see Definition~\ref{defkappascale})?
\label{btoe}
\item[\bf{(c)}] Why is the gravitational constant as defined in~\eqref{wdef} positive?
\item[\bf{(d)}] Why is the local energy condition satisfied?
\item[\bf{(e)}] Why do vacuum measures (see Definition~\ref{defvacuum}) describe a flat spacetime?
\eitem
In order to address these questions, we need to be more specific and focus on
causal fermion systems constructed in a static Lorentzian spacetime,
referred to as {\em{static Dirac systems}}
(for details see~\cite[Section~1.2]{cfs}, \cite{nrstg} or Section~\ref{seclorentz} in the preliminaries).
For such systems, we can answer question~{\bf{(a)}} as follows:
\begin{Thm} \label{thmcorrespond}
For a static Dirac system describing a four-dimensional spacetime which is asymptotically Schwarzschild
(for details see the beginning of Section~\ref{secex}),
the total mass is proportional to the ADM mass
\[ \Mass = c\: \Mass_\text{\rm{ADM}} \:, \]
where the constant~$c$ can be computed from the Lagrangian in the regularized, spherically symmetric
Minkowski vacuum via the formula
\beq \label{cdef}
c = 2\pi \int_{\scrM} |\y|^2\: \L_\kappa\big(0, (t,\y) \big)\: dt\, d^3\mathbf{y} \;>\;0 \:.
\eeq
\end{Thm} \noindent
Our method of proof also explains why for static Dirac systems, the total mass is computable
purely from the knowledge of the metric near infinity.

In order to address the other questions~{\bf{(b)}}--{\bf{(e)}}, one must make use of the
scaling behavior of static Dirac systems as worked out in Section~\ref{secscaling},
based on previous results in~\cite[\S4.2.5]{cfs}, \cite{action} and~\cite[Appendix~A]{jacobson}.
The relevant length scales are given by
\beq \label{lengthscales}
\left\{ \begin{array}{ll}
\text{Compton length} & m^{-1} \\
\text{Planck length} & \:\delta \\
\text{regularization length} & \:\varepsilon \:, \end{array} \right.
\eeq
where~$m$ denotes the mass of the Dirac particles. Moreover, the measure~$\mu$ determines a length scale
as the radius of a ball of $\mu$-volume one (here we fix the freedom in rescaling~$\mu$ by
specifying~$\s$ as well as the parameter~$c$ of the local trace; for details see Section~\ref{secrescale}).
Working on this fixed length scale, we have three dimensionless parameters~$m$, $\delta$ and~$\varepsilon$.
\bitem
\item[\bf{(b)}] For static Dirac systems, the relation~\eqref{wdef} can be derived as follows.
In Definition~\ref{defkappascale} we consider a family~$(\mu_\tau)_{\tau \in (-1,1)}$ of critical
measures for a decreasing value of~$\kappa$, \eqref{lognorm}.
In this family, the parameters in~\eqref{lengthscales} will in general change; we denote them
by~$m(\tau)$, $\delta(\tau)$ and~$\varepsilon(\tau)$. Similarly, the family~$(\tilde{\mu}_\tau)_{\tau \in (-1,1)}$
of measures which describes curved spacetime, involves the parameters~$\tilde{m}(\tau)$, $\tilde{\delta}(\tau)$
and~$\tilde{\varepsilon}(\tau)$. Since these parameters are constant in spacetime,
their $\tau$-dependence can be determined asymptotically near infinity, where spacetime
goes over to Minkowski space. In other words, the $\tau$-dependence of these parameters is the same
in curved and in flat spacetime, i.e.\ $\tilde{m}(\tau)=m(\tau)$, $\tilde{\delta}(\tau)=\delta(\tau)$
and~$\tilde{\varepsilon}(\tau)=\varepsilon(\tau)$.
Therefore, taking the difference of the jets~$\tilde{v}$ and~$v$
which describe the infinitesimal variations in curved and flat spacetime, respectively,
on the right side of~\eqref{wdef} the variation of the parameters~$m$, $\delta$ and~$\varepsilon$
drops out. But the fact that~$\tilde{v}$ changes these parameters in the spacetime described by~$\tilde{\mu}$
has the effect that this spacetime is modified depending on the
distribution of matter and gravity. This can be understood most easily in the case of the
so-called {\em{natural scaling of~$\delta$}} where the dimensionless parameter~$m \delta$ remains constant.
In this case, the system remains unchanged in Planck units where the gravitational constant is fixed.
However, in these units the length scale determined by~$\tilde{\mu}$ changes. In other words, the
gravitational system is fixed, but the volume~$\tilde{\mu}$ changes. Since the $\tilde{\mu}$-volume is fixed in the
definition of the total mass, this implies that in Planck units, the radius of the set~$\tilde{\Omega}$
in~\eqref{osilin2} changes. This in turn gives rise to a change of the total mass, explaining~\eqref{wdef}.
This argument is given in more detail and for more general scalings in Section~\ref{secscaling}.
\item[\bf{(c)}] Working out the above scalings in more detail, in Section~\ref{secscaling} we find
that the gravitational constant has the same sign as
\[ \frac{d}{d\tau} \big( m \delta^2 \big) \big|_{\tau=0} \]
(again for variations normalized according to~\eqref{lognorm}).
Moreover, we know from the general structure of the causal action principle that the parameter~$m$ decreases.
Therefore, the gravitational coupling constant is positive for a natural scaling of~$\delta$.
More generally, the gravitational coupling constant is positive if the scaling of~$\delta$ does not
differ too much from natural scaling. More details are given in Remark~\ref{remattractive}.
\item[\bf{(d)}] One method to deal with the local energy
condition in Definition~\ref{deflec} is to proceed as in general relativity by taking it as a
condition motivated from physical observations which is to be verified case by case for different types
of matter. But in the context of causal fermion systems, one can go a step further and try to
explain the inequality~\eqref{ellmin} from the minimality of the causal action.
In Remark~\ref{remedp}, this argument is worked out for
homogeneous matter distributions. More generally, this argument shows that
the energy conditions hold for all matter densities which are nearly constant on the
Compton scale. However, at present we cannot rule out the possibility that
the energy density might be negative on microscopic scales.
\item[\bf{(e)}] For static Dirac systems, it can be analyzed in detailed what the
vacuum condition~\eqref{ellvac} means. By direct computation, one verifies that
curvature increases~$\ell$. Therefore, the condition~\eqref{ellvac} is satisfied only
for systems in flat Minkowski space. This argument is given in more detail
in Remark~\ref{remflat}.
\eitem

The paper is organized as follows. Section~\ref{secprelim} provides the
necessary preliminaries on causal variational principles and causal fermion systems.
In Section~\ref{seccfsstatic} we specialize causal fermion systems
to the static case and explain how they fit to the general setting of causal variational principles.
In Section~\ref{secnonlin} we prove that the total mass is well-defined
and independent of exhaustions, the inner geometry and the identifications of the
Hilbert spaces of the two causal fermion systems.
In Section~\ref{secpmt} we introduce the equations of linearized gravity and
use them to prove Theorem~\ref{thmpmt}.
In Section~\ref{secex} we compute the total mass in the example of an
asymptotically Schwarzschild spacetime and prove Theorem~\ref{thmcorrespond}.
In Section~\ref{secscaling} we work out the relevant scalings.
Moreover, we discuss the sign of the gravitational coupling
constant (Remark~\ref{remattractive}),
show that homogeneous perturbations of Minkowski space must satisfy the local energy
condition (Remark~\ref{remedp}) and specify the assumptions under which
a vacuum measure describes a flat spacetime (Remark~\ref{remflat}).
In the appendices, we work out the scaling behavior of the total mass for static Dirac systems
(Appendix~\ref{appscalingmass}) and analyze the fermionic projector and the relevant surface layer
integrals for linearized gravity (Appendix~\ref{appgeodesic}).

\section{Preliminaries} \label{secprelim}
We now recall the basics on causal variational principles in the setting
needed here. More details can be found in~\cite{jet, intro}.
We use a slightly different notation in order to get consistency with the causal variational
principle in the static case as will be introduced in Section~\ref{seccfsstatic}.

\subsection{Causal Variational Principles in the Non-Compact Setting} \label{seccvp}
We consider causal variational principles in the non-compact setting as
introduced in~\cite[Section~2]{jet}. Thus we let~$\G$ be a (possibly non-compact)
smooth manifold of dimension~$m \geq 1$
and~$\mu$ a (positive) Borel measure on~$\G$ (the {\em{universal measure}}).
Moreover, we are given a non-negative function~$\L : \G \times \G \rightarrow \R^+_0$
(the {\em{Lagrangian}}) with the following properties:
\begin{itemize}[leftmargin=2em]
\item[(i)] $\L$ is symmetric: $\L(\x,\y) = \L(\y,\x)$ for all~$\x,\y \in \G$.\label{Cond1}
\item[(ii)] $\L$ is lower semi-continuous, i.e.\ for all sequences~$\x_n \rightarrow \x$ and~$\y_{n'} \rightarrow \y$,
\label{Cond2}%
\[ \L(\x,\y) \leq \liminf_{n,n' \rightarrow \infty} \L(\x_n, \y_{n'})\:. \]
\end{itemize}
The {\em{causal variational principle}} is to minimize the action
\beq \label{Sact} 
\Sact (\mu) = \int_\G d\mu(\x) \int_\G d\mu(\y)\: \L(\x,\y) 
\eeq
under variations of the measure~$\mu$, keeping the total volume~$\mu(\G)$ fixed
({\em{volume constraint}}).

If the total volume~$\mu(\G)$ is finite, one minimizes~\eqref{Sact}
over all regular Borel measures with the same total volume.
If the total volume~$\mu(\G)$ is infinite, however, it is not obvious how to implement the volume constraint,
making it necessary to proceed as follows.
We make the following additional assumptions:
\begin{itemize}[leftmargin=2em]
\item[(iii)] The measure~$\mu$ is {\em{locally finite}}
(meaning that any~$\x \in \G$ has an open neighborhood~$U$ with~$\mu(U)< \infty$)
and {\em{regular}} (meaning that the measure of a set can be recovered by approximation from inside
with compact and from outside with open sets). \label{Cond3}
\item[(iv)] The function~$\L(\x,.)$ is $\mu$-integrable for all~$\x \in \G$, giving
a lower semi-continuous and bounded function on~$\G$. \label{Cond4}
\end{itemize}
Given a regular Borel measure~$\mu$ on~$\G$, we vary over all
regular Borel measures~$\tilde{\mu}$ with
\[ 
\big| \tilde{\mu} - \mu \big|(\G) < \infty \qquad \text{and} \qquad
\big( \tilde{\mu} - \mu \big) (\G) = 0 \]
(where~$|.|$ denotes the total variation of a measure).
These variations of the causal action are well-defined.
The existence theory for minimizers is developed in~\cite{noncompact}.
It is shown in~\cite[Lemma~2.3]{jet} that a minimizer
satisfies the {\em{Euler-Lagrange (EL) equations}}
which state that for a suitable value of the parameter~$\s>0$,
the lower semi-continuous function~$\ell : \G \rightarrow \R_0^+$ defined by
\beq \label{ldef}
\ell(\x) := \int_\G \L(\x,\y)\: d\mu(\y) - \s
\eeq
is minimal and vanishes on the support of~$\mu$,
\beq \label{EL}
\ell|_N \equiv \inf_\G \ell = 0 \:.
\eeq
For further details we refer to~\cite[Section~2]{jet}.

\subsubsection{The Weak Euler-Lagrange Equations and Jet Spaces} \label{secwEL}
\hspace*{0.05cm}
We denote the support of~$\mu$ by~$N$,
\beq \label{Ndef}
N:= \supp \mu \;\subset\; \G \:.
\eeq
The EL equations~\eqref{EL} are nonlocal in the sense that
they make a statement on~$\ell$ even for points~$\x \in \G$ which
are far away from~$N$.
It turns out that for the applications in this paper, it is preferable to
evaluate the EL equations locally in a neighborhood of~$N$.
This leads to the {\em{weak EL equations}} introduced in~\cite[Section~4]{jet}.
We here give a slightly less general version of these equations which
is sufficient for our purposes. In order to explain how the weak EL equations come about,
we begin with the simplified situation that the function~$\ell$ is smooth.
In this case, the minimality of~$\ell$ implies that the derivative of~$\ell$
vanishes on~$N$, i.e.\
\beq \label{ELweak}
\ell|_N \equiv 0 \qquad \text{and} \qquad D \ell|_N \equiv 0
\eeq
(where~$D \ell(p) : T_p \G \rightarrow \R$ is the derivative).
In order to combine these two equations in a compact form,
it is convenient to consider a pair~$\u := (a, u)$
consisting of a real-valued function~$a$ on~$N$ and a vector field~$u$
on~$T\G$ along~$N$, and to denote the combination of 
multiplication and directional derivative by
\beq \label{Djet}
\nabla_{\u} \ell(\x) := a(\x)\, \ell(\x) + \big(D_u \ell \big)(\x) \:.
\eeq
Then the equations~\eqref{ELweak} imply that~$\nabla_{\u} \ell(\x)$
vanishes for all~$\x \in N$.
The pair~$\u=(a,u)$ is referred to as a {\em{jet}}.

In the general lower-continuous setting, one must be careful because
the directional derivative~$D_u \ell$ in~\eqref{Djet} need not exist.
Our method for dealing with this problem is to restrict attention to vector fields
for which the directional derivative is well-defined.
Moreover, we must specify the regularity assumptions on~$a$ and~$u$.
To begin with, we always assume that~$a$ and~$u$ are {\em{smooth}} in the sense that they
have a smooth extension to the manifold~$\G$. Thus the jet~$\u$ should be
an element of the jet space
\[ 
\J := \big\{ \u = (a,u) \text{ with } a \in C^\infty(N, \R) \text{ and } u \in \Gamma(N, T\G) \big\} \:, \]
where~$C^\infty(N, \R)$ and~$\Gamma(N,T\G)$ denote the space of real-valued functions and vector fields
on~$N$, respectively, which admit a smooth extension to~$\G$.

Clearly, the fact that a jet~$\u$ is smooth does not imply that the functions~$\ell$
or~$\L$ are differentiable in the direction of~$\u$. This must be ensured by additional
conditions which are satisfied by suitable subspaces of~$\J$
which we now introduce.
First, we let~$\Gdiff$ be those vector fields for which the
directional derivative of the function~$\ell$ exists,
\[ 
\Gdiff = \big\{ u \in C^\infty(N, T\G) \;\big|\; \text{$D_{u} \ell(\x)$ exists for all~$\x \in N$} \big\} \:. \]
This gives rise to the jet space
\beq \label{Jdiffdef}
\Jdiff := C^\infty(N, \R) \oplus \Gdiff \;\subset\; \J \:.
\eeq
For the jets in~$\Jdiff$, the combination of multiplication and directional derivative
in~\eqref{Djet} is well-defined. 
We choose a linear subspace~$\Jtest \subset \Jdiff$ with the property
that its scalar and vector components are both vector spaces,
\[ 
\Jtest = \Ctest(N, \R) \oplus \Gtest \;\subseteq\; \Jdiff \:, \]
and the scalar component is nowhere trivial in the sense that
\beq \label{Cnontriv}
\text{for all~$\x \in N$ there is~$a \in \Ctest(N, \R)$ with~$a(\x) \neq 0$}\:.
\eeq
Then the {\em{weak EL equations}} read (for details cf.~\cite[(eq.~(4.10)]{jet})
\beq \label{ELtest}
\nabla_{\u} \ell|_N = 0 \qquad \text{for all~$\u \in \Jtest$}\:.
\eeq
The purpose of introducing~$\Jtest$ is that it gives the freedom to restrict attention to the portion of
information in the EL equations which is relevant for the application in mind.
For example, if one is interested only in the macroscopic dynamics, one can choose~$\Jtest$
to be composed of jets pointing in directions where the 
microscopic fluctuations of~$\ell$ are disregarded.

We finally point out that the weak EL equations~\eqref{ELtest}
do not hold only for minimizers, but also for critical points of
the causal action. With this in mind, all methods and results of this paper do not apply only to
minimizers, but more generally to critical points of the causal variational principle.
For brevity, we also refer to a measure which satisfies the weak EL equations~\eqref{ELtest}
as a {\em{critical measure}}.

We conclude this section by introducing a few jet spaces 
and specifying differentiability conditions
which will be needed later on. We begin with the spaces~$\J^\ell$, where~$\ell \in \N_0 \cup \{\infty\}$ can be
thought of as the order of differentiability if the derivatives act  simultaneously on
both arguments of the Lagrangian:
\begin{Def} \label{defJvary}
For any~$\ell \in \N_0 \cup \{\infty\}$, the jet space~$\J^\ell \subset \J$
is defined as the vector space of test jets with the following properties:
\begin{itemize}[leftmargin=2em]
\item[\rm{(i)}] For all~$\y \in N$ and all~$\x$ in an open neighborhood of~$N$,
directional derivatives
\beq \label{derex}
\big( \nabla_{1, \v_1} + \nabla_{2, \v_1} \big) \cdots \big( \nabla_{1, \v_p} + \nabla_{2, \v_p} \big) \L(\x,\y)
\eeq
(computed componentwise in distinguished charts around~$\x$ and~$\y$; for details see~\cite[Section~5.2]{banach})
exist for all~$p \in \{1, \ldots, \ell\}$ and all~$\v_1, \ldots, \v_p \in \J^\ell$.
\item[\rm{(ii)}] The functions in~\eqref{derex} are $\mu$-integrable
in the variable~$\y$, giving rise to locally bounded functions in~$\x$. More precisely,
these functions are in the space
\[ L^\infty_\text{\rm{loc}}\Big( M, L^1\big(M, d\mu(y) \big); d\mu(x) \Big) \:. \]
\item[\rm{(iii)}] Integrating the expression~\eqref{derex} in~$\y$ over~$N$
with respect to the measure~$\mu$,
the resulting function (defined for all~$\x$ in an open neighborhood of~$N$)
is continuously differentiable in the direction of every jet~$\u \in \Jtest$.
\end{itemize}
\end{Def} \noindent
Here and throughout this paper, we use the following conventions for partial derivatives and jet derivatives:
\begin{itemize}[leftmargin=2em]
\itemD Partial and jet derivatives with an index $i \in \{ 1,2 \}$, as for example in~\eqref{derex}, only act on the respective variable of the function $\L$.
This implies, for example, that the derivatives commute,
\[ 
\nabla_{1,\v} \nabla_{1,\u} \L(\x,\y) = \nabla_{1,\u} \nabla_{1,\v} \L(\x,\y) \:. \]
\itemD The partial or jet derivatives which do not carry an index act as partial derivatives
on the corresponding argument of the Lagrangian. This implies, for example, that
\[ \nabla_\u \int_\G \nabla_{1,\v} \, \L(\x,\y) \: d\mu(\y) =  \int_\G \nabla_{1,\u} \nabla_{1,\v}\, \L(\x,\y) \: d\mu(\y) \:. \]
\end{itemize}
We point out that (in contrast to the method and conventions used in~\cite{jet})
{\em{jets are never differentiated}}.

In order for all integral expressions to be well-defined, we impose throughout
that the space~$\Jtest$ has the following properties (for details see~\cite[Section~3.5]{osi}).
\begin{Def} \label{defslr}
Let~$m \in \N$. The jet space~$\Jtest$ is {\bf{surface layer regular}}
if~$\Jtest \subset \J^2$ (see Definition~\ref{defJvary}) and
if for all~$\u, \v \in \Jtest$ and all~$p \in \{1, 2\}$ the following conditions hold:
\begin{itemize}[leftmargin=2em]
\item[\rm{(i)}] The directional derivatives
\beq \label{Lderiv1}
\nabla_{1,\u} \,\big( \nabla_{1,\v} + \nabla_{2,\v} \big)^{p-1} \L(\x,\y)
\eeq
exist.
\item[\rm{(ii)}] The functions in~\eqref{Lderiv1} are $\mu$-integrable
in the variable~$\y$, giving rise to locally bounded functions in~$\x$. More precisely,
these functions are in the space
\[ L^\infty_\text{\rm{loc}}\Big( L^1\big(N, d\mu(\y) \big), d\mu(\x) \Big) \:. \]
\item[\rm{(iii)}] The $\u$-derivative in~\eqref{Lderiv1} may be interchanged with the $\y$-integration, i.e.
\[ \int_N \nabla_{1,\u} \,\big( \nabla_{1,\v} + \nabla_{2,\v} \big)^{p-1} \L(\x,\y)\: d\mu(\y)
= \nabla_\u \int_N \big( \nabla_{1,\v} + \nabla_{2,\v} \big)^{p-1} \L(\x,\y)\: d\mu(\y) \:. \]
\end{itemize}
\end{Def}

\subsubsection{The Linearized Field Equations} \label{seclinear}
Usually, linearized fields are obtained by considering a family of
nonlinear solutions and linearizing with respect to a parameter~$\tau$
which describes the field strength.
The analogous notion in the setting of causal fermion systems
is a linearization of a family of measures~$(\tilde{\mu}_\tau)_{\tau \in [0,1)]}$ which all satisfy the weak EL equations~\eqref{ELtest}
(for fixed values of the parameters~$\kappa$ and~$\s$).
It turns out to be fruitful to construct this family of measures by multiplying
a given critical measure~$\mu$ by a weight function~$f_\tau$ and then
``transporting'' the resulting measure with a mapping~$F_\tau$. More precisely, we consider the ansatz
\begin{align} \label{rhoFf}
\tilde{\mu}_\tau = (F_\tau)_* \big( f_\tau \, \mu \big) \:,
\end{align}
where~$f_\tau \in C^\infty(N, \R^+)$ and~$F_\tau \in C^\infty(N, \G)$ are smooth mappings,
and~$(F_\tau)_*\mu$ denotes the push-forward (defined 
for a subset~$\Omega \subset \G$ by~$((F_\tau)_*\mu)(\Omega)
= \mu ( F_\tau^{-1} (\Omega))$; see for example~\cite[Section~3.6]{bogachev}).

The property of the family of measures~$(\tilde{\mu}_\tau)_{\tau \in [0,1)}$ of the form~\eqref{rhoFf}
to satisfy the weak EL equation for all~$\tau$
means infinitesimally in~$\tau$ that the jet~$\v$ defined by
\beq \label{vinfdef}
\v = (b,v) := \frac{d}{d\tau} (f_\tau, F_\tau) \big|_{\tau=0}
\eeq
satisfies the {\em{linearized field equations}}. We now
recall the main step of the construction.
Using the definition of the push-forward measure, we can write the weak EL equations~\eqref{ELtest}
for the measure~$\tilde{\mu}_\tau$ as
\[ \nabla_\u \bigg( \int_N \L\big( F_\tau(\x), F_\tau(\y) \big) \: f_\tau(\y)\: d\mu - \s \bigg) = 0 \:. \]
Since the function~$\ell$ vanishes on the support, we may multiply by~$f_\tau(\x)$ to obtain
\beq \label{lin1}
\nabla_\u \bigg( \int_N f_\tau(\x)\: \L\big( F_\tau(\x), F_\tau(\y) \big) \: f_\tau(\y)\: d\mu - f_\tau(\x)\,\s \bigg) = 0 \:.
\eeq
At this point, the technical complication arise that one must specify the $\tau$-dependence of the jet spaces,
and moreover the last transformation makes it necessary to transform the jet spaces.
Here we do not enter the details but refer instead to the rigorous derivation in~\cite[Section~3.3]{perturb}
or to the simplified presentation in the smooth setting in the textbook~\cite[Chapter~6]{intro}.
Differentiating~\eqref{lin1} with respect to~$\tau$ gives the linearized field equations 
\beq \label{eqlinlip}
\la \u, \Delta \v \ra|_N = 0 \qquad \text{for all~$\u \in \Jtest$} \:,
\eeq
where
\beq \label{Deldef}
\la \u, \Delta \v \ra(\x) := \nabla_{\u} \bigg( \int_N \big( \nabla_{1, \v} + \nabla_{2, \v} \big) \L(\x,\y)\: d\mu(\y) - \nabla_\v \,\s \bigg) \:.
\eeq
We denote the vector space of all solutions of the linearized field equations by~$\Jlin \subset \J^1$.

\subsubsection{A Conserved Surface Layer Integral for Linearized Solutions} \label{secosi}
In the setting of causal fermion systems, the usual integrals over hypersurfaces in spacetime are undefined.
Instead, one considers so-called {\em{surface layer integrals}}, being double integrals of the form
\beq \label{IntrOSI}
\int_\Omega \!d\mu(\x) \int_{N \setminus \Omega} \!\!\!\!\!\!\!d\mu(\y) \:(\cdots)\: \L(\x,\y) \:,
\eeq
where~$\Omega$ is a Borel subset of~$N$, and~$(\cdots)$ stands for a differential operator
acting on the Lagrangian. The structure of such surface layer integrals can be understood most easily 
in the special situation that the Lagrangian is of short range
in the sense that~$\L(\x,\y)$ vanishes unless~$\x$ and~$\y$ are close together.
In this situation, we get a contribution to the double integral~\eqref{IntrOSI} only
if both~$\x$ and~$\y$ are close to the boundary~$\partial \Omega$.
With this in mind, surface layer integrals can be understood as an adaptation
of surface integrals to the setting of causal variational principles
(for a more detailed explanation see~\cite[Section~2.3]{noether}).

Surface layer integrals were first introduced in~\cite{noether} in order to
formulate Noether-like theorems for causal variational principles.
In particular, it was shown that there is a conserved
surface layer integral which generalizes the Dirac current conservation
in relativistic quantum mechanics (see~\cite[Section~5]{noether}).
More recently, in~\cite{jet} another conserved
surface layer integral was discovered which gives rise to a symplectic form on the
solutions of the linearized field equations (see~\cite[Sections~3.3 and~4.3]{jet}).
A systematic study of conservation laws for surface layer integrals is given in~\cite{osi}.
The conservation law which is most relevant for our purposes is summarized in the next lemma.
For a compact subset~$\Omega \subset N$ and a jet~$\u \in \J^1$ we introduce the surface
layer integral
\beq \label{gosidef}
\gamma^\Omega_\mu(\v) := \int_\Omega \!d\mu(\x) \int_{N \setminus \Omega} \!\!\!\!\!\!\!d\mu(\y)\: 
\big( \nabla_{1,\u} - \nabla_{2,\u} \big) \L(\x,\y)
\eeq

\begin{Lemma} \label{lemmaosiconserve}
For every compact~$\Omega \subset N$ and any linearized solution~$\u \in \Jlin$,
\beq
\gamma^\Omega_\mu(\v)
= \int_\Omega \nabla_\u \, \s \: d\mu \:. \label{I1osi}
\eeq
\end{Lemma}
\Proof In view of the anti-symmetry of the integrand,
\[ \int_\Omega d\mu(\x) \int_\Omega d\mu(\y)\: 
\big( \nabla_{1,\u} - \nabla_{2,\u} \big) \L(\x,\y) = 0 \:. \]
Adding this equation to the left side of~\eqref{I1osi}, we obtain
\begin{align*}
&\int_\Omega \!d\mu(\x) \int_{N \setminus \Omega} \!\!\!\!\!\!\!d\mu(\y)\: 
\big( \nabla_{1,\u} - \nabla_{2,\u} \big) \L(\x,\y) \\
&= \int_\Omega \!d\mu(\x) \int_N \!d\mu(\y)\: 
\big( \nabla_{1,\u} - \nabla_{2,\u} \big) \L(\x,\y) \\
&= \int_\Omega d\mu(\x) \bigg( 2\, \nabla_\u \Big(\ell(\x) + \s \Big) 
- \big( \Delta \u \big)(\x) - \nabla_\u \, \s \bigg) \:,
\end{align*}
where in the last line we used the definitions of~$\ell$ and~$\Delta$
(see~\eqref{ldef} and~\eqref{Deldef}). Applying the weak EL equations~\eqref{ELtest}
and the linearized field equations~\eqref{eqlinlip} gives the result.
\QED

\subsubsection{Inner Solutions} \label{secinner}
We again define~$N$ as the support of~$\mu$, \eqref{Ndef}.
Furthermore we make the following simplifying assumption:
\begin{Def} \label{defsms}
Spacetime~$N:= \supp \mu$ has a {\bf{smooth manifold structure}} if
the following conditions hold:
\bitem
\item[\rm{(i)}] $N$ is a $k$-dimensional smooth, oriented and connected submanifold of~$\G$.
\item[\rm{(ii)}] In a chart~$(\x,U)$ of~$N$, the universal measure is absolutely continuous with respect
to the Lebesgue measure with a smooth, strictly positive weight function,
\beq \label{hdef}
d\mu = h(x)\: d^kx \qquad \text{with} \quad h \in C^\infty(N, \R^+) \:.
\eeq
\eitem
\end{Def} \noindent
Let~$v \in \Gamma(N, TN)$ be a vector field. Then, under the above assumptions,
its {\em{divergence}} $\div v \in C^\infty(N, \R)$ can be defined by the relation
\[ \int_N \div v\: \eta(\x)\: d\mu = -\int_N D_v \eta(\x)\: d\mu(\x) \:, \]
to be satisfied by all test functions~$\eta \in C^\infty_0(N, \R)$.
In a local chart~$(\x,U)$, the divergence is computed by
\[ 
\div v = \frac{1}{h}\: \partial_\alpha \big( h\, v^\alpha \big) \]
(where, using the Einstein summation convention, we sum over~$\alpha=1, \ldots, k$).
The jets of the form~$\v:=(\div v, v)$ are of particular significance.
The reason is that, applying the Gauss divergence theorem, integrating its jet derivative
of a compactly supported function gives zero, i.e.\ for for every~$f \in C^1_0(N, \R)$
\[ \int_N \nabla_\v f\: d\mu = 
\int_N \big( \div v\: f + D_v f \big)\: d\mu 
= \int_N \div \big( f v \big)\: d\mu = 0 \:. \]
Integrating by parts formally, one finds that these jets satisfy the linearized field equations,
\begin{align*}
\la \u, \Delta \v \ra_N &= \nabla_\u \bigg( \int_N \big(\nabla_{1,\v} + \nabla_{2,\v} \big) \L(\x,\y)\: d\mu(\y)
- \nabla_\v \,\s \bigg) \\
&= \nabla_\u \bigg( \int_N \nabla_{1,\v} \L(\x,\y)\: d\mu(\y)
- \nabla_\v \,\s \bigg) \\
&= \nabla_\u \nabla_\v \ell(\x) = \nabla_\v \big( \nabla_\u \ell(\x) \big) 
- \nabla_{D_v \u} \ell(\x) = 0 \:.
\end{align*}
In the last step we used that~$\nabla_{D_v \u} \ell(\x)$ vanishes by the EL equations.
Moreover, the function~$\nabla_\u \ell$ vanishes identically
on~$M$ in view of the weak EL equations. Therefore, it is differentiable in the direction
of every vector field on~$M$, and this directional derivative is zero.

The above formal computation has two shortcoming. First,
it is a-priori not clear whether integrating by parts gives boundary terms.
Moreover, we need to be careful because the individual derivatives do not need to exist.
This is why, in order to give the above computation a mathematical meaning,
we need to impose additional technical assumptions. We now specify these assumptions
and prove that the resulting jets are indeed solutions of the linearized field equations.
For technical simplicity, we restrict attention to the case that~$\mu$ has a smooth manifold structure
(see Definition~\ref{defsms}) and one asymptotic
end (see Definition~\ref{defae}); for a more general presentation see~\cite[Section~3]{fockbosonic}.
In preparation, we let~$\Gamma_\x$ be the subspace of the tangent space spanned by the test jets,
\beq \label{Gxdef}
\Gamma_\x := \big\{ u(\x) \:|\: u \in \Gtest \big\} \;\subset\; T_\x\G\:.
\eeq
Similar to our assumption that the scalar components of the test jets
is nowhere trivial~\eqref{Cnontriv}, it is sensible and useful to assume that this
subspace of the tangent space contains all the tangent vectors to~$M$,
\beq \label{Tinclude}
T_\x M \subset \Gamma_\x \qquad \text{for all~$\x \in M$}\:.
\eeq
We introduce a Riemannian metric~$g_\x$ on~$\J_\x$.
This Riemannian metric also induces a pointwise scalar product on the jets. Namely, setting
\beq \label{Jxdef}
\J_\x := \R \oplus \Gamma_\x \:,
\eeq
we obtain the scalar product on~$\J_\x$
\beq
\la \v, \tilde{\v} \ra_\x \,:\, \J_\x \times \J_\x \rightarrow \R \:,\qquad
\la \v, \tilde{\v} \ra_\x := b(\x)\, \tilde{b}(\x) + g_\x \big(v(\x),\tilde{v}(\x) \big) \:. \label{vsprod}
\eeq
We denote the corresponding norm by~$\|.\|_\x$.
We assume that this Riemannian metric is {\em{adapted in the asymptotic end}} in
the sense that its restriction to~$TN$ is equivalent to the Euclidean metric in the asymptotic end, i.e.\
in the chart~$\Phi$ in Definition~\ref{defae} there is a constant~$C>0$ such that
\beq \label{adapted}
\frac{1}{C} \sum_{j=1}^k \big|u^j \big|^2 \leq g_\x(u,u) \leq C \sum_{j=1}^k \big|u^j \big|^2 \qquad
\text{for all~$\x \in N \setminus I$ and~$u \in T_\x N$}\:.
\eeq

\begin{Def} \label{definner} An {\bf{inner solution}} is a jet~$\v$ of the form
\[ \v = (\div v, v) \qquad \text{with} \qquad v \in \Gamma(N, TN) \:. \]
We make the following regularity and decay assumptions:
\bitem
\item[{\rm{(i)}}] The vector field~$v$
can be extended to a vector field~$\tilde{v} \in \Gamma(U, T\G)$ defined in a neighborhood~$U$ of~$N$
such that the directional derivative~$(D_{1,\tilde{v}} + D_{2,\tilde{v}}) \L(\x,\y)$
exists for all~$\x \in U$ and~$\y \in N$ and is integrable in~$\y$, i.e.\
\[ \int_N \Big| \big(D_{1,\tilde{v}} + D_{2,\tilde{v}} \big) \L(\x,\y) \Big| \: d\mu(\y) < \infty \qquad
\text{for all~$\x \in U$}\:. \]
Moreover, the directional derivative~$D_{\tilde{v}} \ell(\x)$ exists for all~$\x \in U$ and is continuous in~$U$.
\item[{\rm{(ii)}}] The integral
\[ \int_N \L(\x,\y)\: \|\v(\y)\|_\y \:d\mu(\y) \]
is finite and bounded locally uniformly in a neighborhood of~$N$
(where~$\|.\|_\y$ is again the norm corresponding to the scalar product~\eqref{vsprod}
adapted in the asymptotic end according to~\eqref{adapted}).
\item[{\rm{(iii)}}] For any test jet~$\u \in \Jtest$, the
directional derivative~$D_v \u$ (computed in the same charts used for computing the higher derivatives in Definition~\ref{defJvary}) is again in~$\Jtest$.
\eitem
The vector space of all inner solutions is denoted by~$\Jin$.
\end{Def} \noindent
Note that~(i) implies that every inner solution is in~$\J^1 \cap \Jdiff$
(see~\eqref{Jdiffdef} and Definition~\ref{defJvary}).

The name ``inner {\em{solution}}'' is justified by the following lemma:
\begin{Lemma} Every inner solution~$\v \in \Jin$ is a solution of the linearized field equations, i.e.\
\[ \la \u, \Delta \v \ra|_N = 0 \qquad \text{for all~$\u \in \Jtest_\mu$} \:. \]
\end{Lemma}
\Proof We choose a monotone decreasing function~$\eta \in C^\infty_0(\R)$
with
\[ \eta_{[0,1]} \equiv 1 \qquad \text{and} \qquad \supp \eta \subset (-2,2) \:. \]
For~$n \in \N$, we introduce the compactly supported cutoff functions
\[ \eta_n(\x) := \bigg\{ \begin{array}{cl}
1 & \text{if~$\x \in I$} \\
\displaystyle \eta\bigg( \frac{\|\Phi(\x)\|_{\R^k} - R}{n} \bigg) & \text{if~$\x \in N \setminus I$}
\end{array} \]
(where~$R$ is the radius of the ball in~\eqref{Phidef}).
By construction, the functions~$\eta_n$ are monotone increasing and exhaust~$N$
in the sense that for any compact set~$K \subset N$ there is~$N$ with~$\eta_n|_K \equiv 1$
for all~$n \geq N$. Moreover, using~\eqref{adapted}, 
the derivatives tend uniformly to zero, i.e.
\beq \label{normadapted}
\lim_{n \rightarrow \infty}\: \sup_{\x \in N} \|D \eta_n(\x)\|_\x = 0 \:.
\eeq

Our first goal is to prove that for any~$\x \in U$,
\beq \label{pint}
\int_N \big( D_{1,\tilde{v}} + \nabla_{2,\v} \big) \L(\x,\y) \: d\mu(\y) = 
D_{\tilde{v}} \big( \ell + \s \big)(\x) \:.
\eeq
Making use of Definition~\ref{definner}~(i), we know
from Lebesgue's dominated convergence theorem that for any~$\x \in U$,
\[ A(\x) := \int_N \big( D_{1,\tilde{v}} + \nabla_{2,\tilde{\v}} \big) \L(\x,\y) \: d\mu(\y) = 
\lim_{n \rightarrow \infty} \int_N \big( D_{1,\tilde{v}} + \nabla_{2,\tilde{\v}} \big) \L(\x,\y)\:
\eta_n(\y)\: d\mu(\y) \:. \]
Now we can integrate by parts to obtain
\beq \label{Aform}
A(\x) =  \lim_{n \rightarrow \infty} \bigg( D_{\tilde{v}} \int_N \L(\x,\y)\: \eta_n(\y)\: d\mu(\y) 
- \int_N \L(\x,\y)\: \big( D_v \eta_n(\y)\big)\: d\mu(\y) \bigg)
\eeq
(here one needs to pull out the derivative~$D_{\tilde{v}}$ before the integral, because
the Lagrangian need not be differentiable; the integral, on the other hand, is well-defined because
the last integral is).
The last integral can be estimated by
\beq \label{intes}
\bigg| \int_N \L(\x,\y) \: \big( D_v \eta_n(\y) \big)\: d\mu(\y) \bigg|
\leq \sup_N \|D_v \eta_n\| \int_N \|v(\y)\|_\y \:\L(\x,\y) \: d\mu(\y) \:.
\eeq
According to Definition~\ref{definner}~(ii), the obtained integral is bounded locally
uniformly in~$\x$. Using~\eqref{normadapted}, we conclude that
the last integral in~\eqref{Aform} tends to zero as~$n \rightarrow 0$, locally uniformly in~$\x$.

As a consequence, also the first integral in~\eqref{Aform} converges as~$n \rightarrow 0$, locally uniformly in~$\x$.
In order to prove~\eqref{pint}, it remains to show that this limit is given by
\beq \label{limn}
\lim_{n \rightarrow \infty} \bigg( D_{\tilde{v}} \int_N \L(\x,\y)\: \eta_n(\y)\: d\mu(\y) \bigg) =
D_{\tilde{v}} \big( \ell + \s \big)(\x) \:.
\eeq
Assume conversely that this equation does not hold for all~$\x \in U$. Then by continuity
(note that the left side of~\eqref{limn} is continuous as a locally uniform limit of continuous
function, as is the right side by Definition~\ref{definner}~(i)), the equation~\eqref{limn} is violated in an open set.
Therefore, we may
choose a path~$\gamma : [t_0, t_1] \rightarrow U$ along the integral curves of~$\tilde{v}$ such that
\[ \int_{t_0}^{t_1} \lim_{n \rightarrow \infty} \bigg( D_{\tilde{v}} \int_N \L\big(\gamma(t),\y \big)\: \big(1-\eta_n(\y)\big)\: d\mu(\y) \bigg)\: dt
\neq 0 \:. \]
Due to the locally uniform convergence, we may interchange the integral and the limit to conclude that
\[ \lim_{n \rightarrow \infty} \int_N \L(\x,\y)\: \big(1-\eta_n(\y)\big)\: d\mu(\y) \Big|^{\x=\gamma(t_1)}_{\x=\gamma(t_0)} 
\neq 0 \:. \]
On the other hand, using assumption~(iv) on page~\pageref{Cond4}, the limits on the left
vanish using Lebesgue's dominated convergence theorem. This is a contradiction.
Hence~\eqref{limn} holds. This concludes the proof of~\eqref{pint}.

We rewrite~\eqref{pint} as
\[ \Delta \tilde{\v}(\x) = \nabla_{\tilde{\v}} \ell(\x) \qquad \text{for all~$\x \in U$} \]
(where the scalar component of~$\v$ can be extended to~$U$ arbitrarily).
The next and final step is to show that for any~$\u \in \Jtest$ and~$\x \in N$, the jet derivative~$\nabla_\u$
of this equation exists and vanishes. To this end, we write the jet derivative of the right side as
\[ \nabla^2\ell|_\x(\u,\v) = \nabla_{\v(\x)} \big( \nabla_\u \ell(\x)\big) - \nabla_{D_v \u} \ell(\x) \]
(where the first summand on the right is an iterated directional derivative).
The last summand vanishes because of the weak EL equations,
using that~$D_v \u \in \Jtest$ (see Definition~\ref{definner}~(iii)).
In order to treat the first summand, we note that the function~$\nabla_\u \ell$ vanishes identically
on~$N$ by the weak EL equations. Therefore, this function is differentiable in the direction
of every vector field on~$N$, and this directional derivative is zero.
This concludes the proof.
\QED

We next show that for any function~$a$ on~$N$ one can find an inner solution
whose scalar component coincides with~$a$. If~$N$ were compact, the analogous
statement would be the infinitesimal version of Moser's theorem (see for
example~\cite[Section~XVIII, \S2]{langDG}).
Here we give a detailed proof if~$N$ is non-compact with one asymptotic end,
based on~\cite[Theorem~1.2 in Section~XVIII]{langDG}. 

\begin{Lemma} \label{lemmamoser}
Assume that~$N:= \supp \nu$ has a smooth manifold structure
(see Definition~\ref{defsms}) and one asymptotic end
(see Definition~\ref{defae}).
Then for any~$a \in C^\infty(N)$ there is a vector field~$v \in \Gamma(N, TN)$
such that~$\div v = a$.
\end{Lemma}
\Proof We choose a partition of unity~$(\phi_n)_{n \in \N}$ of~$N$ supported in
annuli of our coordinate system~$\Phi$. More precisely,
\begin{align*}
\supp \phi_1 &\subset I \cup \Phi^{-1} \big( B_{R+2} \big) \qquad \text{and} \\
\supp \phi_n &\subset \Phi^{-1} \big( B_{R+n+1} \setminus \overline{B_{R+n-1}} \,\big) \qquad
\text{for~$n \geq 2$} \:.
\end{align*}
Due to the smoothness assumption~\eqref{hdef},
the measure~$\mu$ can be represented by a volume form~$\psi \in \Lambda^k(N)$, i.e.
\[ \mu(U) = \int_U \psi \qquad \text{for all compact~$U \subset N$}\:. \]
Likewise, the measure~$a \mu$ can be represented by a volume form~$\omega \in \Lambda^k(N)$, i.e.
\[ \int_U a(\x)\: d\mu(\x) = \int_U \omega  \]
(again valid for all compact~$U \subset N$). We now proceed inductively: We choose a real number~$c_1$ such that
\[ \int_N \big( \phi_1 \,\omega - c_1\: \phi_1\, \psi \big) = 0 \:. \]
According to~\cite[Theorem~1.2 in Section~XVIII]{langDG},
there is a compactly supported $(k-1)$-form~$\eta_1 \in \Lambda^{k-1}_0(N)$ such that
\[ \phi_1 \,\omega - c_1\: \phi_1\, \psi = d\eta_1 \:. \]
In the induction step from~$n$ to~$n+1$ we choose~$c_{n+1} \in \R$ such that
\[ \int_N \: \Big( \phi_{n+1}\, \omega + c_n\, \phi_n\, \psi - c_{n+1}\, \phi_{n+1}\,\psi \big) = 0 \:. \]
Then there exists~$\eta_{n+1} \in \Lambda^{k-1}_0(N)$ with
\[ \phi_{n+1}\, \omega + c_n\, \phi_n\, \psi - c_{n+1}\, \phi_{n+1}\,\psi = d\eta_{n+1} \:. \]
By applying~\cite[Theorem~1.2 in Section~XVIII]{langDG} on the manifold chosen as the above annuli,
one sees that the support of the~$\eta_n$ can be arranged to also lie in these annuli, i.e.\
\begin{align*}
\supp \eta_1 &\subset I \cup \Phi^{-1} \big( B_{R+2} \big) \qquad \text{and} \\
\supp \eta_n &\subset \Phi^{-1} \big( B_{R+n+1} \setminus \overline{B_{R+n-1}} \,\big) \qquad
\text{for~$n \geq 2$}
\end{align*}
(here we make use of the fact that all the annuli are connected).

Being locally finite, we can carry out the sum over~$n$.
The summands involving~$\phi_n \psi$ cancel. We conclude that
\[ \eta := \sum_{n=1}^\infty \eta_n \qquad \text{satisfies} \qquad
d\eta = \sum_{n=1}^\infty \phi_n\, \omega = \omega \:. \]

Finally, we need to identify~$\eta$ with a vector field such that~$\div v\: d\mu = d\eta$.
To this end, we choose a Riemannian metric~$g$ on~$N$. By a conformal transformation
we can arrange that the corresponding volume form coincides with the measure~$\mu$.
Now we choose the vector field as
\[ v^\alpha = g^{\alpha \beta}\: (* \,\eta )_\beta \:, \] 
where~$* : \Lambda^{k-1}(N) \rightarrow \Lambda^1(N)$ is the Hodge star. This concludes the proof.
\QED
We point out that in general, the vector field~$v$ constructed in this lemma is not compactly supported,
even if the function~$a$ is. In particular, if the integral of~$a$ is non-zero, then the Gauss
divergence theorem implies that the flux of~$v$ through large coordinate spheres
must be non-zero.

In what follows, we always assume that the vector field constructed in this lemma satisfies all the
regularity and decay assumptions in Definition~\ref{definner}. We then obtain a corresponding inner solution
\[ \v := (a, v) \in \Jin \:. \]
In view of Lemma~\ref{lemmamoser}, given any jet we can arrange by adding a suitable
inner solution that the scalar component of the jet vanishes. 
With this in mind, in what follows
we may always restrict attention to jets with vanishing scalar component.
However, one must keep in mind that the resulting vector components will not be zero near infinity.
Indeed, the asymptotics of these vector fields will encode the total mass.

\subsubsection{A Nonlinear Surface Layer Integral} \label{secosinonlin}
We finally mention another surface layer integral which was first introduced in~\cite{fockbosonic}.
Instead of working with linearized solutions, we directly compare the perturbed measure~$\tilde{\mu}$,
which takes into account the nonlinear interaction, with a vacuum measure~$\mu$. 

\begin{Def} \label{defosinl}
We let~$\mu$ and~$\tilde{\mu}$ be two Borel measures on~$\G$ and set~$N := \supp \mu,
\tilde{N}:= \supp \tilde{\mu}$.
Given compact subsets~$\Omega \subset N$ and~$\tilde{\Omega} \subset \tilde{N}$,
the {\bf{nonlinear surface layer integral}}
$\gamma^{\tilde{\Omega}, \Omega}(\tilde{\mu}, \mu)$ is defined by
\[ 
\gamma^{\tilde{\Omega}, \Omega}(\tilde{\mu}, \mu) =
\int_{\tilde{\Omega}} \!d\tilde{\mu}(\x) \int_{N\setminus \Omega} \!\!\!\!\!\!\!d\mu(\y)\: \L(\x,\y)
- \int_{\Omega} \!d\mu(\x) \int_{\tilde{N} \setminus \tilde{\Omega}}  \!\!\!\!\!\!\!\!d\tilde{\mu}(\y)\: \L(\x,\y) \:. \]
\end{Def} \noindent
In~\cite[Section~4 and Appendix~A]{fockbosonic} a conservation law for this nonlinear surface layer integral
was derived, and it was used in order to rewrite the dynamics with a 
norm-preserving linear operator on Fock spaces.
This nonlinear surface layer integral was also our starting point when searching for the
right definition of the total mass. Indeed, the total mass in~\eqref{massgen}
can be written in the short form
\beq \label{Mshort}
\Mass(\tilde{\mu}, \mu) := \lim_{\Omega \nearrow N} \;\lim_{\tilde{\Omega} \nearrow \tilde{N}} 
\bigg( -\s \Big( \tilde{\mu}(\tilde{\Omega}) - \mu(\Omega) \Big) 
+ \gamma^{\tilde{\Omega}, \Omega}(\tilde{\mu}, \mu) \bigg) \:.
\eeq

\subsection{Causal Fermion Systems and the Causal Action Principle} \label{seccfs}
We now recall the basic definitions of a causal fermion system and the causal action principle.
The connection to causal variational principles will be made in the static setting
in Section~\ref{secregular}.

\begin{Def} \label{defcfs} (causal fermion system) {\em{ 
Given a separable complex Hilbert space~$\H$ with scalar product~$\la .|. \ra_\H$
and a parameter~$n \in \N$ (the {\em{``spin dimension''}}), we let~$\F \subset \Lin(\H)$ be the set of all
symmetric\footnote{Here by a symmetric operator~$A$ we mean that~$\la A u | v \ra_\H =
\la u | A v \ra_\H$ for all~$u,v \in \H$. Representing the operator in
an orthonormal basis, the resulting matrix is Hermitian.
For bounded operators as considered here,
the notions ``symmetric'' and ``self-adjoint'' coincide.} 
operators on~$\H$ of finite rank, which (counting multiplicities) have
at most~$n$ positive and at most~$n$ negative eigenvalues. On~$\F$ we are given
a positive measure~$\rho$ (defined on a $\sigma$-algebra of subsets of~$\F$), the so-called
{\em{universal measure}}. We refer to~$(\H, \F, \rho)$ as a {\em{causal fermion system}}.
}}
\end{Def} \noindent
A causal fermion system describes a spacetime together
with all structures and objects therein.
In order to single out the physically admissible
causal fermion systems, one must formulate physical equations. To this end, we impose that
the universal measure should be a minimizer of the causal action principle,
which we now introduce. For any~$x, y \in \F$, the product~$x y$ is an operator of rank at most~$2n$. 
However, in general it is no longer a symmetric operator because~$(xy)^* = yx$,
and this is different from~$xy$ unless~$x$ and~$y$ commute.
As a consequence, the eigenvalues of the operator~$xy$ are in general complex.
We denote these eigenvalues counting algebraic multiplicities
by~$\lambda^{xy}_1, \ldots, \lambda^{xy}_{2n} \in \C$
(more specifically,
denoting the rank of~$xy$ by~$k \leq 2n$, we choose~$\lambda^{xy}_1, \ldots, \lambda^{xy}_{k}$ as all
the non-zero eigenvalues and set~$\lambda^{xy}_{k+1}, \ldots, \lambda^{xy}_{2n}=0$).
We introduce the Lagrangian and the causal action by
\begin{align}
\text{\em{Lagrangian:}} && \L(x,y) &= \frac{1}{4n} \sum_{i,j=1}^{2n} \Big( \big|\lambda^{xy}_i \big|
- \big|\lambda^{xy}_j \big| \Big)^2 \label{Lagrange} \\
\text{\em{causal action:}} && \Sact(\rho) &= \iint_{\F \times \F} \L(x,y)\: d\rho(x)\, d\rho(y) \:. \label{Sdef}
\end{align}
The {\em{causal action principle}} is to minimize~$\Sact$ by varying the measure~$\rho$
under the following constraints:
\begin{align}
\text{\em{volume constraint:}} && \rho(\F) = \text{const} \quad\;\; & \label{volconstraint} \\
\text{\em{trace constraint:}} && \int_\F \tr(x)\: d\rho(x) = \text{const}& \label{trconstraint} \\
\text{\em{boundedness constraint:}} && \iint_{\F \times \F} 
|xy|^2
\: d\rho(x)\, d\rho(y) &\leq C \:, \label{Tdef}
\end{align}
where~$C$ is a given parameter, $\tr$ denotes the trace of a linear operator on~$\H$, and
the absolute value of~$xy$ is the so-called spectral weight,
\[ |xy| := \sum_{j=1}^{2n} \big|\lambda^{xy}_j \big| \:. \]
This variational principle is mathematically well-posed if~$\H$ is finite-dimensional.
For the existence theory and the analysis of general properties of minimizing measures
we refer to~\cite{discrete, continuum, lagrange}.
In the existence theory one varies in the class of regular Borel measures
(with respect to the topology on~$\Lin(\H)$ induced by the operator norm),
and the minimizing measure is again in this class. With this in mind, here we always assume that
\beq \label{regular}
\text{$\rho$ is a regular Borel measure}\:.
\eeq

Let~$\rho$ be a {\em{minimizing}} measure. {\em{Spacetime}}
is defined as the support of this measure,
\[ 
M := \supp \rho \:. \]
Thus the spacetime points are symmetric linear operators on~$\H$.
These operators contain a lot of additional information which, if interpreted correctly,
gives rise to spacetime structures like causal and metric structures, spinors
and interacting fields. We refer the interested reader to~\cite[Chapter~1]{cfs}.

The only results on the structure of minimizing measures
which will be needed in what follows concern the treatment of the
trace constraint and the boundedness constraint.
As a consequence of the trace constraint, for any minimizing measure~$\rho$
the local trace is constant in spacetime, i.e.\
there is a real constant~$c \neq 0$ such that (see~\cite[Theorem~1.3]{lagrange} or~\cite[Proposition~1.4.1]{cfs})
\beq \label{trc}
\tr x = c \qquad \text{for all~$x \in M$} \:.
\eeq
Restricting attention to operators with fixed trace, the trace constraint~\eqref{trconstraint}
is equivalent to the volume constraint~\eqref{volconstraint} and may be disregarded.
The boundedness constraint, on the other hand, can be treated with a Lagrange multiplier.
More precisely, in~\cite[Theorem~1.3]{lagrange} it is shown that for every minimizing measure~$\rho$, 
there is a Lagrange multiplier~$\kappa>0$ such that~$\rho$ is a critical point
of the causal action
with the Lagrangian replaced by
\[ \L_\kappa(x,y) := \L(x,y) + \kappa\, |xy|^2 \:, \]
leaving out the boundedness constraint.

\subsection{Constructing Causal Fermion Systems in Static Lorentzian Spacetimes} \label{seclorentz}
\subsubsection{Construction in Globally Hyperbolic Spacetimes}
We now recall how, starting from a globally Lorentzian spacetime,
one can construct corresponding causal fermion systems. The general method is to
choose~$\H$ as a subspace of the solution space of the Dirac equation, as we now
explain (for more details see~\cite{nrstg}).
Let~$(\scrM, g)$ be a smooth, globally hyperbolic Lorentzian manifold of dimension~$k \geq 2$.
For the signature of the metric we use the convention~$(+ ,-, \ldots, -)$.
As proven in~\cite{bernal+sanchez}, $\scrM$ admits a smooth foliation~$(\scrN_t)_{t \in \R}$
by Cauchy hypersurfaces. Thus~$\scrM$ is topologically the product of~$\R$ with a $k-1$-dimensional manifold.
In the case~$k=4$ of a four-dimensional spacetime, this implies that~$\scrM$ is spin
(for details see~\cite{baum, lawson+michelsohn}).
For a general spacetime dimension we need to impose that~$\scrM$ is spin.
We let~$S\scrM$ be the spinor bundle on~$\scrM$ and denote the smooth sections of the spinor bundle
by~$C^\infty(\scrM, S\scrM)$. Similarly, $C^\infty_0(\scrM, S\scrM)$ denotes the smooth sections with compact support.
The sections of the spinor bundle are also referred to as wave functions.
The fibres~$S_x\scrM$ are endowed with an inner product of signature~$(n,n)$
with~$n=2^{[k/2]-1}$ (where~$[\cdot]$ is the Gau{\ss} bracket; for details see
again~\cite{baum, lawson+michelsohn}),
which we denote by~$\Sl .|. \Sr_x$. The Lorentzian metric induces a Levi-Civita connection
and a spin connection, which we both denote by~$\nabla$.
Every vector of the tangent space acts on the corresponding spinor space by Clifford multiplication.
Clifford multiplication is related to the Lorentzian metric via the anti-commutation relations.
Denoting the mapping from the
tangent space to the linear operators on the spinor space by~$\gamma$, we thus have
\[ \gamma \::\: T_x\scrM \rightarrow \Lin(S_x\scrM) \qquad
\text{with} \qquad \gamma(u) \,\gamma(v) + \gamma(v) \,\gamma(u) = 2 \, g(u,v)\,\1_{S_x(\scrM)} \:. \]
We also write Clifford multiplication in components with the Dirac matrices~$\gamma^j$.
The connections, inner products and Clifford multiplication satisfy Leibniz rules and compatibility
conditions; we refer to~\cite{baum, lawson+michelsohn} for details.

Combining the spin connection with Clifford multiplication gives the geometric Dirac operator
denoted by
\[ 
\Dir := i \gamma^j \nabla_j \::\: C^\infty(\scrM, S\scrM) \rightarrow C^\infty(\scrM, S\scrM)\:. \]
Given a real parameter~$m$ (the ``rest mass''), the Dirac equation reads
\beq \label{dirac}
(\Dir - m) \,\psi = 0 \:.
\eeq
We mainly consider solutions in the class~$\Cisc(\scrM, S\scrM)$ of smooth sections
with spatially compact support. On such solutions, one has the scalar product
\beq \label{print}
(\psi | \phi)_m = 2 \pi \int_\scrN \Sl \psi \,|\, \gamma(\nu)\, \phi \Sr_x\: d\mu_\scrN(x) \:,
\eeq
where~$\scrN$ denotes any Cauchy surface and~$\nu$ its future-directed normal
(due to current conservation, the scalar product is
in fact independent of the choice of~$\scrN$; for details see~\cite[Section~2]{finite}).
Forming the completion gives the Hilbert space~$(\H_m, (.|.)_m)$.

Next, we choose a closed subspace~$\H \subset \H_m$
of the solution space of the Dirac equation.
The induced scalar product on~$\H$ is denoted by~$\la .|. \ra_\H$.
There is the technical difficulty that the wave functions in~$\H$ are in general not continuous,
making it impossible to evaluate them pointwise.
For this reason, we need to introduce an {\em{ultraviolet regularization}} on
the length scale~$\varepsilon$, described mathematically by a linear
\beq \label{regop}
\text{\em{regularization operator}} \qquad {\mathfrak{R}}_\varepsilon \::\: \H_m \rightarrow C^0(\scrM, S\scrM) \:.
\eeq
In the simplest case, the regularization can be realized by a convolution
on a Cauchy surface or in spacetime (for details see~\cite[Section~4]{finite}
or~\cite[Section~\S1.1.2]{cfs}). For us, the regularization is not just a technical tool,
but it realizes the concept that we want to change the geometric structures on the microscopic
scale. With this in mind, we always consider the regularized quantities as those having mathematical and
physical significance. Different choices of regularization operators realize different
microscopic spacetime structures.

Given~${\mathfrak{R}}_\varepsilon$, for any spacetime point~$x \in \scrM$ we consider the sesquilinear form
\[ b_x \::\: \H \times \H \rightarrow \C\:,\quad
b_x(\psi, \phi) = -\Sl ({\mathfrak{R}}_\varepsilon\psi)(x) | ({\mathfrak{R}}_\varepsilon \phi)(x) \Sr_x \:. \]
This sesquilinear form is well-defined and bounded because~${\mathfrak{R}}_\varepsilon$ 
maps to the continuous wave functions and because evaluation at~$x$ gives a linear operator of finite rank.
Thus for any~$\phi \in \H$, the anti-linear form~$b_x(.,\phi) : \H \rightarrow \C$
is continuous. By the Fr{\'e}chet-Riesz theorem,
there is a unique~$\chi \in \H$
such that~$b_x(\psi,\phi) = \la \psi | \chi \ra_\H$ for all~$\psi \in \H$.
The mapping~$\phi \mapsto \chi$ is linear and bounded. We thus obtain a unique bounded linear
operator~$F^\varepsilon(x)$ on~$\H$ which is characterized by the relation
\beq \label{Fepsdef}
(\psi \,|\, F^\varepsilon(x)\, \phi) = -\Sl ({\mathfrak{R}}_\varepsilon\psi)(x) | 
({\mathfrak{R}}_\varepsilon \phi)(x) \Sr_x \qquad \text{for all~$\psi, \phi
\in \H$} \:.
\eeq
Taking into account that the inner product on the Dirac spinors at~$x$ has signature~$(n,n)$,
the local correlation operator~$F^\varepsilon(x)$ is a symmetric operator on~$\H$
of rank at most~$2n$, which (counting multiplicities) has at most~$n$ positive and at most~$n$
negative eigenvalues. Varying the spacetime point, we obtain a mapping
\[ F^\varepsilon \::\: \scrM \rightarrow \F \subset \Lin(\H)\:, \]
where~$\F$ again denotes all symmetric operators of
rank at most four with at most two positive and at most two negative eigenvalues.
Finally, we introduce the
\[ \text{\em{universal measure}} \qquad d\rho := (F^\varepsilon)_* \,d\mu_\scrM \]
as the push-forward of the volume measure on~$\scrM$ under the mapping~$F^\varepsilon$
(thus~$\rho(\Omega) := \mu_\scrM((F^\varepsilon)^{-1}(\Omega))$).

In this way, we obtain a measure~$\rho$ on the set~$\F \subset \Lin(\H)$ of
linear operators on a Hilbert space~$\H$.
The basic concept is to work exclusively with these objects, but to drop all
other structures (like the Lorentzian metric~$g$, the structure of the spinor bundle~$S\scrM$ and
the manifold structure of~$\scrM$). This leads us to the structure of a causal fermion system of spin
dimension~$n$, as defined abstractly in Definition~\ref{defcfs} above.

For clarity, we close with a few comments on the underlying physical concepts.
The vectors in the subspace~$\H \subset \H_m$ have the interpretation
as those Dirac wave functions which are realized in the physical system under
consideration. Therefore, the vectors in~$\H$ are referred to as the
{\em{physical wave functions}}. 
If we describe for example a system of one electron,
then the wave function of the electron is contained in~$\H$. Moreover, $\H$ includes all the wave functions
in~$\H_-$ which form the so-called Dirac sea (for an explanation of this point
see for example~\cite{srev}). The name causal {\em{fermion}} system
is motivated by the fact that Dirac particles are fermions.
According to~\eqref{Fepsdef}, 
the local correlation operator~$F^\varepsilon(x)$ describes 
densities and correlations of the physical
wave functions at the spacetime point~$x$.
Working exclusively with the local correlation operators and the
corresponding push-forward measure~$\rho$ means in particular
that the geometric structures are encoded in and must be retrieved from the physical wave functions.
Since the physical wave functions describe the distribution of
matter in spacetime, one can summarize this concept
by saying that {\em{matter encodes geometry}}.

\subsubsection{Construction in Static Spacetimes} \label{secstaticdirac}
By a {\em{static}} spacetime we here mean a globally hyperbolic spacetime where
the foliation~$(\scrN_t)_{t \in \R}$ can be chosen such that the timelike
vector field~$\partial_t$ is a Killing field which is orthogonal to the hypersurfaces~$\scrN_t$.
A typical example is the Schwarzschild geometry.
Since all Cauchy surfaces~$\scrN_t$ are isometric, we may restrict attention to one of them
and omit the subscript~$t$. 
In a static spacetime, it is most convenient to write the Dirac equation~\eqref{dirac} in the Hamiltonian form
\beq \label{DirHamilton}
i \partial_t \psi = H \psi \:,
\eeq
where~$H$, the {\em{Dirac Hamiltonian}}, is an elliptic operator acting on the spatial
sections~$\Gamma(\scrN, S\scrM)$. For convenience, we choose its domain as the
smooth, compactly supported spinors,
\[ \D(H) = C^\infty_0(\scrN, S\scrM) \:. \]
Identifying the initial data on~$\scrN$ with the corresponding solution of the Cauchy problem,
this domain is a subspace of the Hilbert space~$\H_m$ introduced after~\eqref{print}.
Moreover, the scalar product~\eqref{print} can be expressed in terms of the functions on~$\scrN$ by
\beq \label{printN}
(\psi | \phi)_\scrN = 2 \pi \int_\scrN \Sl \psi \,|\, \gamma(\nu)\, \phi \Sr_x\: d\mu_\scrN(x) \:.
\eeq
As a consequence of current conservation, the Dirac Hamiltonian~$H$ with domain~$\D(H)$
is a symmetric operator on~$\H_m$. Using finite propagation speed,
the general method by Chernoff~\cite{chernoff73} yields that the Dirac Hamiltonian is essentially
self-adjoint (for details in the more general case with boundary conditions see~\cite{chernoff}).
We denote the unique self-adjoint extension again by~$H$.
The spectral theorem gives a decomposition
\beq \label{Dirspec}
H = \int_{\sigma(H)} \omega\: dE_\omega \:,
\eeq
where~$E$ is a projection-valued measure. 
In simple terms, the spectral parameter~$\omega \in \R$ is the frequency of the
usual separation with a plane wave ansatz,
\[ \psi(t, \vec{x}) = e^{-i \omega t}\: \psi_\omega(\vec{x}) \:. \]
Consequently, the solution space~$\H_m$ splits into the direct sum of the solutions of
positive and negative frequency,
\[ \H_m = \H_+ \oplus \H_- \qquad \text{with} \qquad
\H_+ := E_{[0, \infty)}(\H_m),\; \H_- := E_{(-\infty, 0)}(\H_m)\:. \]
In this paper we always choose the subspace~$\H$ used in the above construction of
the causal fermion system as a subspace which differs from~$\H_-$
by a finite-dimensional subspace. More precisely, we assume that there are
finite-dimensional subspaces~$\H_\text{p} \subset \H_+$
and~$\H_\text{ap} \subset \H_- \cap \H^\perp$ such that
\beq \label{HHminus}
\H \oplus \H_\text{ap} = \H_- \oplus \H_\text{p} \:.
\eeq
We refer to the resulting causal fermion systems as {\em{static Dirac systems}}.
We remark that the choice~\eqref{HHminus}
means that we consider a system involving a finite number of particles
and anti-particles (whose states span~$\H_\text{p}$ and~$\H_\text{a}$, respectively).

We finally point out that in general, the static causal fermion systems obtained in this way
are {\em{not}} minimizers of the causal action. But, as worked out in detail in~\cite{cfs},
they are critical points of the causal action in a limiting case where~$\varepsilon \rightarrow 0$,
referred to as the {\em{continuum limit}}, provided that the classical field equations
(Maxwell, Yang-Mills and Einstein equations) hold. Therefore, static Dirac systems
are suitable examples for understanding the connection between the total mass of
static causal fermion systems and the ADM mass.

\section{The Causal Action Principle in the Static Case} \label{seccfsstatic}
We now specialize the setting of causal fermion systems to the static case (Section~\ref{secstatic}).
Adapting the causal action principle to static causal fermion systems and imposing a
regularity condition, we get into the setting of causal variational principles (Sections~\ref{secSstatic}
and~\ref{secregular}). We finally explain the scaling freedom for static causal fermion systems
(Section~\ref{secrescale}).

\subsection{Static Causal Fermion Systems} \label{secstatic}
In this paper, we shall restrict attention to causal fermion systems which are time independent
in the following sense.
\begin{Def} \label{defstatic}
Let~$(\scrU_t)_{t \in \R}$ be a strongly continuous one-parameter group of unitary transformations 
on the Hilbert space~$\H$ (i.e.\ $s$-$\lim_{t' \rightarrow t} \scrU_{t'}= \scrU_t$ and~$\scrU_t \scrU_{t'} = \scrU_{t+t'}$).
The causal fermion system~$(\H, \F, \rho)$ is {\bf{static with respect to~$(\scrU_t)_{t \in \R}$}}
if it has the following properties:
\begin{itemize}[leftmargin=2em]
\item[\rm{(i)}] Spacetime $M:= \supp \rho \subset \F$ is a topological product,
\[ M = \R \times N \:. \]
We write a spacetime point~$x \in M$ as~$x=(t,\x)$ with~$t \in \R$ and~$\x \in N$.
\item[\rm{(ii)}] The one-parameter group~$(\scrU_t)_{t \in \R}$ leaves
the universal measure invariant, i.e.
\[ \rho\big( \scrU_t \,\Omega\, \scrU_t^{-1} \big) = \rho(\Omega) \qquad \text{for all
	$\rho$-measurable~$\Omega \subset \F$} \:. \]
Moreover,
\[ \scrU_{t'}\: (t,\x)\: \scrU_{t'}^{-1} = (t+t',\x)\:. \]
\end{itemize}
\end{Def} \noindent
Before going on, we point out that we here restrict attention to spacetimes of {\em{infinite lifetime}}.
Alternatively, one could also consider static and time-periodic spacetimes, in which case~$M$
would be the topological product~$S^1 \times N$.
We also remark that our definition of ``static'' even
applies to spacetimes like the Kerr geometry which are not static but {\em{stationary}}. In fact, in the
above generality without a Lorentzian metric, it is a-priori not clear how to distinguish between static and
stationary spacetimes.
Since in this paper, we have static spacetimes like the Schwarzschild geometry in mind,
it is more appropriate and more modest to refer to our spacetimes as being static. But clearly, the same definition
could also be used when extending our work to stationary spacetimes.

Given a static causal fermion system, we also consider the set of operators
\[ N := \{ (0, \x) \} \subset \F \:. \]
The universal measure induces a measure~$\mu$ on~$N$ defined by
\[ \mu(\Omega) := \rho\big( [0,1] \times \Omega \big) \:. \]
The fact that the causal fermion system is static implies that~$\rho([t_1,t_2] \times \Omega) = (t_2-t_1)\,
\mu(\Omega)$, valid for all~$t_1<t_2$. This can be expressed more conveniently as
\beq \label{fubini}
d\rho = dt \,d\mu \:.
\eeq

\subsection{The Causal Action Principle in the Static Setting} \label{secSstatic}
The causal action principle can be formulated in a straightforward manner
for static causal fermion systems. The only point to keep in mind is that, when considering
families of measures, these measures should all be static with respect to the same
group~$(\scrU_t)_{t \in \R}$ of unitary operators (see Definition~\ref{defstatic}).
In order to make this point clear, right from the beginning we
choose a group of unitary operators~$(\scrU_t)_{t \in \R}$ on~$\H$.
We denote the equivalence classes of~$\F$ under the action of the one-parameter group by
\[ \F/\R := \{ \scrU_t \,x\, \scrU_t^{-1} \:|\: x \in \F, t \in \R \} \:. \]
We denote the elements of~$\F/\R$ just as the spatial points by~$\x$ and~$\y$.
Next, we define the following functions:
\begin{align}
\text{\em{static Lagrangian}} &&
\L(\x, \y) &:= \int_{-\infty}^\infty \L \big((0,\x), (t,\y) \big) \: dt \label{Ltime} \\
\text{\em{static boundedness function}}  && \T(\x, \y) &:= \int_{-\infty}^\infty \big| (0,\x)\, (t,\y) \big|^2\: dt \label{Ttime} \\
\text{\em{static $\kappa$-Lagrangian}}  && \L_\kappa(\x, \y) &:= \L(\x, \y) + \kappa\, \T(\x, \y) \:. \label{kLtime}
\end{align}
Due to the unitary invariance of the Lagrangian, this definition does not depend on the
choice of representatives. Moreover, the static Lagrangian is again symmetric because
\begin{align}
\L(\x, \y) &= \int_{-\infty}^\infty \L \big((0,\x), \scrU_t \,(0,\y)\, \scrU_t^{-1} \big) \: dt
= \int_{-\infty}^\infty \L \big(\scrU_t \,(0,\y)\, \scrU_t^{-1},\, (0,\x) \big) \: dt \notag \\
&= \int_{-\infty}^\infty \L \big(\scrU_t^{-1} \,\scrU_t \,(0,\y)\, \scrU_t^{-1}\, \scrU_t,\, \scrU_t^{-1} \,(0,\x)\,
\scrU_t \big) \: dt \notag \\
&= \int_{-\infty}^\infty \L \big((0,\y),\, \scrU_t^{-1} \,(0,\x)\,\scrU_t \big) \: dt
= \int_{-\infty}^\infty \L \big((0,\y),\, (-t,\x) \big) \: dt \notag \\
&= \Big\{ t \rightarrow -t \Big\} 
= \int_{-\infty}^\infty \L \big((0,\y), (t,\x) \big) \: dt = \L(\y, \x) \:, \label{sLsymm}
\end{align}
and similarly for~$\T(x,y)$.
For a measure~$\rho$ which is static with respect to~$(\scrU_t)_{t \in \R}$, we introduce the
\begin{align}
\text{\em{static causal action:}} && \Sact(\rho) &= \int_{\F/\R} d\mu(\x) \int_{\F/\R} d\mu(\y) \:\L_\kappa(\x, \y) \:.\label{Sdefstatic}
\end{align}
The {\em{static causal action principle}} is to minimize~$\Sact$ by varying the measure~$\rho$
with in the class of measures which are static with respect to~$(\scrU_t)_{t \in \R}$
under the following constraints:
\begin{align}
\text{\em{volume constraint:}} && \mu(\F/\R) = \text{const} \quad\;\; & \label{volconstraintstatic} \\
\text{\em{trace constraint:}} && \int_{\F/\R} \tr(\x)\: d\mu(\x) = \text{const} \:. & \label{trconstraintstatic}
\end{align}
Note that the boundedness constraint is taken into account by the Lagrange multiplier term~$\kappa \T(x,y)$
in~\eqref{kLtime}.

\subsection{The Regular Setting as a Causal Variational Principle} \label{secregular}
We now explain how to get to the setting of causal variational principles introduced
in Section~\ref{seccvp}. The main differences between the static causal action principle
and the setting of causal variational principles is that the set~$\F/\R$ does not need to be a manifold
and that there is the additional trace constraint~\eqref{trconstraintstatic}.
As explained after~\eqref{trc}, the trace constraint can be treated by restricting attention
to operators of fixed trace. In order to give the set of operators a manifold structure,
we assume that~$\rho$ is {\em{regular}} in the sense that all operators in its support
have exactly~$n$ positive and exactly~$n$ negative eigenvalues.
This leads us to introduce the set~$\F^\text{reg}$ 
as the set of all operators~$F$ on~$\H$ with the following properties:
\begin{itemize}[leftmargin=2em]
\item[(i)] $F$ is symmetric, has finite rank and (counting multiplicities) has
exactly~$n$ positive and~$n$ negative eigenvalues. \\[-0.8em]
\item[(ii)] The trace is constant, i.e.~$\tr(F) = c>0$.
\end{itemize}
If~$\H$ is finite-dimensional, the set~$\F^\text{reg}$ has a smooth manifold structure
(see the concept of a flag manifold in~\cite{helgason} or the detailed construction
in~\cite[Section~3]{gaugefix}).
Assuming that the action of the group~$(\scrU_t)_{t \in \R}$ on~$\G$ is proper and has no fixed points,
the quotient is again a manifold. Thus setting
\[ \G = \F^\text{reg} / \R \:, \]
we get into the setting of causal variational principles as introduced in Section~\ref{seccvp}.
We now verify that also the technical assumptions are satisfied.
\begin{Prp} \label{prpreg} Assume that~$\H$ is finite-dimensional and that the action of the
group~$(\scrU_t)_{t \in \R}$ on~$\G$ is proper and has no fixed points.
Then for every minimizing measure, the conditions~{\rm{(i)}}--{\rm{(iv)}}
in Section~\ref{seccvp} on page~\pageref{Cond1} are satisfied for the static $\kappa$-Lagrangian.
\end{Prp}
\Proof The symmetry property~(i) was already verified for the
static Lagrangian~\eqref{kLtime} in~\eqref{sLsymm}.
Since the eigenvalues of a matrix depend continuously on the matrix entries
(see for example~\cite[\S II.1]{kato}), the Lagrangian~$\L_\kappa(x,y)$ is continuous
in both arguments. However, this does not mean that the static Lagrangian~\eqref{kLtime}
is also continuous. But Fatou's lemma yields
\begin{align*}
\L_\kappa(\x,\y) &= \int_{-\infty}^\infty \L_\kappa\big( (0,\x), (t, \y) \big) \: dt
= \int_{-\infty}^\infty \lim_{\x' \rightarrow \x} \L_\kappa\big( (0,\x'), (t, \y) \big) \: dt \\
&\leq \liminf_{\x' \rightarrow \x} \int_{-\infty}^\infty \L_\kappa\big( (0,\x'), (t, \y) \big) \: dt
= \liminf_{\x' \rightarrow \x} \L_\kappa(\x', \y) \:,
\end{align*}
showing that the static $\kappa$-Lagrangian is lower semi-continuous. This proves~(ii).

In order to prove~(iii) and~(iv), we make use of the EL equations~\eqref{EL}, which
we write as
\beq \label{ELstar}
\int_N \L_\kappa(\x,\y)\: d\mu(\y) \leq \s \qquad \text{for all~$x \in \G$}\:.
\eeq
The regularity of~$\mu$ follows immediately from the regularity of~$\rho$ in~\eqref{regular}.
Using that the trace of~$x$ is non-zero, the Lagrangian~$\L_\kappa(x, x)$ is strictly positive on the
diagonal (for details see~\cite[Proposition~4.3]{discrete}). By continuity of the Lagrangian,
also the static Lagrangian is strictly positive on the diagonal,~$\L_\kappa(\x,\x)>0$.
By lower semi-continuity of the static Lagrangian, there is~$\varepsilon>0$ and
an open neighborhood~$U$ of~$x$ where~$\L_\kappa(\x,.)$ is larger than~$\varepsilon$.
Combining this fact with the positivity of the Lagrangian, the inequality~\eqref{ELstar} gives rise to
the estimate
\[ \s \geq \int_U \L_\kappa(\x,\y)\: d\mu(\y) \geq \varepsilon\: \mu(U) \:, \]
showing that~$\mu$ is locally finite. This proves~(iii).
Next, it is obvious from~\eqref{ELstar} that~$\L_\kappa(\x, .)$ is $\mu$-integrable and that the
resulting function is bounded. It remains to prove that this function is lower semi-continuous.
To this end, we can argue similar as in the proof of~(ii) above:
\begin{align*}
\int_\G \L_\kappa(\x, \y) \: d\mu(\y) &\leq
\int_\G \lim_{\x' \rightarrow \x} \L_\kappa(\x', \y) \: d\mu(\y)
\leq \lim_{\x' \rightarrow \x} \int_\G \L_\kappa(\x', \y) \: d\mu(\y)
\end{align*}
where in the first step we used that the static $\kappa$-Lagrangian is lower semi-continuous
and applied again Fatou's lemma. This concludes the proof of~(iv).
\QED

In the infinite-dimensional setting, the set~$\F^\text{reg}$ is an infinite-dimensional
Banach manifold (for details see~\cite{banach}). 
For technical simplicity, here we shall not enter the details of the infinite-dimensional analysis.
Instead, our method is to restrict attention to a finite-dimensional submanifold of~$\F^\text{reg}$.
Clearly, this submanifold must contain the supports of both measures~$\rho$ and~$\tilde{\rho}$,
and the unitary group~$(\scrU_t)_{t \in \R}$ must map the submanifold to itself.
Moreover, the vector fields of the jets needed for the analysis must all be
tangential to this submanifold. Restricting the Lagrangian to this submanifold,
all the results of Proposition~\ref{prpreg} must hold.
Then we simply choose~$\G$ as the equivalence classes of this submanifold
under the action of the group~$(\scrU_t)_{t \in \R}$. This procedure will be illustrated
in Section~\ref{secex} in the example of static Dirac systems in asymptotically Schwarzschild spacetimes. 

\subsection{Freedom in Rescaling Solutions of the Euler-Lagrange Equations} \label{secrescale}
Let~$\rho$ be a critical measure of the static causal action principle for given
values of the parameters~$c$ and~$\kappa$. Then for a suitable Lagrange multiplier~$\s>0$,
the equations~\eqref{trc} and~\eqref{EL} hold. For clarity, we add a subscript~$\kappa$ to~$\ell$
and write~\eqref{ldef} as
\[ 
\ell_\kappa(\x) := \ell(\x) + \kappa \,\mathfrak{t}(\x)\:, \]
where
\begin{align}
\ell(\x) &:= \int_M \L_\kappa(\x,\y)\: d\rho(\y)  - \s \label{elldef} \\
\mathfrak{t}(\x) &:= \int_M \T(\x,\y)\: d\rho(\y) \:. \label{tdef}
\end{align}

There is a two-parameter family of rescalings which again give critical measures.
Indeed, the new measure~$\hat{\mu}$ defined by
\beq \label{tilrho}
\hat{\mu}(\Omega) = \sigma\: \mu\Big( \frac{\Omega}{\lambda} \Big) \qquad \text{with~$\lambda, \sigma>0$}
\eeq
again satisfies the EL equation with new Lagrange multipliers
\[ \hat{c} = \lambda\, c \quad \text{and} \qquad \hat{\s} = \sigma\, \lambda^4\: \s\:. \]
This rescaling freedom could be fixed for example by imposing that
\[ c = \s = 1 \:. \]
Note that the Lagrange multiplier~$\kappa$ remains unchanged; it is a dimensionless
parameter which characterizes the solution independent of the values of~$c$ and~$\s$.

\section{General Properties of the Total Mass} \label{secnonlin}
In this section we work out a few general properties of the total mass as introduced
in Definition~\ref{defmassgen}.

\subsection{Independence of the Exhaustion} \label{secexhaust}
We first verify that the total mass does not depend on the choice of the exhaustions of~$N$ and~$\tilde{N}$.
\begin{Prp} \label{prpindepend}
Assume that~$\mu$ and~$\tilde{\mu}$ are asymptotically close (see Definition~\ref{defasyclose}).
Then the limits~$\Omega \nearrow N$ and~$\tilde{\Omega} \nearrow \tilde{N}$ in~\eqref{massgen}
exist and are independent of the exhaustions. The total mass is finite. It can also be written as
the difference of the spatial integrals
\beq \label{Mint}
\Mass(\tilde{\mu}, \mu) = \int_{\tilde{N}} \big( \tilde{n}(\x) - \s \big) \: d\tilde{\mu}(\x) - \int_N \big( n(\x) - \s \big) \: d\mu(\x) \:.
\eeq
\end{Prp}
\Proof 
Using that the Lagrangian is symmetric,
the integral expression in Definition~\ref{defmassgen}
can be rewritten with the help of the correlation measures~\eqref{nuintro} as
\begin{align*}
&\int_{\tilde{\Omega}} \!d\tilde{\mu}(\x) \int_{N\setminus \Omega} \!\!\!\!\!\!\!d\mu(\y)\: \L_\kappa(\x,\y)
- \int_{\Omega} \!d\mu(\x) \int_{\tilde{N} \setminus \tilde{\Omega}} \!\!\!\!\!\!\!d\tilde{\mu}(\y)\: \L_\kappa(\x,\y) \\
&=\int_{\tilde{\Omega}} \!d\tilde{\mu}(\x) \int_{N} \!d\mu(\y)\: \L_\kappa(\x,\y)
- \int_{\Omega} \!d\mu(\x) \int_N  \!d\tilde{\mu}(\y)\: \L_\kappa(\x,\y) = \tilde{\nu} \big( \tilde{\Omega} \big) - \nu(\Omega) \:,
\end{align*}
where in the last step we used the definition of the correlation measures~\eqref{nuintro}.
We thus obtain the compact formula for the total mass
\begin{align}
\Mass(\tilde{\mu}, \mu) &= \lim_{\Omega \nearrow N} \;\lim_{\tilde{\Omega} \nearrow \tilde{N}} 
\Big( -\s \big( \tilde{\mu}(\tilde{\Omega}) - \mu(\Omega) \big) + \tilde{\nu} \big( \tilde{\Omega} \big) - \nu(\Omega) \Big)
\notag \\
&= \lim_{\Omega \nearrow N} \;\lim_{\tilde{\Omega} \nearrow \tilde{N}} 
\bigg( \int_{\tilde{\Omega}} \big( \tilde{n}(\x) - \s \big) \: d\tilde{\mu}(\x) - \int_\Omega \big( n(\x) - \s \big) \: d\mu(\x) \bigg)
\label{ntn}
\end{align}
with~$n$ and~$\tilde{n}$ as defined in~\eqref{ndef}. Since~$\mu$ and~$\tilde{\mu}$ are asymptotically close,
the integrands in~\eqref{ntn} are in~$L^1$. Therefore, the limits~$\Omega \nearrow N$ and~$\tilde{\Omega}
\nearrow \tilde{N}$ exist by Lebesgue's dominated convergence theorem. We thus obtain~\eqref{Mint}.
\QED

We next recall the definition of non-atomic measures (see for example~\cite[Section~40]{halmosmt}).
\begin{Def} \label{defnonatomic} A Borel set~$\Omega \subset \M$ is called an {\bf{atom}} of
the Borel measure~$\mu$ if $\mu(\Omega)>0$ and if every Borel
subset~$K \subset \Omega$ with~$\mu(K) < \mu(\Omega)$
has measure zero. A Borel measure is said to be
{\bf{non-atomic}} if it has no atoms.
\end{Def}

\begin{Prp} \label{prpnonatomic} If~$\mu$ or~$\tilde{\mu}$ is non-atomic, then the total 
mass (see Definition~\ref{defmassgen}) can be written equivalently in the form~\eqref{massintro}.
\end{Prp}
\Proof Clearly, for every exhaustions~$(\Omega_n)_{n \in \N}$ of~$N$
and~$(\tilde{\Omega}_n)_{N \in \N}$ of~$\tilde{N}$ which satisfy the
condition~$\mu(\Omega_n)=\tilde{\mu}(\tilde{\Omega}_n)<\infty$ for all~$n \in \N$,
the formula~\eqref{massgen} reduces to~\eqref{massintro}.
In view of the independence of the choice of exhaustions (Proposition~\ref{prpindepend}),
it remains to show that there are exhaustions~$(\Omega_n)_{n \in \N}$ and~$(\tilde{\Omega}_n)_{n \in \N}$
with~$\mu(\Omega_n)=\tilde{\mu}(\tilde{\Omega}_n)<\infty$ for all~$n$.

To this end, assume for example that~$\mu$ is non-atomic.
Let~$(U_n)_{n \in \N}$ and~$(\tilde{U}_n)_{n \in \N}$ be exhaustions
of~$N$  and~$\tilde{N}$, respectively, by sets of finite volume.
In view of~\eqref{nuinf}, the volumes of these sets tends to infinity.
Therefore, we can choose subsequences (for simplicity we again denoted
by~$(U_n)_{n \in \N}$ and~$(\tilde{U}_n)_{n \in \N}$) such that
\[ \mu(U_1) \leq \tilde{\mu}(\tilde{U}_1) \leq \mu(U_2) \leq \tilde{\mu}(\tilde{U}_2) \leq \cdots \:. \]
Since~$\mu$ is a non-atomic and regular Borel measure, we can find measurable sets~$V_n$ with
\[ U_n \subset V_n \subset U_{n+1} \qquad \text{and} \qquad \mu(V_n) = \tilde{\mu}(\tilde{U}_n) \:. \]
Clearly, the sequence~$(V_n)_{n \in \N}$ is again an exhaustion of~$N$ by sets of finite measure.
Therefore, the sets~$\Omega_n := V_n$ and~$\tilde{\Omega}_n := \tilde{U}_n$ have
all the required properties.
\QED

\subsection{Independence of the Inner Geometry} \label{secind}
We now want to verify and make precise that the total mass in Definition~\ref{defmassgen}
depends only on the geometry near infinity and on the inner volume, but not on the
inner geometry. To this end, we shall introduce a more general notion of total mass
which only involves the geometry outside an arbitrarily chosen compact set
(see Definition~\ref{defmass}). Then we shall prove that this notion of mass
coincides with the total mass of Definition~\ref{defmassgen}.

We describe the ``inner regions'' of our spacetimes by 
relatively compact open subsets~$I \subset M$ and~$\tilde{I} \subset \tilde{M}$.
In order to disregard the geometry of these subsets, we modify the surface layer integral
of Definition~\ref{defosinl} to
\[ 
\gamma^{\tilde{\Omega}, \Omega}_{\tilde{I}, I}(\tilde{\mu}, \mu) :=
\int_{\tilde{\Omega}} \!d\tilde{\mu}(\x) \int_{N\setminus (I \cup \Omega)} \!\!\!\!\!\!\!\!\!\!\!\!\!\!\!\!d\mu(\y)\: \L(\x,\y)
- \int_{\Omega} \!d\mu(\x) \int_{\tilde{N} \setminus (\tilde{I} \cup \tilde{\Omega})} 
\!\!\!\!\!\!\!\!\!\!\!\!\!\!\!\!d\tilde{\mu}(\y)\: \L(\x,\y) \:, \]
where~$\Omega$ and~$\tilde{\Omega}$ are now subsets of~$N \setminus I$
and~$\tilde{N} \setminus \tilde{I}$, respectively.
 
\begin{Def} \label{defmass}
The {\bf{total mass~$\Mass$ on~$\tilde{N} \setminus \tilde{I}$ relative to~$N \setminus I$}} 
is defined in generalization of~\eqref{Mshort} by
\beq \label{defmassgenI}
\Mass_{\tilde{I}, I} \big(\tilde{\mu}, \mu \big)
:= \lim_{\Omega \nearrow N \setminus I} \;\lim_{\tilde{\Omega} \nearrow \tilde{N} \setminus \tilde{I}} 
\bigg( -\s \Big( \tilde{\mu}(\tilde{\Omega}) - \mu(\Omega) \Big) 
+ \gamma^{\tilde{\Omega}, \Omega}_{\tilde{I}, I}(\tilde{\mu}, \mu) \bigg) \:,
\eeq
where the sets~$\Omega$ and~$\tilde{\Omega}$ form exhaustions
of~$N \setminus I$ and~$\tilde{N} \setminus \tilde{I}$ by sets of finite volume, respectively.
\end{Def}

This version of total mass can again be rewritten in terms of correlation measures.
Indeed, introducing the {\em{correlation measures}} $\nu_{\tilde{I}}$ on~$N \setminus I$ and~$\tilde{\nu}_I$
on~$\tilde{N} \setminus \tilde{I}$ by
\[ d\nu_{\tilde{I}}(\x) = n_{\tilde{I}}(\x)\: d\mu(\x) \qquad \text{and} \qquad
d\tilde{\nu}_I(\x) = \tilde{n}_I(\x)\: d\tilde{\mu}(\x) \]
with functions
\beq \label{nIdef}
\left\{ \begin{array}{rl} 
n_{\tilde{I}} \::\: N \rightarrow \R^+_0 \cup \{\infty\}\:,\qquad
n_{\tilde{I}}(\x) &\!\!\!\!= \displaystyle \int_{\tilde{N} \setminus \tilde{I}} \L(\x,\y)\: d\tilde{\mu}(\y) \\[1em]
\tilde{n}_I \::\: \tilde{N} \rightarrow \R^+_0 \cup \{\infty\}\:,\qquad
\tilde{n}_I(\x) &\!\!\!\!= \displaystyle \int_{N \setminus I} \L(\x,\y)\: d\mu(\y) \:,
\end{array} \right.
\eeq
a short computation yields
\[ \gamma^{\tilde{\Omega}, \Omega}_{\tilde{I}, I}(\tilde{\mu}, \mu)
= \tilde{\nu}_I(\tilde{\Omega}) - \nu_{\tilde{I}}(\Omega) \:. \]
We now generalize the result of Proposition~\ref{prpindepend}:
\begin{Prp} \label{prpindependI}
Assume that~$\mu$ and~$\tilde{\mu}$ are asymptotically close.
Then the limits~$\Omega \nearrow N \setminus I$ and~$\tilde{\Omega} \nearrow \tilde{N} \setminus \tilde{I}$
in~\eqref{defmassgenI} exist and are independent of the exhaustions. The total mass
on~$\tilde{N} \setminus \tilde{I}$ relative to~$N \setminus I$ is finite. It can be written as
\beq \label{MintI}
\Mass_{\tilde{I}, I}(\tilde{\mu}, \mu)
= \int_{\tilde{N} \setminus \tilde{I}} \big( \tilde{n}_I(\x) - \s \big) \: d\tilde{\mu}(\x) -
\int_{N \setminus I} \big( n_{\tilde{I}}(\x) - \s \big) \: d\mu(\x) \:.
\eeq
\end{Prp}
\Proof Following the proof of Proposition~\ref{prpindepend}, it suffices to show that
the functions~$n_{\tilde{I}} - \s$ and~$\tilde{n}_I - \s$ are integrable.
We only consider the latter function, because the proof for the first function is analogous.
Thus our task is to show that
\[ \int_{N \setminus I} \big| n_{\tilde{I}} - \s \big| \:d\mu < \infty \:. \]
Comparing~\eqref{nIdef} with~\eqref{ndef}, we can rewrite the function in the integrand as
\beq \label{lastint}
n_{\tilde{I}}(\x) - \s = \big[ n(\x) -s \big] - \int_{\tilde{I}} \L(\x,\y)\: d\tilde{\mu}(\y) \:.
\eeq
The square bracket is integrable because~$\mu$ and~$\tilde{\mu}$ are asymptotically
close (see Definition~\ref{defasyclose}). Integrating the last summand, the fact that the
Lagrangian is non-negative makes it possible to apply Tonelli's theorem,
\[ \int_N \bigg| \int_{\tilde{I}} \L(\x,\y)\: d\tilde{\mu}(\y) \bigg|\: d\mu(\x)
= \int_{\tilde{I}} \bigg( \int_N \L(\x,\y)\: \: d\mu(\x) \bigg)\: d\tilde{\mu}(\y) \;\in\; \R^+_0 \cup \{\infty\}\:. \]
According to condition~(iv) in Section~\ref{seccvp} on page~\pageref{Cond4},
the $\x$-integration gives a lower semi-continuous and bounded function on~$\G$.
Since~$\tilde{I}$ is compact and~$\tilde{\mu}$ is locally finite, the $\y$-integral exists and is finite. 
We conclude that the last summand in~\eqref{lastint} is integrable over~$N$,
concluding the proof.
\QED

We can now state and prove the main result of this section.
\begin{Thm} \label{thmindepend}
Assume that~$\mu$ and~$\tilde{\mu}$ are asymptotically close.
Then the total mass depends only on the volumes
of the relatively compact open subsets~$I \subset N$ and~$\tilde{I} \subset \tilde{N}$. More precisely,
\[ \Mass_{\tilde{I}, I} (\tilde{\mu}, \mu)
= \Mass \big(\tilde{\mu}, \mu \big) + \s\, \Big( \tilde{\mu}(\tilde{I}) - \mu(I) \Big) \:. \]
\end{Thm}
\Proof We rewrite~\eqref{MintI} as follows,
\begin{align*}
\Mass_{\tilde{I}, I} (\tilde{\mu}, \mu) 
&= \int_{\tilde{N} \setminus \tilde{I}} \big( \tilde{n}(\x) - \s \big) \: d\tilde{\mu}(\x) -
\int_{N \setminus I} \big( n(\x) - \s \big) \: d\mu(\x) \\
&\quad\, -\int_{\tilde{N} \setminus \tilde{I}} d\tilde{\mu}(\x) \int_I d\mu(\y)\: \L(\x,\y)
+ \int_{N \setminus I} d\mu(\x) \int_{\tilde{I}} d\tilde{\mu}(\y) \: \L(\x,\y) \\
&= \Mass(\tilde{\mu}, \mu) - 
\int_{\tilde{I}} \big( \tilde{n}(\x) - \s \big) \: d\tilde{\mu}(\x) + \int_I \big( n(\x) - \s \big) \: d\mu(\x) \\
&\quad\, -\int_{\tilde{N} \setminus \tilde{I}} d\tilde{\mu}(\x) \int_I d\mu(\y)\: \L(\x,\y)
+ \int_{N \setminus I} d\mu(\x) \int_{\tilde{I}} d\tilde{\mu}(\y) \: \L(\x,\y) \:. 
\end{align*}
We now transform the last line: Since the Lagrangian is symmetric and non-negative, by Tonelli's theorem we may
interchange the orders of integration. Moreover, using again the symmetry of the Lagrangian,
we may replace the integration range~$\tilde{N} \setminus \tilde{I}$
by~$\tilde{N}$ and~$N \setminus I$ by~$N$. Then the last line becomes
\[ - \int_{I} n(\y)\: d\mu(\y) + \int_{\tilde{I}} \tilde{n}(\y)\: d\tilde{\mu}(\y) \:. \]
Collecting all the terms gives the result.
\QED

\subsection{Independence of the Identification of Hilbert Spaces} \label{secindident}
In Section~\ref{secintro}, the total mass was introduced for two measures~$\rho$ and~$\tilde{\rho}$
defined on a set of linear operators~$\F$ on a Hilbert space~$\rho$.
In most applications, however, the two spacetimes are described by two causal fermion
systems~$(\H, \F, \rho)$ and~$(\tilde{\H}, \tilde{\F}, \tilde{\rho})$ which
are defined on two different Hilbert spaces~$\H$ and~$\tilde{\H}$. 
Both spacetimes are static in the sense that~$\rho$ is static with respect to a group
unitary transformations~$(\scrU_t)_{t \in \R}$ on~$\H$ (see Definition~\ref{defstatic}),
whereas~$\tilde{\rho}$ is static with respect to a unitary group~$(\tilde{\scrU}_t)_{t \in \R}$ on~$\tilde{\H}$.
In order to get into the setting of the introduction, the two
Hilbert spaces must be identified by a unitary transformation~$V \::\: \H \rightarrow \tilde{\H}$,
in such a way that the spacetimes become jointly static, i.e.\
\beq \label{UVcomm}
\tilde{\scrU}_t = V\, \scrU_t\, V^{-1} \qquad \text{for all~$t \in \R$}\:.
\eeq
In Section~\ref{secidentify}, this construction will be explained in more detail in the
example of the Schwarzschild geometry. Here our point of interest is that the condition~\eqref{UVcomm}
does not determine~$V$ uniquely. Indeed, $V$ is unique only up to the transformations~$V \rightarrow V W$,
where~$W$ is a unitary transformation on~$\H$ which is {\em{static}} in the sense that
it commutes with the time evolution,
\beq \label{Wstatic}
\scrU_t\,W = W\, \scrU_t \qquad \text{for all~$t \in \R$}\:.
\eeq

We now prove that, within the class of asymptotically flat spacetimes,
the total mass does not depend on the choice of~$W$.
To this end, we let~$(\H, \F, \rho)$ and~$(\H, \F, \tilde{\rho})$ be two
causal fermion systems which are both static with respect to a group~$(\scrU_t)_{t \in \R}$
of unitary transformations of~$\H$ (see Definition~\ref{defstatic}). 
Given a unitary transformation~$W$ on~$\H$ which is static~\eqref{Wstatic},
the unitarily transformed static measure~$W \tilde{\rho}$ defined by
\beq \label{Wrhodef}
(W \tilde{\rho})(\Omega) := \tilde{\rho} \big( W^{-1} \,\Omega\, W \big) 
\quad \text{for~$\Omega \subset \F$}
\eeq
is again static, making it possible to decompose it similar to~\eqref{fubini} as
\[ 
d(W \tilde{\rho}) = dt\, d(W \tilde{\mu}) \:. \]

\begin{Thm} \label{thmindident} If both~$\tilde{\mu}$ and~$W \tilde{\mu}$ are asymptotically flat with respect to
the vacuum measure~$\mu$ (see Definition~\ref{defvacuum} and~\ref{defasyflat}),
then the total masses of~$\tilde{\mu}$ and~$W \tilde{\mu}$ coincide,
\[ \Mass(\tilde{\mu}, \mu) = \Mass \big( W \tilde{\mu}, \mu \big) \:. \]
\end{Thm}

The proof of this theorem is based on the unitary invariance of the causal action.
In preparation of the proof, we must introduce the necessary concepts.
The causal action principle is {\em{unitarily invariant}} in the following sense. Let~$W \in \U(\H)$ be a
unitary transformation. Given a measure~$\rho$ on~$\F$, we can unitarily transform the measure 
similar to~\eqref{Wrhodef} by setting~$(W \rho)(\Omega) := \rho(W^{-1} \,\Omega\, W)$.
Since the eigenvalues of an operator are invariant under unitary transformations,
a universal measure~$\rho$ is a minimizer or critical point of the causal action principle if and only
if~$W \rho$ is. Next, we specialize again to the setting that all objects are static with respect to
a unitary group~$(\scrU_t)_{t \in \R}$. Then, as explained in Sections~\ref{secstatic} and~\ref{secSstatic}, we can work
with the static causal action principle. If also~$W$ is static (in the sense~\eqref{Wstatic}),
the resulting static measure~$W \mu$ defined by~$d(W \rho) = dt\: d (W \mu)$ is a critical point of the
static causal action if and only if~$\mu$ is. This fact can be used to construct solutions of the linearized field equations, 
as we now explain.
Let~$\mu$ be a critical static measure. Moreover, let~$(W_s)_{s \in [0, \tau_{\max}]}$ be a
smooth and strongly continuous family of static unitary transformations with generator
\[ \scrA := -i \frac{d}{ds}\, W_s \big|_{s=0} \:. \]

\begin{Lemma} \label{lemmacomm}  Assume that the jet
\beq \label{judef}
\u :=(0,u) \qquad \text{with} \qquad u(\x) := i \big[\scrA, \x \big]
\eeq
has the property
\beq \label{DuComm}
D_v u \in \Jtest \qquad \text{for all~$\v \in \Jtest$}
\eeq
(where the directional derivatives are computed in the distinguished charts mentioned
in Definition~\ref{defJvary}). 
Then the jet~$\u$ is a solution of the linearized field equations, i.e.
\beq \label{lincomm}
\nabla_\v \int_N (D_{1,u} + D_{2,u} )\, \L(\x,\y)\: d\mu(\y) = 0 \qquad \text{for all~$\v \in \Jtest$}\:.
\eeq
\end{Lemma}
\Proof One method of proof would be to differentiate through the EL equations.
However, this would involve a transformation of the space of test jets
(similar as explained for example in~\cite[Section~3.1]{osi}).
Here we prefer to show that the integrand of~\eqref{lincomm} vanishes identically. Indeed,
due to the unitary invariance of the Lagrangian,
\[ \L\big( \scrU_\tau \,\x\, \scrU_\tau^{-1},\: \scrU_\tau \,\y\, \scrU_\tau^{-1} \big) = \L(\x,\y)\:. \]
Differentiating with respect to~$s$ gives
\[ (D_{1,u} + D_{2,u} )\, \L(\x,\y)\: d\rho(\y) = 0 \:. \]
Hence the integrand in~\eqref{lincomm} vanishes for all~$\x, \y \in \F/\R$. As a consequence,
the integral in~\eqref{lincomm} vanishes for all~$x \in \F/\R$.
Consequently, also its derivative in the direction of~$u$ vanishes.
Using our convention that the
jet derivatives act only on the Lagrangian (see the end of Section~\ref{seccvp}),
the directional derivative differs from the derivative by the term~$D_{D_v u} \ell(\x)$.
This term vanishes in view of~\eqref{DuComm} and the weak EL equations~\eqref{ELtest}.
\QED
Due to the commutator in~\eqref{judef}, we refer to jets of this form
as {\bf{static commutator jets}}. For more details on commutator jets we refer to~\cite{fockfermionic}.

\Proof[Proof of Theorem~\ref{thmindident}] Since both~$\tilde{\mu}$ and~$W \tilde{\mu}$
are asymptotically flat with respect to~$\mu$, their total mass can be expressed via~\eqref{osilin2}
in terms of a vector field~$w$ near infinity.
The freedom to perform static unitary transformations means for the jets that~$w$ can
be changed by infinitesimal unitary transformations, which by~\eqref{judef} correspond to
static commutator jets. Therefore, the freedom to perform static unitary transformations of~$\tilde{\mu}$
means that in~\eqref{osilin2} we have the freedom to transform~$w$ according to~$w \rightarrow w + u$
with $u$ a static commutator jet.
Since~$u$ is a linearized solution without scalar component (see Lemma~\ref{lemmacomm}),
we can apply the conservation law of Lemma~\ref{lemmaosiconserve} to conclude
that~$u$ does not contribute to the surface layer integral~\eqref{osilin2}.
\QED

\section{The Positive Mass Theorem} \label{secpmt}
The goal of this section is to prove Theorem~\ref{thmpmt}. In preparation, we need
to derive and analyze the equations of linearized gravity which were already mentioned in words in the introduction.

\subsection{The Equations of Linearized Gravity} \label{seclininhom}
We saw in Section~\ref{seclinear} that, given a family~$(\tilde{\mu}_\tau)_{\tau \in [0,1)}$ of solutions of the EL equations
of the form~\eqref{rhoFf} for fixed values of the parameters~$\kappa$ and~$\s$, the infinitesimal generator~$\v$ 
of this family is a solution of the linearized field equations (see~\eqref{vinfdef} and~\eqref{eqlinlip}).
Keeping the parameter~$\s$ fixed is a matter of convenience, because the rescaling freedom in Section~\ref{secrescale}
makes it possible to give this parameter an arbitrary value.
However, the situation is different for the parameter~$\kappa$, which is dimensionless and scaling invariant.
For this reason, it is interesting to also consider families~$(\tilde{\mu}_\tau)_{\tau \in [0,1)}$ where~$\kappa(\tau)$
depends on~$\tau$. It is most convenient to arrange by a reparametrization of~$\tau$ that
\beq \label{kappatau}
\frac{d}{d\tau} \log \kappa(\tau) \big|_{\tau=0} = -1 \:.
\eeq
Moreover, as will become clear below, it is preferable to also choose~$\s$ as a function of~$\tau$.
Writing the EL equations~\eqref{lin1} for the $\kappa$-Lagrangian in~\eqref{kLtime},
\beq \label{lin2}
\begin{split}
\nabla_\u \bigg( \int_N f_\tau(\x)\: 
\Big( \L_\kappa\big( F_\tau(\x), F_\tau(\y) \big) + \kappa(\tau)\: \T(\x,\y) \Big) \: f_\tau(\y)\: d\mu(\y) \\
- f_\tau(\x)\,\s(\tau) \bigg) &= 0 \:,
\end{split}
\eeq
differentiating with respect to~$\tau$ and evaluating at~$\tau=0$ gives in generalization of~\eqref{eqlinlip}
the equation
\[ 
\la \u, \Delta \v \ra|_N = -\kappa'(0)\, \nabla_\u \mathfrak{t} 
+ \s'(0)\: \nabla_\u 1 \qquad \text{for all~$\u \in \Jtest$} \:, \]
where~$\Delta$ is defined by~\eqref{Deldef}, but for the $\kappa$-Lagrangian,
\[ 
\la \u, \Delta \v \ra(\x) := \nabla_{\u} \bigg( \int_N \big( \nabla_{1, \v} + \nabla_{2, \v} \big) \L_\kappa(\x,\y)\: d\mu(\y) - \nabla_\v \,\s \bigg) \:, \]
and with~$\mathfrak{t}$ as in~\eqref{scrtintro}.

By assumption, the measure~$\mu$ is asymptotically flat
and satisfies the EL equations. Using these properties, we know
according to Definition~\ref{defasyflat}~(ii)
that the function~$\ell$ has a limit at infinity.
Moreover, the EL equations~\eqref{ELintro} and~\eqref{ellkappa} imply that
the same is true for the function~$\mathfrak{t}$. We set
\[ \mathfrak{t}_\infty :=  \lim_{\x \rightarrow \infty} \mathfrak{t}(\x) > 0 \:. \]
We apply~\eqref{kappatau} and choose~$\s'(0)$ as
\[ 
\s'(0) = -\kappa(0)\: \mathfrak{t}_\infty < 0 \:. \]
We thus obtain the inhomogeneous linearized field equations
\[ \la \u, \Delta \v \ra|_N = \kappa\, \nabla_\u \big( \mathfrak{t} - \mathfrak{t}_\infty \big) \big|_N \qquad \text{for all~$\u \in \Jtest$} \:. \]
Using the EL equations, we can rewrite the inhomogeneity in terms of~$\ell$,
\[ \la \u, \Delta \v \ra|_N = -\nabla_\u \big( \ell - \ell_\infty \big) \big|_N \qquad \text{for all~$\u \in \Jtest$} \:. \]
By adding a suitable inner solution (see Section~\ref{secinner}) we can arrange that the jet~$\v=(0,v)$
has no scalar component. We refer to the resulting equation
\beq \label{inhom}
\la \u, \Delta v \ra|_N = -\nabla_\u \big( \ell- \ell_\infty \big) \big|_N \qquad \text{for all~$\u \in \Jtest$}
\eeq
as the {\em{equations of linearized gravity}}.
We point out that the solution~$v$ of the equations of linearized gravity will in general not be
unique. For example, one can add to~$v$ any divergence-free vector field on~$N$.
But this has no effect on the following conservation law:

\begin{Prp} \label{prposinhom}
For a jet~$v \in \J^1$ without scalar component which satisfies the equations of linearized gravity~\eqref{inhom},
the surface layer integral~\eqref{gosidef} (again for the~$\kappa$-La\-gran\-gian) is computed for an exhaustion of~$N$ by
\[ 
\lim_{\Omega \nearrow N} \gamma^\Omega_\mu(v) = \int_N \big( \mathfrak{\ell} - \mathfrak{\ell}_\infty \big) \: d\mu \:. \]
\end{Prp}
\Proof A direct computation using~\eqref{inhom} yields
\begin{align*}
\gamma^\Omega_\mu(v) &= \int_{\Omega} \!d\mu(\x) \int_{M \setminus \Omega} \!\!\!\!\!\!\!\!d\mu(\y)\:
\big( D_{1,v} - D_{2,v} \big) \L_\kappa(\x, \y) \\
&=\int_{\Omega} \!d\mu(\x) \int_M \!d\mu(\y)\:
\big( D_{1,v} - D_{2,v} \big) \L_\kappa(\x, \y) \\
&= \int_\Omega \Big( 2 D_v \ell(\x) - (\Delta \v)(\x) \Big) d\mu(\x) 
= \int_\Omega \big( \ell - \ell_\infty \big) \: d\mu \:.
\end{align*}
Taking the limit~$\Omega \nearrow N$ gives the result.
\QED

\subsection{Proof of the Positive Mass Theorem} \label{secproofthmpmt}
We are now in the position to prove our positive mass theorem.

\Proof[Proof of Theorem~\ref{thmpmt}] Using~\eqref{wdef} in~\eqref{osilin2} gives
\[ \Mass(\tilde{\mu}, \mu)
= g \lim_{\Omega \nearrow N} \gamma^\Omega_\mu \big(\tilde{v}-v \big) \:, \]
where we used the notation~\eqref{gosidef}.
For the computation of the surface layer integral~$\gamma^\Omega_\mu(\tilde{v})$ at infinity,
we can work just as well with the measure~$\tilde{\mu}$, i.e.\
\[ \Mass(\tilde{\mu}, \mu)
= g \lim_{\Omega \nearrow N} \Big( \gamma^{F(\Omega)}_{\tilde{\mu}} \big(\tilde{v} \big)
- \gamma^\Omega_\mu(v) \Big)\:. \]
This makes it possible to apply Proposition~\ref{prposinhom} to obtain
\[ \Mass(\tilde{\mu}, \mu)
= g \int_{\tilde{N}} \big( \tilde{\ell} - \tilde{\ell}_\infty \big)\: d\tilde{\mu}
- g \int_{N} \big( \ell - \ell_\infty \big)\: d\mu \:. \]
The last integral vanishes because~$\mu$ is a vacuum measure (see Definition~\ref{defvacuum}~(ii)).
This concludes the proof.
\QED

\section{Example: Asymptotically Schwarzschild Spacetimes} \label{secex}
In this section we establish the correspondence between the total mass and the ADM mass
by proving Theorem~\ref{thmcorrespond}. 
\subsection{Construction of Static Dirac Systems}
Let~$(\scrM, g)$ be a four-dimensional
static globally hyperbolic Lorentzian manifold~$\scrM$ with one asymptotic end
which is asymptotically Schwarzschild. Thus we assume that the manifold is the topological product
\[ \scrM = \R \times \scrN \:. \]
Moreover, denoting the coordinates by~$x = (t, \x)$ with~$t \in \R$ and~$\x \in \scrN$, we assume that
the metric takes the form
\[ ds^2 = L(\x)^2\: dt^2 - g_\scrN \:, \]
where~$L(\x)$ is the lapse function and~$g_\scrN$ is a complete Riemannian metric on~$\scrN$.
Finally, we assume that
outside a compact set~$K \subset \scrN$, the metric coincides with the Schwarzschild metric.
Thus choosing polar coordinates~$\x = (r, \vartheta, \varphi)$ on~$\scrN \setminus K$, the line
element becomes
\beq \label{schwarzschild}
ds^{2} = \left(1-\frac{2M_\text{S}}{r} \right) dt^{2}-
\left(1-\frac{2M_\text{S}}{r} \right)^{-1} dr^2 -r^2\: \big( d\theta^{2}+\sin^{2}\theta\,d\varphi^{2} \big)\:.
\eeq

We now recall a few basics on the Dirac equation in the Schwarzschild geometry
(for details and explicit formulas see~\cite{FSYperiodic} or~\cite{platzer}).
For our purpose, it is most convenient to write the Dirac equation in the Hamiltonian form~\eqref{DirHamilton}
and to take its spectral decomposition~\eqref{Dirspec}.
The essential spectrum is determined from the
asymptotic behavior of the metric at infinity. Therefore,
\[ \sigma_\text{ess}(H) = (-\infty, -m] \:\cup\: [m, \infty) \:. \]
In addition, there could be a point spectrum in~$[-m,m]$ describing bound states of the system.

Following the general procedure in Section~\ref{seclorentz}, we now construct 
static Dirac systems. The {\em{Dirac sea}} is defined as the negative spectral subspace of the essential spectrum,
\[ \H_- := E_{(-\infty, m]}(\H_m) \:. \]
Following the general procedure in~\eqref{HHminus}, we choose~$\H$ as a subspace
which differs from~$\H_-$ by a finite-dimensional subspace. As a consequence, the
kernel of the fermionic projector of our system differs from that of the Dirac sea by smooth
contributions,
\[ P(x,y) = P^\text{sea}(x,y) + (\text{smooth contributions}) \:. \]
Moreover, describing bound states, these smooth contributions decay rapidly near spatial infinity.
Due to this rapid decay, the smooth contributions drop out when computing the surface layer integral
near infinity. With this in mind, we may disregard the smooth contributions and simply choose
\beq \label{Hsea}
\H=\H_- \:.
\eeq
Moreover, for technical simplicity we choose the regularization operator~\eqref{regop}
as a spatial convolution operator in the asymptotic end, i.e.\ for all~$\x \in \scrN \setminus K$,
\beq \label{Reps}
{\mathfrak{R}}_\varepsilon \::\: \H_m \rightarrow C^0(\scrM, S\scrM) \cap \H_m\:,
\qquad ({\mathfrak{R}}_\varepsilon \psi)(\x) = \int_\scrN
h_\varepsilon \big( |\x-\y| \big)\: \psi(\y)\: d\mu(\y) \:,
\eeq
where~$h_\varepsilon \in C^\infty_0(\R^3, \R^+_0)$ is a mollifier and~$|\x-\y|$ 
denotes the Euclidean norm in our coordinate system near infinity
(for the general context of this regularization method
see~\cite[Section~4]{finite}). As already mentioned at the end of Section~\ref{secstaticdirac},
the resulting static causal fermion systems are critical points of the causal action
principle in the continuum limit, provided that the Einstein equations are satisfied.
With this in mind, in what follows we assume that the static causal fermion systems
satisfy the weak EL equations~\eqref{ELtest} (for the $\kappa$-Lagrangian).

\subsection{Identifying the Dirac Solution Spaces} \label{secidentify}
In preparation of the computation of the total mass, we need to identify the Dirac solution spaces
in Minkowski space and in our asymptotically flat spacetime.
This needs to be done in such a way that both spacetimes are jointly static
(meaning that they are both static with respect to the same one-parameter group~$(\scrU_t)_{t \in \R}$
in Definition~\ref{defstatic}).
In order to distinguish the spacetimes, we again denote all objects of the asymptotically
flat spacetime with a tilde, whereas the objects without a tilde refer to Minkowski space.
Clearly, the corresponding one-parameter groups are the time evolution operators, i.e.\
\[ \scrU_t = e^{-i t H} \::\: \H \rightarrow \H \qquad \text{and} \qquad
\tilde{\scrU}_t = e^{-i t \tilde{H}}\::\: \tilde{\H} \rightarrow \tilde{\H} \:, \]
where~$\H$ and~$\tilde{\H}$ are both chosen as the respective Dirac seas~\eqref{Hsea}.
Since both Hamiltonians have the same essential spectrum and no point spectrum, they
can be mapped to each other by a unitary transformation, i.e.\ there is a
\beq \label{Vunit}
{\text{unitary }} V \,:\, \H \rightarrow \tilde{\H} \qquad \text{with} \qquad
\tilde{H} = V H V^{-1} \:.
\eeq
Identifying~$\H$ and~$\tilde{\H}$ by this unitary transformation,
the one-parameter groups~$(\scrU_t)_{t \in \R}$ and~$(\tilde{\scrU}_t)_{t \in \R}$ are also mapped
to each other. In this way, the corresponding causal fermion systems become
jointly static. Dividing out the group action, we get into the setting of causal variational principles.

We point out that the above unitary transformation~$V$ is
not unique, because it involves the freedom in unitarily transforming the Dirac solutions for every fixed energy.
However, we proved in Theorem~\ref{thmindident} that the total mass does not depend on the choice of~$V$.
With this in mind, in what follows we may choose the unitary transformation~$V$ in a convenient way.
A particularly convenient choice of identification is obtained by a perturbative treatment, as we now explain
(for other identifications of the Hilbert spaces see~\cite{platzer}).

\subsection{Perturbative Description Near Infinity} \label{secschperturb}
Since the metric~$\tilde{g}$ is asymptotically flat, its effect can be treated
asymptotically in first order perturbation theory. To this end, we decompose the Dirac Hamiltonian
in the gravitational field as
\[ \tilde{H} = H + \Delta H \]
and treat~$\Delta H$ as a static perturbation in the Dirac equation for fixed energy~$\omega$.
One must keep in mind that changing the metric also modifies the spatial integration measure
in the scalar product~\eqref{printN}. Compensating for this fact by a a local rescaling of the
Dirac wave functions, one can work in a fixed Hilbert space (for details on this procedure
see~\cite[Appendix~A]{firstorder}). This has the advantage
that we get a natural identification of~$\H$ and~$\tilde{\H}$.
Then the above-mentioned non-uniqueness of the identification operator~$V$ reduces to the
gauge freedom of the Dirac operator (for details see~\cite{u22}).
For a specific gauge, the Dirac operator is given explicitly in~\cite{FSYperiodic}.
For the details of the perturbative treatment of the operator~$\Delta H$ we refer
again to~\cite[Appendix~A]{firstorder} or, more generally, to~\cite[Appendix~F]{cfs}.
The perturbation expansion is gauge covariant, meaning that gauge invariant quantities like the
closed chain do not depend on the choice of the gauge. 
The methods in~\cite{firstorder} or, more generally, in~\cite{light}
(for an introduction see~\cite[Section~2.2]{cfs}) also give explicit formulas for the
kernel of the fermionic projector. Non-perturbatively, these formulas correspond do the
Hadamard expansion of the bi-distribution~$P(x,y)$ (see for example the textbook~\cite{baer+ginoux}).
In the presence of a regularization~\eqref{Reps},
the resulting {\em{regularized Hadamard expansion}} is worked out in~\cite{reghadamard}.

In order to illustrate these results, we now state a few formulas which will be needed later on
and make the connection to the jet formalism.
Working with jets corresponds to linear perturbations by gravity. Thus we consider a metric~$g$ of the form
\[ g_{ij} = \eta_{ij} + h_{ij} \:, \]
where~$\eta$ is the Minkowski metric and~$h_{ij}$ is the linear perturbation.
The gravitational field modifies the light cones. The corresponding modification of the
singularities of the unregularized kernel~$P(x,y)$ on the light cone is described by the formula
\beq \label{Pform}
\Delta P(x,y) = \frac{1}{2} \int_0^1 d\alpha\: h^i_j|_{\alpha y + (1-\alpha)\, x} \:\xi^j \:\frac{\partial}{\partial y^i} P(x,y) \:,
\eeq
where we set~$\xi=y-x$. This formula is derived in~\cite[Appendix~A]{firstorder}.
A more geometric way of understanding this formula is to integrate the geodesic equation;
for details see Appendix~\ref{appgeodesic}. In the static and spherically symmetric situation,
the formula~\eqref{Pform} remains valid for the regularized kernel (for details see again Appendix~\ref{appgeodesic}),
\beq \label{Pformstatic}
\Delta P^\varepsilon(x,y) = \frac{1}{2} \int_0^1 d\alpha\: h^i_j|_{\alpha y + (1-\alpha)\, x} \:\xi^j \:\frac{\partial}{\partial y^i} P^\varepsilon(x,y) \:.
\eeq
In addition to this effect of the ``deformation of the light cone,'' there are effects by
curvature. This becomes apparent in the formulas of the light cone expansion by
terms which involve the Riemann tensor and its derivatives. Here we do not need the detailed form
of these expressions. It suffices to keep in mind that these contributions are less singular on the light cone
than~\eqref{Pformstatic}.

In the jet formalism, the linear perturbation by gravity is described by a jet~$\v$.
Note that the volume form in the Schwarzschild geometry in Schwarzschild coordinates
does not depend on the mass, because
\beq \label{volindM}
d\tilde{\rho} = \sqrt{|\det \tilde{g}|}\: d^4x = dt\: d\mu \qquad \text{with} \qquad d\mu := r^2\, dr\, d\omega \:,
\eeq
where~$d\omega := d\varphi \:d\cos \vartheta$ is the volume measure on the unit sphere.
Therefore, the jet describing the linear perturbation by gravity has no scalar component,
\beq \label{vdef}
\v = (0, v) \:.
\eeq
The perturbation in~\eqref{Pformstatic} is obtained by perturbing the wave functions at both points~$x$
and~$y$ in the same way, i.e.
\[ \Delta P^\varepsilon(x,y) = \big(D_{1,v} + D_{2,v} \big) \,P^\varepsilon(x,y) \:. \]
The surface layer integral needed for the computation of the total mass (see~\eqref{osilin2}) has a different
form, because it involves the difference~$D_1 - D_2$ of jet derivatives
(this comes about because the arguments~$\x$ and~$\y$ in the formula for the total
mass~\eqref{massgen} lie in different spacetimes, one with and one without gravitational field).
Therefore, one must extend the perturbative treatment to the case where the wave functions are
perturbed only at~$x$, but not at~$y$ or vice versa. This case is treated in~\cite[Appendix~F]{pfp}
and~\cite[Section~5.1]{action}. It amounts to replacing the bounded line integral in~\eqref{Pform} by
suitable unbounded line integrals. In the static and spherically symmetric case, the resulting formulas
simplify to (for more details see Appendix~B)
\beq \label{D12P}
\begin{split}
D_{1,v} P^\varepsilon(x,y) &= -\frac{1}{4} \int_{-\infty}^\infty d\alpha\: \epsilon(\alpha)\:
h^i_j|_{\alpha y + (1-\alpha)\, x} \:\xi^j \:\frac{\partial}{\partial x^i} P^\varepsilon(x,y) \\
D_{2,v} P^\varepsilon(x,y) &= -\frac{1}{4}  \int_{-\infty}^\infty d\alpha\: \epsilon(\alpha-1)\:
h^i_j|_{\alpha y + (1-\alpha)\, x} \:\xi^j \:\frac{\partial}{\partial y^i} P^\varepsilon(x,y) \:,
\end{split}
\eeq
where~$\epsilon$ is the sign function.
Clearly, there are again additional contributions involving curvature and its derivatives,
but they are all less singular on the light cone.

Let us come back to the freedom in identifying the Hilbert spaces~$\H$ and~$\tilde{\H}$ as
already mentioned after~\eqref{Vunit}. The above perturbative procedure gives a canonical way to identify
the Hilbert spaces in linearized gravity. However, it does {\em{not}} give a canonical 
identification of the Hilbert space~$\H$ and~$\tilde{\H}$ for two spacetimes whose gravitational
fields coincide only asymptotically. Namely, in this case, the perturbative treatment is
admissible only near spatial infinity. Consequently, it only gives an identification of the Hilbert spaces
in the asymptotic ends. In general, this identification does not extend canonically to a unitary transformation
which also preserves the Hamiltonians~\eqref{Vunit}.
In view of Theorem~\ref{thmindident}, the resulting non-uniqueness of the identifications
has no effect on the total mass.

In more technical terms, the just-mentioned non-uniqueness of the identification of~$\H$ and~$\tilde{\H}$
becomes manifest in the fact that the unbounded line integrals in~\eqref{D12P} are well-defined only
in the asymptotic region where gravity can be treated linearly. One method of extending~\eqref{D12P}
to the spacetime regions with strong gravity is to replace the line integrals by integrals along null geodesics
and~$h^i_j$ by a first order variation of the metric. But this procedure is not canonical. Here we do not need
to worry about this issue because we already know from Theorem~\ref{thmindident} that the total
mass does not depend on the identification of Hilbert spaces. Correspondingly, we shall see in
Proposition~\ref{prpnoinner} below that the unbounded line integrals will drop out of our computations.

\subsection{Implementing the Volume Constraint} \label{secincorporate}
After the above preparations, we can enter the computation of the total mass.
It is most convenient to work with the formula~\eqref{osilin2}.
Thus we consider exhaustions~$\Omega_n$ of~$N$ and~$\tilde{\Omega}_n$ of~$\tilde{N}$,
subject to the volume constraint in~\eqref{massintro}
\beq \label{volOtO}
\mu(\Omega_n) = \tilde{\mu}(\tilde{\Omega}_n) < \infty \:.
\eeq
Due to the independence of the choice of the exhaustion (Proposition~\ref{prpindepend})
we may choose the sets as coordinate balls, i.e.
\beq \label{Rnballs}
\Omega_n = I \cup \Phi^{-1} \big( B_{R_n} \big) \qquad \text{and} \qquad 
\tilde{\Omega}_n = \tilde{I} \cup \tilde{\Phi}^{-1} \big( B_{\tilde{R}_n} \big)
\eeq
(where we used the notation in Definition~\ref{defae}).
For convenience, we choose the coordinates in the asymptotic end as the spatial part of the Schwarzschild
coordinates~\eqref{schwarzschild} with mass~$M_S=0$ (for~$\mu$) and~$M_S \neq 0$
(for~$\tilde{\mu}$).

Since the inner volumes~$\mu(I)$ and~$\tilde{\mu}(\tilde{I})$ will in general be different,
the volume constraint~\eqref{volOtO} forces us to also choose the radii differently, i.e.\
$R_n \neq \tilde{R}_n$. But, using that the volume form is independent of the mass (see~\eqref{volindM}),
the radii will coincide asymptotically in the sense that their difference decays quadratically,
\[ |R_n - \tilde{R}_n| \lesssim \frac{1}{R_n^2}\:. \]
Working again with the identification of the spacetimes used in the perturbative
description in the previous section, the mapping from~$\Omega_n$ to~$\tilde{\Omega}_n$
can be described near infinity by an infinitesimal volume-preserving diffeomorphism,
i.e.\ by a divergence-free vector field~$u$. We combine the corresponding inner solution~$\u=(0, u)$
with the jet~$v$ in~\eqref{vdef},
\beq \label{wdef2}
w := v + u \:.
\eeq
The resulting jet~$w$ describes the change of the metric near infinity,
taking into account the volume constraint~\eqref{volOtO}. Therefore, it can be used to
compute the total mass via~\eqref{osilin2}.
We note for clarity that, while the jet~$v$ is given explicitly by~\eqref{D12P},
the inner solution~$u$ is {\em{not}} known, because it depends on the difference of the
inner volumes of~$\mu$ and~$\tilde{\mu}$. We will come back to this issue in Section~\ref{seccompute}.

\subsection{Reduction of the Surface Layer Integral to a Spacetime Integral} \label{secreduce}
Our task is to compute the surface layer integral in~\eqref{osilin2} for the jet~$w$ in~\eqref{wdef2}.
We write this surface layer integral in the short form
\begin{align}
\Mass(\tilde{\mu}, \mu)
&= \lim_{R \rightarrow \infty} \int_{R_{\min}}^R dr \int_R^\infty dr'\: A(r, r') \qquad \text{with} \label{Mosi} \\
A(r, r') &= r^2\, r'^2\int_{S^2} d\omega \int_{S^2} d\omega' \; \big(D_{1,w} - D_{2,w} \big) \L(\x, \y) \:,
\end{align}
where~$R_{\min}$ is any radius for which the sphere is contained in the domain of the coordinate
chart at infinity (for example, one could choose~$R_{\min}=R_1$ with~$R_1$ as in~\eqref{Rnballs}).
Since the jet~$w$ is a solution of the linearized field equations without scalar component,
the surface layer integral~\eqref{Mosi} does not depend on~$R$.
Using this conservation law, we can follow the procedure in~\cite[Lemma~4.2]{action}
(a similar method was first used in~\cite[proof of Lemma~5.5]{noether}) to rewrite the surface layer integral
as an average of integrals over all of space:
\begin{Lemma} The surface layer integral satisfies the relation
\[ \int_{R_{\min}}^R dr \int_R^\infty dr\: A(r, r')
= \lim_{L \rightarrow \infty} \frac{1}{2 L} \int_R^{R+L} dr \int_{R_{\min}}^\infty dr'\: (r'-r)\: A(r,r') \:. \]
\end{Lemma}
\Proof We closely follow~\cite[proof of Lemma~5.2]{action}.
Since the surface layer integral does not depend on~$R$, we may take an average over~$R$,
\begin{align*}
\int_{R_{\min}}^R & dr \int_R^\infty dr'\: A(r, r') = \frac{1}{L} \int_0^L\: d\ell \int_{R_{\min}}^{R+\ell} dr \int_{R+\ell}^\infty dr'\: A(r, r') \\
&= \frac{1}{L} \int_0^L\: d\ell \int_{R_{\min}}^\infty dr \int_{R_{\min}}^\infty dr'\: 
\Theta\big(R+\ell-r)\: \Theta(r'-R-\ell)\: A(r, r') \\
&= \frac{1}{L} \int_{R_{\min}}^\infty dr \int_{R_{\min}}^\infty dr'\: A(r, r')\: \bigg( 
\int_0^L \Theta\big(R+\ell-r)\: \Theta(r'-R-\ell)\: d\ell \bigg) \:.
\end{align*}
Computing the integral of the Heaviside functions involves different cases.
A straightforward calculation yields
\begin{align}
\int_{R_{\min}}^R dr \int_R^\infty dr'\: A(r, r') 
&= \frac{1}{L} \int_{R_{\min}}^R dr \int_R^{R+L} dr'\: (r'-R)\: A(r, r') \label{t1} \\
&\quad\:+ \frac{1}{L} \int_{R_{\min}}^R dr \int_{R+L}^\infty dr'\: L\: A(r, r') \label{t2} \\
&\quad\:+ \frac{1}{L} \int_R^{R+L} dr \int_r^{R+L} dr'\: (r'-r)\: A(r, r') \label{t3} \\
&\quad\:+ \frac{1}{L} \int_{R}^{R+L} dr \int_{R+L}^\infty dr'\: (L-r+R)\: A(r, r') \:. \label{t4}
\end{align}
At this point, we make use of the fact that the causal Lagrangian is of short range in the
sense that~$A(r,r')$ decays at least cubically in~$r'-r$, i.e.\
\[ \big|  A(r,r') \big| \leq \frac{c}{|r'-r|^3} \qquad \text{for all~$r,r'>R$} \]
and a suitable constant~$c>0$ (for details see~\cite{action} and~\cite[Appendix~A]{jacobson}).
As a consequence, the double integrals in~\eqref{t1}, \eqref{t2} and~\eqref{t4} are bounded
uniformly in~$L$. Therefore, in the limit~$L \rightarrow \infty$ only the summand~\eqref{t3}
contributes, i.e.\
\beq \label{e1}
\int_{R_{\min}}^R dr \int_R^\infty dr'\: A(r, r')
= \lim_{L \rightarrow \infty} \frac{1}{L} \int_R^{R+L} dr \int_r^{R+L} dr'\: (r'-r)\: A(r, r') \:.
\eeq
Rewriting the boundaries of integration in the last integral with the help
of a Heaviside function, we may exchange the integrals and use that the integrand is symmetric,
\begin{align*}
&\int_R^{R+L} dr \int_r^{R+L} dr'\: (r'-r)\: A(r, r')
= \int_R^{R+L} dr \int_R^{R+L} dr'\: \Theta(r'-r)\: (r'-r)\: A(r, r') \\
&= \int_R^{R+L} dr' \int_R^{R+L} dr\: \Theta(r'-r)\: (r'-r)\: A(r, r')
= \int_R^{R+L} dr \int_R^r dr'\: (r'-r)\: A(r, r') \:.
\end{align*}
Using this relation, we can write~\eqref{e1} as
\[ 
\int_{R_{\min}}^R dr \int_R^\infty dr\: A(r, r')
= \lim_{L \rightarrow \infty} \frac{1}{2L} \int_R^{R+L} dr \int_R^{R+L} dr'\: (r'-r)\: A(r, r') \:. \]
Finally, changing the integration range of the last integral to~$(R_{\min}, \infty)$
modifies the integrals only by a contribution which vanishes in the limit~$L \rightarrow \infty$.
This gives the result.
\QED

In view of this lemma, it suffices to compute the following expression for large~$R$,
\begin{align*}
{\Mass}(R) \,\,&\!\!:= \frac{1}{2} \int_{R_{\min}}^\infty (r-R)\: A(R,r)\: dr \\
&= 2 \pi R^2 \int_{R_{\min}}^\infty r^2\, dr \int_{S^2} d\omega\: (r-R)\: \big(D_{1,w} - D_{2,w} \big) \L(\x, \y) \:,
\end{align*}
where, choosing polar coordinates, the points~$\x$ and~$\y$ have the form
\beq \label{xyform}
\x = (R,N) \qquad \text{and} \qquad \y = (r,\omega) \:,
\eeq
and~$N \in S^2$ is the north pole. Including the time integral, we obtain
\beq \label{MRdef}
{\Mass}(R) = 2 \pi R^2 \int_\scrM (r-R)\:
\big(D_{1,w} - D_{2,w} \big) \L\big(x, y \big)\: d\rho(y)  \:,
\eeq
where~$x=(0,\x)$ and~$y=(t,\y)$. Our findings are summarized as follows:

\begin{Prp} \label{prplimit}
Assume that integral expression~${\Mass}(R)$ in~\eqref{MRdef}
exists and converges in the limit~$R \rightarrow \infty$. Then it coincides with the total mass,
\[ {\Mass} = \lim_{R \rightarrow \infty} {\Mass}(R) \:. \]
\end{Prp}

\subsection{Computation of the Spacetime Integral} \label{seccompute}
The remaining task is to compute the spacetime integral in~\eqref{MRdef}.
The first order variation of the kernel of the fermionic projector in 
the presence of a gravitational field was computed
in~\cite[Appendix~B]{firstorder}. These formulas also apply in our setting.
Since the curvature tensor involves second derivatives of the metric, its components
decay at least cubically,
\[ \text{Riem} = \O\big( R^{-3} \big) \:. \]
As a consequence, the contributions to~\eqref{MRdef} involving curvature or
derivatives of curvature vanish in the limit~$R \rightarrow \infty$.
Therefore, it suffices to consider the contributions by the infinitesimal diffeomorphism
as given in~\eqref{D12P}. Moreover, we must also take into account the inner solution~$u$
in~\eqref{wdef2}.

\begin{Prp} \label{prpnoinner}
The volume constraint~\eqref{volOtO} can be taken into account in the
computation of the spacetime integral in~\eqref{MRdef}
by setting the inner solution~$u$ to zero and by
computing the jet-derivatives instead of~\eqref{D12P} by\footnote{
{\textsf{Footnote added in October 2023:}} These equations are not quite correct
because the scaling argument does not capture all the contributions of higher order in~$\varepsilon$.
As a consequence, the obtained formula for the total mass~\eqref{Mform}
is correct only up to a prefactor. A fully convincing treatment of this point
is given in \href{https://arxiv.org/abs/2310.07544}{\textrm{arXiv:2310.07544 [math-ph]}}, Appendix~A.}
\begin{align*} 
D_{1,w} P^\varepsilon(x,y) &= 0 \\
D_{2,w} P^\varepsilon(x,y) &= -\frac{1}{2}  \int_0^1 d\alpha\:
h^i_j|_{\alpha y + (1-\alpha)\, x} \:\xi^j \:\frac{\partial}{\partial y^i} P^\varepsilon(x,y) \:.
\end{align*}
\end{Prp}
\Proof We use a scaling argument which is based on the observation that
contributions by the inner solution~$u$ to the surface layer integral~\eqref{Mosi} involve a scaling factor~$\s$.
This can be seen in various ways: One way is to note that inner solutions describe
infinitesimal diffeomorphisms and that the corresponding volume change shows up in~\eqref{massgen}
with a prefactor~$\s$. Another way is to evaluate the conservation law of Lemma~\ref{lemmaosiconserve}
for an inner solution~$\u=(\div u, u)$,
\[ \int_\Omega \!d\mu(\x) \int_{N \setminus \Omega} \!\!\!\!\!\!\!\!d\mu(\y)\: 
\big( \nabla_{1,\u} - \nabla_{2,\u} \big) \L(\x,\y) 
= \s \int_\Omega \div u \: d\mu \:. \]
Alternatively, this surface layer integral can be computed using integration by parts
(for details see~\cite[eq.~(3.3) in Proposition~3.4]{fockbosonic}).
Such an integration-by-parts method can also be applied in~\eqref{MRdef}, showing
that the contribution of the inner solution to~$\mathfrak{M}(R)$ again involves a factor~$\s$.

Using that the EL equations~\eqref{EL} hold for all~$x \in N$, we can compute~\eqref{ldef}
asymptotically near infinity to obtain
\[ \s = \int_{\scrM} \L_\kappa\big(0, (t,\y) \big)\: dt\, d^3\mathbf{y} \:, \]
where~$\L_\kappa$ is the Lagrangian in the Minkowski vacuum.
Comparing with the constant~$c$ in~\eqref{cdef}, one sees that the total mass is by
at least one scaling factor~$\varepsilon$ {\em{smaller}} than the 
contribution by the inner solution. This will also become clear in the computations leading to~\eqref{Mform} below,
whereas a general scaling argument is given in Appendix~\ref{appscalingmass}.

Combining these results, we conclude that the inner solutions compensate precisely all the contributions
by the jet~$v$ in~\eqref{D12P} to~$\mathfrak{M}(R)$ which involve a scaling factor~$\s$.
Since these jets describe an infinitesimal diffeomorphism, their contribution to~\eqref{MRdef} can be written as
\beq \label{MR2}
{\Mass}(R) \asymp 2 \pi R^2 \int_\scrM (r-R)\:
\bigg(v^j(x,y)\: \frac{\partial}{\partial x^j} - v^j(y,x)\: \frac{\partial}{\partial y^j} \bigg) \L\big(x, y\big)\: d\rho(y) \:,
\eeq
where~$v^j(x,y)$ is given by the line integral in~\eqref{D12P}.
We now expand~$v^j(x,.)$ and~$v^j(.,x)$ in Taylor series about~$x$ for~$y$ along a null line through~$x$.
A direct computation shows that the zero order term of this expansion gives
precisely the contributions to~\eqref{MR2} which involve a scaling factor~$\s$.
Subtracting these contributions gives the result.
\QED
We remark that in Proposition~\ref{prpdifflin},
it is shown by explicit computation that the unbounded line integrals in~\eqref{D12P} do not contribute
to the surface layer integrals, provided that the linear perturbation of the metric does not change the volume form.
This gives an alternative, more computational proof of the above proposition for linearized gravity.

In view of this result, we may keep the point~$x$ fixed.
The change of the metric, however, can be described by the following change of coordinates:
\begin{Lemma} Choosing in Minkowski space the coordinates
\[ \tilde{t} = t + t\: \frac{M_\text{S}}{R} \:,\quad
\tilde{r} = r - (r-R)\: \frac{M_\text{S}}{R} \:,\qquad \tilde{\vartheta} = \vartheta \:,\qquad \tilde{\varphi}=\varphi \:, \]
the new metric coincides with the Schwarzschild metric near~$x$.
\end{Lemma}
\Proof A short computation gives
\begin{align*}
t &= \tilde{t} - \tilde{t}\: \frac{M_\text{S}}{R} + \O\big( M_\text{S}^2 \big) \:,&
r &= \tilde{r} + (\tilde{r}-R)\: \frac{M_\text{S}}{R} + \O\big( M_\text{S}^2 \big) \\
\frac{\partial t}{\partial \tilde{t}} &= 1- \frac{M_\text{S}}{R} \:,&
\frac{\partial r}{\partial \tilde{r}} &= 1+ \frac{M_\text{S}}{R} \\
\tilde{g}_{00} &= 1- \frac{2 M_\text{S}}{R} \:,& 
\tilde{g}_{11} &= 1+ \frac{2 M_\text{S}}{R} \:,
\end{align*}
giving agreement with the Schwarzschild metric linearly in~$M_\text{S}$.
\QED

In order to clarify the signs, we write this coordinate transformation as a diffeomorphism,
\[ \Phi \::\: (t,r,\omega) \mapsto (\tilde{t}, \tilde{r}, \omega)\:. \]
Then
\[ \tilde{g}_{ij}\, \tilde{u}^i\, \tilde{u}^j = \eta_{ij}\, u^i u^j \:, \]
where~$\tilde{g}$ and~$\eta$ are the Schwarzschild and Minkowski metrics, respectively. Therefore, the Lagrangian
in the Schwarzschild metric can be written as~$\L(x,\Phi^{-1}(y))$, where~$\L$ denotes
is the Lagrangian of the vacuum. Hence
\[ D_{2,w} \L(x,y) = \L\big(x, \Phi^{-1}(y) \big) - \L(x,y)  + \O \big( M_S^2 \big)\:, \]
and using that
\[ \Phi^{-1}(t,r,\omega) -(t,r,\omega) = \frac{M_S}{R} \: \big(-t, (r-R), 0 \big) + \O \big( M_S^2 \big) \:, \]
we conclude that
\[ D_{2,w} \L(x, y) = \frac{M_\text{S}}{R} \;
\Big( -t\: \frac{\partial}{\partial t} + (r-R)\: \frac{\partial}{\partial r} \Big) \L(x,y) + \mathscr{O} \big( M_S^2 \big)\:. \]
Hence, to first order in~$M_\text{S}$,
\begin{align*}
{\Mass}(R) 
&= 2 \pi R^2 \int_\scrM (r-R)\: \big(D_{1,w} - D_{2,w} \big) \L\big(x, y \big)\: d\rho(y) \\
&= -2 \pi R^2 \int_\scrM (r-R)\: D_{2,w} \L\big(x, y \big)\: d\rho(y) \\
&= -2 \pi R M_\text{S}
\int_{-\infty}^\infty \!dt \int_0^\infty \!dr\: (r-R)\: r^2 \int_{S^2} \!d\omega\:
\Big( -t\: \frac{\partial}{\partial t} + (r-R)\: \frac{\partial}{\partial r} \Big) \L\big(x, (t,r,\omega) \big) \\
&\overset{(*)}{=} -2 \pi R M_\text{S}
\int_{-\infty}^\infty dt \int_0^\infty dr \int_{S^2} d\omega \:
\L\big(x, (t,r,\omega) \big) \\
&\qquad \qquad \times \Big( \frac{\partial}{\partial t} \big( t\: (r-R)\: r^2 \big)
- \frac{\partial}{\partial r} \big( (r-R)^2\: r^2 \big) \Big) \\
&= 2 \pi R M_\text{S} \int_{-\infty}^\infty dt \int_0^\infty dr \int_{S^2} d\omega\:
\L\big(x, (t,r,\omega) \big) \:r\: (r-R)\:\Big( 2\, (r - R) + r \Big) \:,
\end{align*}
where in~$(*)$ we integrated by parts. Carrying out the time integration according to~\eqref{Ltime}
and again choosing polar coordinates~\eqref{xyform}, we obtain
\[ {\Mass}(R) = 2 \pi R M_\text{S} \int_0^\infty dr \int_{S^2} d\omega\:
\L(\x, \y) \:r\: (r-R)\:\Big( 2\, (r - R) + r \Big) \:. \]

From now on, we can work in Minkowski space.
Assuming spherical symmetry, the Lagrangian depends only on the Euclidean distance,
which with the law of cosines is given by
\[ |\x-\y|^2 = R^2 - r^2 - 2 R r\, \cos \vartheta = (r-R)^2 + 2 R r\, \big(1 - \cos \vartheta \big) \:. \]
In order to obtain a clean expansion in powers of~$1/R$, it is useful to transform to the new coordinates
\[ \ell(r) = r-R \qquad \text{and} \qquad \sigma(\vartheta) = \sqrt{2 R r\, \big(1 - \cos \vartheta \big)} \:. \]
Then
\[ d\ell = dr \qquad \text{and} \qquad \sigma\, d\sigma = R r \, d\cos \vartheta \:. \]
Using that~$d\omega = 2 \pi \,d\cos \vartheta$, the integral transforms to
\begin{align*}
{\Mass}(R)
&= 2 \pi R M_\text{S} \int_{-\infty}^\infty d\ell \int_0^\infty \frac{2 \pi \sigma}{R r}\, d\sigma\;
\L(\x, \y) \:r\: (r-R)\:\Big( 2\, (r - R) + r \Big) \\
&=2 \pi M_\text{S} \int_{-\infty}^\infty d\ell \int_0^\infty 2 \pi \sigma\, d\sigma\;
\L\big[ \ell^2 + \sigma^2 \big] \:\ell\:\Big( 2\, \ell + (\ell + R) \Big) \:,
\end{align*}
where the square brackets indicate that the Lagrangian depends only on~$\ell^2 + \sigma^2$.
The linear term in~$\ell$ drops out by symmetry,
\[ {\Mass}(R)
= 2 \pi M_\text{S} \int_{-\infty}^\infty d\ell \int_0^\infty 2 \pi \sigma\, d\sigma\;
\L\big[ \ell^2 + \sigma^2 \big] \:3 \ell^2 \:. \]
Now we can regard~$\ell$ and~$\sigma$ as cylindrical coordinates in~$\R^3$.
Again using spherical symmetry, we obtain
\[ {\Mass}(R) = 2 \pi M_\text{S} \int_{\R^3} |\y|^2\, \L\big[ \y^2 \big] \:d^3\y \:. \]
Again writing out the time integration and applying Proposition~\ref{prplimit}, we conclude that
the total mass is given by
\beq \label{Mform}
\Mass = 2 \pi M_\text{S} \int_{\scrM} |\y|^2\: \L\big(0, (t,\y) \big)\: d^4y \:.
\eeq
This concludes the proof of Theorem~\ref{thmcorrespond}.

\section{Scaling Behavior of Dirac Systems} \label{secscaling}
In this section we work out the scaling behavior for static Dirac systems
and analyze the questions~{\bf{(b)}}--{\bf{(e)}} posed in the introduction
on page~\pageref{btoe}. We begin by discussing the length scales
in~\eqref{lengthscales} in more detail. First of all, these length scales
are related to each other by
\[ \varepsilon \lesssim \delta \ll \frac{1}{m} \:. \]
At first sight, it might seem natural to identify the regularization length~$\varepsilon$ with
the Planck scale~$\delta$. However, as is explained in detail in~\cite[Chapter~4]{cfs},
it is preferable to regard~$\varepsilon$ and~$\delta$ as two different parameters,
where typically~$\varepsilon \ll \delta$.

We first recall the scalings in the {\em{Minkowski vacuum}}
as worked out in~\cite{action} and~\cite[Appendix~A]{jacobson}. The scaling of the function~${\mathfrak{t}}$
in~\eqref{tdef} can be obtained in a straightforward way from a dimensional argument,
\[ \mathfrak{t}(x) \simeq \frac{\sigma\, \lambda^4}{\varepsilon^8}\:, \]
where~$\sigma$ and~$\lambda$ are the parameters describing the rescaling freedom
(see~\eqref{tilrho} in Section~\ref{secrescale}). The contributions to the Lagrangian
are less obvious because of our limited knowledge about the structure of physical spacetime
on the Planck scale. Nevertheless, the Lagrangian can be written as
\[ \ell(x) + \s \simeq \sigma \,\lambda^4
\bigg( \frac{(\varepsilon m)^p}{\varepsilon^8} + \frac{1}{\delta^8}
\: \Big(\frac{\delta}{\varepsilon}\Big)^{\hat{s}} \bigg) \]
with parameters~$p$ and~$\hat{s}$, which from general considerations are known to be in the range
\[ 
p \geq 5 \qquad \text{and} \qquad \hat{s} \in \{0,2\}\:. \]

We next explain how to choose the parameters~$\sigma$ and~$\lambda$ used in the rescaling.
It is most convenient to work in the units of length determined by the measure~$\tilde{\mu}$.
In order to keep this length scale fixed, we must not rescale the volume, meaning that~$\sigma$
must be kept fixed. For simplicity, we choose
\[ \sigma = 1 \qquad \text{for all~$\tau$}\:. \]
Moreover, the local trace must be kept constant~\eqref{trc}.
As is worked out in detail in~\cite[Section~2.5]{cfs} and~\cite[Appendix~A.3]{jacobson}, this trace
scales like
\[ \tr(x) = \Tr \big( P(x,x) \big) \simeq \frac{\lambda m}{\varepsilon^2} \:. \]
This leads us to choose~$\lambda \simeq \varepsilon^2/m$. In this way, the freedom in scaling
the measure is exhausted. Moreover, the above formulas for~${\mathfrak{t}}(x)$
and~$\ell(x) + \s$ simplify to
\beq \label{scaleMink}
\mathfrak{t}(x) \simeq \frac{1}{m^4}
\qquad \text{and} \qquad
\ell(x) + \s \simeq 
\bigg( \frac{(\varepsilon m)^p}{m^4} + \frac{1}{m^8}
\: \Big(\frac{\varepsilon}{\delta}\Big)^{8-\hat{s}} \bigg) \:.
\eeq

In {\em{curved spacetime}}, the functions~$\ell(x)$ and~${\mathfrak{t}}(x)$ vary in
spacetime, but the parameters in~\eqref{lengthscales} as well as the Lagrange parameters~$\s$, $\kappa$
and the parameter~$c$ in~\eqref{trc} are still constants. Moreover, we always consider critical
measures. The resulting weak EL equations~\eqref{ELtest} (again for the $\kappa$-Lagrangian;
see also~\eqref{lin2}) read
\[ \nabla_\u \big( \ell + \kappa \, \mathfrak{t} \big)|_N = 0 \:. \]
In Definition~\ref{defkappaextend} we consider a family~$(\mu_\tau)_{\tau \in (-1,1)}$
of critical static measures for a {\em{de}}creasing value of~$\kappa$.
The crucial question is how the parameters~$m$ and~$\delta$
change when we decrease~$\kappa$ for a measure describing the Minkowski vacuum.
The parameter~$\kappa$ is the Lagrange multiplier corresponding to the boundedness
constraint. Therefore, decreasing~$\kappa$ corresponds to weakening the boundedness
constraint by increasing the parameter~$C$ in~\eqref{Tdef}.
This has the effect that the functional in~\eqref{Tdef} becomes larger, whereas the
causal action~\eqref{Sdef} becomes smaller. For a translation invariant system like Minkowski space,
this means that~${\mathfrak{t}}(x)$ {\em{in}}creases, whereas~$\ell(x)+\s$ {\em{de}}creases. In view of the
left side of~\eqref{scaleMink}, this means that the mass of the Dirac particles decreases,
\beq \label{mscale}
\frac{d}{d\tau} \log m \Big|_{\tau=0} < 0 \:.
\eeq
The scaling behavior of~$\delta$ is less obvious because of the unknown parameters~$p$ and~$\hat{s}$
in~\eqref{scaleMink}. In view of the fact that the parameter~$\delta$
was introduced in~\cite[Chapter~4]{cfs} in order to compensate for the fact that the neutrino masses
differ from the masses of the charged leptons, it is natural to assume that it scales in the same
way as~$1/m$, i.e.\
\beq \label{mdscale}
\frac{d}{d\tau} \log \big(m\, \delta \big) \Big|_{\tau=0} = 0 \:.
\eeq
This {\em{natural scaling of~$\delta$}} means that when~$\kappa$ is decreased, 
the Compton length and the Planck length are increased by the same factor.
When considering a gravitating system, the natural scaling of~$\delta$ has the convenient property
that the interaction remains unchanged, only the size of the whole system changes.
Keeping in mind that the length scale is determined by the measure,
one can also say that, assuming natural scaling of~$\delta$, only the measure~$\tilde{\mu}$
and the regularization length~$\varepsilon$ are rescaled, but otherwise the system remains unchanged.
Although the natural scaling of~$\delta$ seems reasonable and sensible, it is not compelling, neither for mathematical
nor for physical reasons. With our present knowledge, it is conceivable that for families of
minimizers of the causal action, the parameter~$\delta$ might have a different
or more complicated scaling behavior. 

Having the above scalings in mind, the relation~\eqref{wdef} has a direct meaning:
The vector field~$v$ changes the parameters~$m$, $\delta$ and~$\varepsilon$
for the measure describing the Minkowski vacuum. Likewise, the vector field~$\tilde{v}$
describes a variation of the gravitating system. Since asymptotically at infinity,
the gravitating system goes over to Minkowski space, the parameters~$m$, $\delta$ and~$\varepsilon$
of the gravitating system change just as in Minkowski space. As a consequence, the effect of the change of these
parameters drops out when taking the {\em{difference}} of~$v$ and~$\tilde{v}$ in~\eqref{wdef}.
What remains is the change of the gravitating system, which in turn
changes the strength of the gravitational field at infinity as described by the vector field~$w$
in~\eqref{wdef}. This consideration also explains why static Dirac systems are $\kappa$-scalable
in the sense of Definition~\ref{defkappascale}, thus answering question~{\bf{(b)}}
on page~\pageref{btoe}.

The above scaling analysis also makes it possible to address the questions~{\bf{(c)}}--{\bf{(e)}}
on page~\pageref{btoe}. Making these considerations mathematically precise goes beyond the
scope of this paper. Instead, we merely discuss these questions in remarks which explain connections
to be explored in more detail in the future.

\begin{Remark} {\bf{Why is the gravitational force attractive?}} \label{remattractive} {\em{
Having explained how the relation~\eqref{wdef} comes about, we can now
consider the sign of the gravitational coupling constant. As explained above, the
jet~$\tilde{v}$ describes a change of the gravitating system due to a variation of
the parameters~$m$, $\varepsilon$ and~$\delta$,
on the length scale determined by the measure~$\tilde{\mu}$.
Clearly, in view of~\eqref{osilin2} we only need to be concerned about the behavior near spatial infinity.
Thus the question is how the strength of the gravitational field changes
near infinity if the parameters~$m$, $\varepsilon$ and~$\delta$ are varied infinitesimally.
Clearly, increasing~$m$ makes the gravitational field stronger.
In general relativity, the strength of the gravitational field is given by~$G m$ (having the dimension of length).
Keeping in mind that~$G \sim \delta^2$, in the causal fermion system the corresponding quantity
is~$m\, \delta^2$. Thus for a static spacetime which is asymptotically Schwarzschild,
the change of the gravitational field is given by the quantity
\[ 
\frac{d}{d\tau} \log \big(m\, \delta^2 \big) \Big|_{\tau=0} \:. \]
Consequently, the gravitational constant~$g$ in~\eqref{wdef} has the same sign as this quantity.
If~$\delta$ has the natural scaling, then this sign is {\em{positive}}
in view of~\eqref{mscale} and~\eqref{mdscale}.
More generally, the gravitational constant is positive, provided that
the parameter~$m \delta$ does not decrease too fast if~$\kappa$ is decreased.
This seems a very sensible and natural assumption. However, exactly as explained after~\eqref{mdscale}
for the natural scaling, there seems no compelling mathematical argument
which explains the sign of~$g$.
}} \QEDrem
\end{Remark}

\begin{Remark} {\bf{Why is the local energy condition satisfied?}} \label{remedp} {\em{
It is sensible to assume that the measure~$\mu$ describing the Minkowski vacuum is
a minimizer of the causal action principle. Writing the action as
\[ \Sact = \int_N d\mu(\x) \int_M d\rho(y)\: \L(\x,y) = \int_N \big( \ell(\x) + \s \big) \: d\rho(\x) \:, \]
the function~$\ell(\x)+\s$ can be regarded as the action per spatial volume. Likewise,
in an asymptotically flat spacetime, the function~$\tilde{\ell}(\x)-\tilde{\ell}_\infty$ tells about how
the action per volume at~$\x$ differs from the action per volume in the vacuum.
With this in mind, the positivity of this function as imposed by the local energy condition in Definition~\ref{deflec}
seems to be a direct consequence of the minimality of the causal action in the vacuum.
Although being correct in principle, it seems that this argument cannot be made precise
in a simple way. The basic difficulty is that comparing the actions per volume
in a rigorous way makes sense only in the homogeneous setting where these actions per volume
are constant. Using a rescaling method where one ``zooms into'' spacetime on smaller and smaller
scales, one can construct a homogeneous measure for which~$\tilde{\ell}$ coincides with~$\tilde{\ell}(\x)$
of our curved spacetime. But in this limiting case, the rest mass~$m$ tends to zero.
As a consequence, the contribution by matter to the function~$\tilde{\ell}$ also tends to zero,
which also means that the information on the sign of the energy density gets lost.
This consideration shows that, in order to relate the sign of~$\tilde{\ell}(\x)-\tilde{\ell}_\infty$ to the minimality
of the causal action in the vacuum, one needs to enter a quantitative analysis of the scaling behavior
of different contributions to~$\tilde{\ell}(\x)$.

In view of these difficulties, here we are content with the following weaker statement:
\begin{quote}
Assume that a contribution by matter to~$\tilde{\ell}(\x)$ has the property
that the same contribution can be arranged for a {\em{homogeneous}} physical system
(i.e.\ for a system where~$\tilde{\ell}(\x)$ is constant). Then this contribution to~$\tilde{\ell}(\x)-\tilde{\ell}_\infty$
is necessarily non-negative.
\end{quote}
For the resulting homogeneous systems, the minimality of~$\ell$ for the vacuum immediately
gives the result. Typical examples for contributions to~$\tilde{\ell}$ which can be ``homogenized''
in the above sense are the energy-momentum tensor of Dirac particles (where in the corresponding
homogeneous system one replaces the Dirac wave functions of matter by a plane wave)
or the energy-momentum tensor for a Maxwell field (in which case for the homogeneous system one
takes a plane electromagnetic wave). In this formulation, our argument explains why the local energy condition
is satisfied for classical matter on large scales. But we cannot exclude the possibility that there
might be quantum fluctuations on microscopic scales which violate the local energy condi\-tion. \\
}} \hspace*{1cm} \QEDrem
\end{Remark}

\begin{Remark} {\bf{Why do vacuum measures describe flat spacetime?}} \label{remflat} {\em{
For static Dirac systems, there is a direct way of understanding why a vacuum measure
according to Definition~\ref{defvacuum} describes Minkowski space.
Indeed, as is shown explicitly in~\cite[Appendix~A]{firstorder}, the curvature tensor enters the kernel of the
fermionic projector~$P(x,y)$. As a consequence, the curvature tensor has an effect
on the eigenvalues~$\lambda^{xy}_i$ of the closed chain which enter the Lagrangian~\eqref{Lagrange}.
The contributions involving the Ricci tensor are worked out in more detail in~\cite[\S4.5.2]{cfs}.
Even without working out the detailed form of the resulting contributions, it is clear
from the formulas in~\cite[Appendix~A]{firstorder} that also the components of the Weyl tensor
come into play. Since the Lagrangian is non-negative and vanishes in the continuum limit, the
resulting curvature contributions are strictly positive. Therefore, the condition~\eqref{ellvac}
is satisfied if and only if the curvature tensor vanishes identically.
}} \QEDrem
\end{Remark}

\appendix
\section{Scaling Behavior of the Total Mass for Static Dirac Systems} \label{appscalingmass}
The goal of this section is to show that for static Dirac systems, the total mass scales like
\[ \mathfrak{M}(\tilde{\mu}, \mu) \lesssim \frac{\varepsilon}{l_{\text{\tiny{macro}}}}\: \s \:, \]
where~$l_{\text{\tiny{macro}}}$ denotes the length scale on which the gravitational field changes.
In order to derive this scaling behavior, we deform spacetime such as to obtain a spacetime
with zero total mass. Then we analyze how the total mass changes along the path describing the
deformation. Before entering the construction, we point out that the spacetimes along the
deformed spacetimes are in general not static. Therefore, this appendix is the only place in this
paper where we consider the time-dependent setting as introduced in Section~\ref{seccfs}.

\begin{figure}
%
\psscalebox{1.0 1.0} 
{
\begin{pspicture}(0.3,-0.9479111)(14.955736,0.9479111)
\definecolor{colour0}{rgb}{0.8,0.8,0.8}
\pspolygon[linecolor=colour0, linewidth=0.02, fillstyle=solid,fillcolor=colour0](10.305245,-0.68427676)(10.845245,-0.6642767)(11.4352455,-0.62927675)(11.865246,-0.5992767)(12.115246,-0.57427675)(12.135245,-0.6042767)(12.200245,-0.6042767)(12.220245,-0.58927673)(12.345245,-0.57427675)(12.405246,-0.58927673)(12.440246,-0.56427675)(12.595245,-0.5492767)(12.860246,-0.62927675)(13.315246,-0.6792767)
\pspolygon[linecolor=colour0, linewidth=0.02, fillstyle=solid,fillcolor=colour0](12.225245,-0.39927673)(12.370245,-0.40927672)(12.485246,-0.45427674)(12.600245,-0.56427675)(12.4352455,-0.5442767)(12.420245,-0.48927674)(12.385245,-0.46427673)(12.335245,-0.50427675)(12.335245,-0.57927674)(12.225245,-0.56427675)(12.205245,-0.50427675)(12.170245,-0.48927674)(12.135245,-0.5292767)(12.110246,-0.6042767)(11.980246,-0.57927674)(12.120245,-0.44927675)
\psbezier[linecolor=black, linewidth=0.02, fillstyle=solid,fillcolor=colour0](9.215089,-0.6529436)(9.977128,-0.7116009)(11.386855,-0.7002764)(12.036971,-0.6991437791361181)(12.687088,-0.69801116)(13.975498,-0.73890966)(14.895134,-0.6868016)
\pspolygon[linecolor=colour0, linewidth=0.02, fillstyle=solid,fillcolor=colour0](0.05024536,-0.6642767)(8.280246,-0.68427676)(7.400245,-0.62427676)(6.960245,-0.52427673)(6.610245,-0.38427675)(6.0802455,-0.06427673)(5.3902454,-0.044276733)(5.2802453,-0.09427673)(5.170245,-0.14427674)(4.9102454,-0.18427673)(4.7202454,-0.10427673)(4.6402454,0.0057232664)(4.6002455,0.17572327)(4.5802455,0.37572327)(3.8602455,0.31572327)(3.8102453,0.17572327)(3.7402453,0.0057232664)(3.6302454,-0.11427674)(3.4202454,-0.15427673)(3.2002454,-0.17427674)(2.7302454,0.13572326)(2.5902452,-0.044276733)(2.3402452,-0.26427674)(2.1302454,-0.37427673)(1.8002454,-0.52427673)(1.3702453,-0.5942767)(0.78024536,-0.63427675)
\pspolygon[linecolor=colour0, linewidth=0.02, fillstyle=solid,fillcolor=colour0](3.6552453,0.8857233)(4.1952453,0.91572326)(4.6152453,0.9007233)(5.005245,0.87072325)(5.400245,0.7207233)(5.6652455,0.47572327)(5.7902455,0.27072325)(5.9752455,0.055723265)(6.150245,-0.09927674)(5.295245,-0.044276733)(5.2202454,0.10072327)(5.1352453,0.21572326)(5.0152454,0.32072327)(4.860245,0.36072326)(4.7152452,0.35572326)(4.5952454,0.34072328)(3.8352454,0.33572328)(3.8202453,0.26072326)(3.7202454,0.24572326)(3.5752454,0.18572326)(3.4702454,0.12072327)(3.3802454,0.015723266)(3.3252454,-0.07927673)(3.3102453,-0.13927673)(2.9652452,-0.15427673)(2.7352455,0.14572327)(2.8752453,0.38572326)(3.0052454,0.5957233)(3.2102454,0.7607233)(3.4302454,0.8457233)
\rput[bl](7.5802455,0.40572327){\normalsize{$F_\tau$}}
\psbezier[linecolor=black, linewidth=0.04, arrowsize=0.05291667cm 4.0,arrowlength=1.4,arrowinset=0.0]{->}(6.7702456,-0.054276735)(7.3302455,0.41572326)(8.570246,0.18572326)(9.0602455,-0.0842767333984375)
\rput[bl](2.2102454,0.25572327){\normalsize{$\tilde{N}$}}
\psbezier[linecolor=black, linewidth=0.04](9.215245,-0.64731246)(10.861139,-0.67206985)(11.963539,-0.61048585)(12.025083,-0.5426695905412907)(12.086626,-0.4748533)(12.186235,-0.38356328)(12.325228,-0.38927674)(12.464219,-0.3949902)(12.495013,-0.4834946)(12.6153965,-0.5671339)(12.73578,-0.65077317)(13.988632,-0.70747757)(14.955245,-0.68427676)
\psbezier[linecolor=black, linewidth=0.02](3.3056798,-0.159792)(3.4125028,0.16684057)(3.674839,0.20425065)(3.814811,0.2512385425925072)
\psbezier[linecolor=black, linewidth=0.02](4.6002455,0.32572326)(4.9502454,0.38572326)(5.1002455,0.32572326)(5.3002453,-0.0542767333984375)
\psbezier[linecolor=black, linewidth=0.04](3.1802454,-0.16427673)(3.6202455,-0.11427674)(3.7002454,-0.19427674)(3.8502455,0.3457232666015625)
\psbezier[linecolor=black, linewidth=0.04](4.5902452,0.40572327)(4.5902452,-0.39427674)(5.130245,-0.15427673)(5.3902454,-0.0242767333984375)
\psbezier[linecolor=black, linewidth=0.02, fillstyle=solid,fillcolor=colour0](0.04024536,-0.6542767)(1.2117486,-0.8590868)(3.380252,-0.9306448)(4.380245,-0.9342767333984375)(5.3802385,-0.93790865)(6.908211,-0.91802883)(8.300245,-0.68427676)
\psbezier[linecolor=black, linewidth=0.04](0.0,-0.6373125)(1.3302472,-0.62206984)(1.730882,-0.66048586)(2.3902454,-0.17766959054128903)(3.0496087,0.3051467)(2.687982,0.89143676)(3.9502454,0.91572326)(5.2125087,0.94000983)(5.430293,0.82150537)(5.7702456,0.30286613)(6.1101975,-0.21577315)(6.894213,-0.6974776)(8.290245,-0.6742767)
\psbezier[linecolor=black, linewidth=0.02](12.140245,-0.5342767)(12.090245,-0.63927674)(12.212467,-0.60245854)(12.215245,-0.5792767333984375)(12.218023,-0.55609494)(12.190246,-0.42927673)(12.140245,-0.5342767)
\psbezier[linecolor=black, linewidth=0.02](12.355246,-0.51427674)(12.305245,-0.61927676)(12.427467,-0.58245856)(12.430245,-0.5592767333984375)(12.433023,-0.5360949)(12.405246,-0.40927672)(12.355246,-0.51427674)
\rput[bl](9.455245,-0.42927673){\normalsize{$\tilde{N}_\tau$}}
\end{pspicture}
}
\caption{Deformation of the initial data.}
\label{figdeform}
\end{figure}
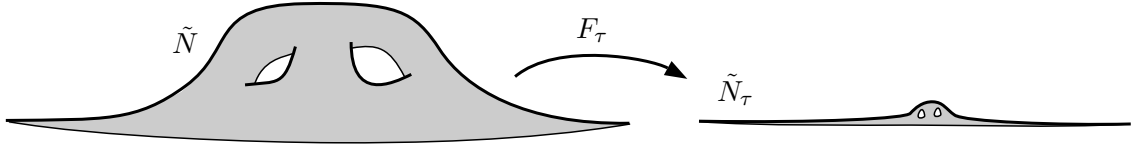

We let~$\tilde{\rho} = dt \,d\tilde{\mu}$ be the static spacetime of interest.
We assume that the corresponding spacetime is of the form~$M = \R \times N$,
where~$N$ is a surface of constant time. Typically, the measure~$\tilde{\rho}$
will be a causal fermion system constructed from a static Lorentzian spacetime
as explained in Section~\ref{seclorentz}, but this is not essential for the following construction.
We first deform the measure~$\tilde{\mu}$ such as to obtain a family of measures~$(\tilde{\mu}_\tau)_{\tau \in [0,1)}$
with the following properties:
\bitem
\item[(i)] For~$\tau=0$, we get back the static measure describing our spacetime of interest, i.e.
\[ \tilde{\mu}_0 = \tilde{\mu} \:. \]
\item[(ii)] In the limit~$\tau \nearrow 1$ and outside a compact set, the measures~$\tilde{\mu}_\tau$
should go over to a measure describing a spatial hyperplane in Minkowski space.
\item[(iii)] The deformation should be smooth in the sense that
\[ \tilde{N}_\tau = F_\tau(N) \qquad \text{with} \qquad F \in C^\infty([0,1) \times \tilde{N}, \F)\:. \] 
\end{itemize}
This deformation is illustrated in Figure~\ref{figdeform}.
The next step is to extend the measures~$\tilde{\mu}_\tau$ to
critical  measures~$\tilde{\rho}_\tau$  in spacetime.
To this end, one can proceed in two alternative ways:
One method is to solve the classical field equations (Einstein equations coupled to Dirac and possibly other fields)
with initial data~$\tilde{N}_\tau$, and then to define~$\tilde{\rho}_\tau$ as in Section~\ref{seclorentz}
as the push-forward of the volume measure under the local correlation map.
Alternatively, one can also solve directly the EL equations of the causal action
using the results and methods in~\cite{linhyp}: 
Generalizing~\eqref{rhoFf} to the time-dependent setting, we make the ansatz
\[ 
\tilde{\rho}_\tau = (F_\tau)_* \big( f_\tau \, \rho \big) \:, \]
where~$f$ and~$F$ are smooth,
\[ f \in C^\infty\big([0,1) \times M \rightarrow \R^+ \big) \qquad \text{and} \qquad
F \in C^\infty\big([0,1) \times M \rightarrow \F \big) \:. \]
Next, using that
\[ \tilde{\rho}_\tau = \tilde{\rho}_0 + \int_0^\tau \dot{\tilde{\rho}}_s\: ds \:, \]
it suffices to solve the linearized field equations for each~$\tau$ for the jet
\[ \v_\tau(x) = \frac{d}{d\tau} \big(f_\tau, F_\tau \big) \:. \]
As shown in~\cite{linhyp}, this Cauchy problem can be solved with energy methods.
Before going on, we point out that the above procedure does {\em{not}} work if the
solutions develop singularities, as is the case for example if a black hole forms.
Thus, our deformation method requires that the deformed initial data admits solutions
of the Cauchy problem which exist for a time which is sufficiently large for the surface layer integrals
to be well-defined. This requirement seems sensible for most applications in mind.

Our next step is to incorporate the volume constraint by arranging that the weight function~$f_\tau$
vanishes identically, i.e.\
\[ \tilde{\rho}_\tau = (F_\tau)_* \tilde{\rho} \qquad \text{with} \qquad F \in C^\infty([0,1) \times M, \F) \:. \]
This can be arranged by Moser's theorem (i.e.\ by integrating the infinitesimal version
of Lemma~\ref{lemmamoser} or similarly~\cite[Proposition~3.6]{fockbosonic} in the hyperbolic
setting).

After these preparation, we can analyze how the total mass changes along the curve~$(\tilde{\rho}_\tau)_{\tau \in [0,1]}$.
At~$\tau=0$, we can compute the total mass via~\eqref{osilin2}, which in the spacetime setting we write as
\[ {\mathfrak{M}}
= \lim_{V \nearrow M} \int_{V \cap N} \!\!\!\!\!\!\!\!d\mu(x) \int_{M \setminus V} \!\!\!\!\!\!\!\!d\rho(y) \big( D_{1,w} - D_{2,w} \big) \L(x,y) \:, \]
and~$V$ is chosen of the form~$\R \times \Omega$ with~$\Omega \subset N$.
Since the system is static, we can incorporate the time integral equivalently in the first argument. We thus obtain
the alternative formula for the total mas
\[ {\mathfrak{M}}
= \frac{1}{2} \:\lim_{V \nearrow M} \bigg( \int_{V \cap N} \!\!\!\!\!\!\!\!d\mu(x) \int_{M \setminus V} \!\!\!\!\!\!\!\!d\rho(y) 
+ \int_V \!\!d\rho(x) \int_{N \setminus V} \!\!\!\!\!\!\!\!d\mu(y) \bigg) \big( D_{1,w} - D_{2,w} \big) \L(x,y) \:, \]
which is indeed preferable for the following consideration.
Applying the fundamental theorem of calculus and using that the total mass vanishes in the limit~$\tau \nearrow 1$,
we obtain
\begin{align*}
{\mathfrak{M}} = -\frac{1}{2} \:\lim_{V \nearrow M} \int_0^1 d\tau
\bigg( \int_{V \cap N}& \!\!\!\!\!\!\!\!d\mu(x) \int_{M \setminus V} \!\!\!\!\!\!\!\!d\rho(y) 
+ \int_V \!\!d\rho(x) \int_{N \setminus V} \!\!\!\!\!\!\!\!d\mu(y) \bigg) \\
&\qquad \times \big( D_{1,v_\tau} - D_{2,v_\tau} \big) \L(x,y) \:,
\end{align*}
where~$v_\tau$ is the vector field~$v_\tau = dw/d\tau$.
Since only the asymptotics near spatial infinity enters, this surface layer integral can be computed just as well using the perturbed measures,
\begin{align*}
{\mathfrak{M}}(V) = -\frac{1}{2} \:\lim_{\tilde{V} \nearrow \tilde{M}} \int_0^1 d\tau
\bigg( \int_{\tilde{V} \cap \tilde{N}}& \!\!\!\!\!\!\!\!d\tilde{\mu}_\tau(x) \int_{M \setminus V} \!\!\!\!\!\!\!\!d\tilde{\rho}_\tau(y) 
+ \int_{\tilde{V}} \!\!d\tilde{\rho}_\tau(x) \int_{\tilde{N} \setminus \tilde{V}} \!\!\!\!\!\!\!\!d\tilde{\mu}_\tau(y) \bigg) \\
&\qquad \times \big( D_{1,v_\tau} - D_{2,v_\tau} \big) \L\big( F_\tau(x), F_\tau(y)\big) \:.
\end{align*}
This has the advantage that the jets~$v_\tau$ are defined globally.
Using that the integrand is anti-symmetric, we obtain
\begin{align*}
\bigg( &\int_{\tilde{V} \cap \tilde{N}} \!\!\!\!\!\!\!\!d\tilde{\mu}_\tau(x) \int_{M \setminus V} \!\!\!\!\!\!\!\!d\tilde{\rho}_\tau(y) 
+ \int_{\tilde{V}} \!\!d\tilde{\rho}_\tau(x) \int_{\tilde{N} \setminus \tilde{V}} \!\!\!\!\!\!\!\!d\tilde{\mu}_\tau(y) \bigg) \big( D_{1,v_\tau} - D_{2,v_\tau} \big) \L\big( F_\tau(x), F_\tau(y)\big) \\
&= \bigg( \int_{\tilde{V} \cap \tilde{N}} \!\!\!\!\!\!\!\!d\tilde{\mu}_\tau(x) \int_{M} \!\!d\tilde{\rho}_\tau(y) 
+ \int_{\tilde{V}} \!\!d\tilde{\rho}_\tau(x) \int_{\tilde{N}} \!\!d\tilde{\mu}_\tau(y) \bigg) \big( D_{1,v_\tau} - D_{2,v_\tau} \big) \L\big( F_\tau(x), F_\tau(y)\big) \\
&= \bigg( -\int_{\tilde{V} \cap \tilde{N}} \!\!\!\!\!\!\!\!d\tilde{\mu}_\tau(x) \int_{M} \!\!d\tilde{\rho}_\tau(y) 
+ \int_{\tilde{V}} \!\!d\tilde{\rho}_\tau(x) \int_{\tilde{N}} \!\!d\tilde{\mu}_\tau(y) \bigg) \big( D_{1,v_\tau} - D_{2,v_\tau} \big) \L\big( F_\tau(x), F_\tau(y)\big)\:,
\end{align*}
where in the last step we used the EL equations and the linearized field equations.
The last expression is anti-symmetrized in the time integrals. As a consequence, we only get
a contribution if either~$v_\tau$ or~$F_\tau$ are differentiated with respect to time. This gives the desired
scaling factor~$\varepsilon/l_{\text{\tiny{macro}}}$.

\section{Explicit Treatment of a Linearized Gravitational Field} \label{appgeodesic}
We now explain geometrically how the formulas~\eqref{Pform}, \eqref{Pformstatic}
and~\eqref{D12P} for the kernel of the fermionic projector in the presence of linearized gravity
come about. Moreover, we analyze how the line integrals of the metric perturbation enter the surface layer integral
which defines the total mass. Clearly, in curved spacetime the boundary of the light cone is
generated by null geodesics. Perturbing the geodesic equation
\[ \ddot{\gamma}^i(\alpha) = -\Gamma^i_{jk} \big(\gamma(\alpha) \big)
\:\dot{\gamma}^j(\alpha)\: \dot{\gamma}^k(\alpha) \]
to first order about the straight line~$\alpha y + (1-\alpha)\,x$ gives
\[ \ddot{\gamma}^i(\alpha) = -\Gamma^i_{jk}|_{\alpha y + (1-\alpha)\, x}\: \xi^j \xi^k \:, \]
with the linearized Christoffel symbols given by
\beq \label{Glin}
\Gamma^i_{jk} = \frac{1}{2}\: \eta^{il} \big( \partial_j h_{lk} + \partial_k h_{lj} - \partial_l h_{jk} \big)
\eeq
(where we again set~$\xi=y-x$).
Integrating by parts twice and keeping the geodesic fixed at $\alpha=-\infty$, one sees that
\begin{align}
\Delta \gamma^i|_x &= -\int_{-\infty}^0 \alpha\: \ddot{\gamma}^i(\alpha)\: d\alpha
= \int_{-\infty}^0 \alpha\: \Gamma^i_{jk} \big|_{\alpha y + (1-\alpha)\, x}\: \xi^j \xi^k\: d\alpha \notag \\
&= \frac{1}{2} \int_{-\infty}^0 \alpha\: \big( 2\,\partial_j h^i_{\;k} - \partial^i h_{jk} \big)
\big|_{\alpha y + (1-\alpha)\, x}\: \xi^j \xi^k\: d\alpha \:, \label{Delgam}
\end{align}
where in the last step we used~\eqref{Glin}.

The transformation~$x \mapsto x + \Delta \gamma|_x$ describes an infinitesimal
diffeomorphism. Therefore, in a a suitable gauge and ignoring curvature terms,
the transformation of the kernel of the fermionic projector
is described by an infinitesimal coordinate transformation,
\begin{align}
\Delta P(x,y) &= \bigg( \Delta \gamma^i|_x \frac{\partial}{\partial x^i}
+ \Delta \gamma^i|_y  \frac{\partial}{\partial y^i} \bigg) P(x,y) \notag \\
&= -\bigg( \Delta \gamma^i|_x - \Delta \gamma^i|_y \bigg) \: \frac{\partial}{\partial y^i}  P(x,y) \:, \label{DelP}
\end{align}
where in the last step we used that the kernel~$P(x,y)$ in the Minkowski vacuum depends only
on the difference vector~$y-x$. 

The above formulas can be further simplified. We begin with the {\em{unregularized kernel}}.
In this case, we know from Lorentz invariance that the partial derivatives of~$P(x,y)$ are
proportional to the vector~$\xi$. Contracting~\eqref{Delgam} with~$\xi$, we obtain
\beq \label{contract}
\Delta \gamma^i|_x \:\xi_i = \frac{1}{2} \int_{-\infty}^0 \alpha\: \partial_j h^i_{\;k}
\big|_{\alpha y + (1-\alpha)\, x}\: \xi^j \xi^k \xi_i \: d\alpha \:.
\eeq
Rewriting the directional derivative~$\xi^j \partial_j$ as an $\alpha$-derivative, we can
integrate by parts to obtain
\beq \label{DelPunbounded}
\Delta \gamma^i|_x \:\xi_i= -\frac{1}{2} \int_{-\infty}^0 h^i_{\;k} 
\big|_{\alpha y + (1-\alpha)\, x}\: \xi^k\: \xi_i\: d\alpha \:.
\eeq
Using this formula in~\eqref{DelP} gives
\begin{align}
\Delta P(x,y) &= -\frac{1}{2} \:
\bigg( \int_{-\infty}^0 - \int_{-\infty}^1 \bigg)\, d\alpha \;h^i_{\;k} 
\big|_{\alpha y + (1-\alpha)\, x}\: \xi^k\: \frac{\partial}{\partial y^i} P(x,y) \notag \\
&= \frac{1}{2} \int_0^1 d\alpha \;h^i_{\;k} 
\big|_{\alpha y + (1-\alpha)\, x}\: \xi^k\: \frac{\partial}{\partial y^i} P(x,y) \:.
\label{DelPeich}
\end{align}
We thus obtain~\eqref{Pform}.

Before going on, we make a few remarks. Clearly, keeping the geodesic fixed at~$\alpha=-\infty$
was an arbitrary choice. If instead we had kept the geodesic fixed at~$\alpha=+\infty$,
the unbounded line integrals would have to be modified according to the replacement rules
\[ \int_{-\infty}^0 \rightarrow -\int_0^\infty \qquad \text{and} \qquad 
\int_{-\infty}^1 \rightarrow -\int_1^\infty \:. \]
However, this has no effect on expressions involving bounded line integrals like~\eqref{DelPeich}.
We also remark that the appearance of unbounded line integrals in~\eqref{DelPunbounded}
motivates the line integrals in~\eqref{D12P}. The only additional ingredient needed in order to derive~\eqref{D12P}
is that the way causality is incorporated in the so-called causal perturbation expansion
(see~\cite{sea} or the more recent paper~\cite{norm}) has the consequence that
one must always take the {\em{arithmetic mean}} of the expressions obtained in the cases
when the geodesic is fixed at~$\alpha=-\infty$ and~$\alpha=+\infty$.

We next consider the case {\em{with regularization}}.
Since the regularized kernel~$P^\varepsilon(x,y)$ is not Lorentz invariant, the method used after~\eqref{contract}
no longer applies. But we can derive similar results in the {\em{static}} and {\em{spherically symmetric}}
situation as follows. The first summand in the integrand in~\eqref{Delgam} contains a
derivative~$\xi^j \partial_j$ which we can again rewrite as an $\alpha$-derivative and integrate by parts.
The second summand in~\eqref{Delgam}, however, is not of this form.
In order to study this summand in more detail, we consider its contribution to
one of the terms in~\eqref{DelP},
\beq \label{DelPeps}
\Delta \gamma^i|_x\: \frac{\partial}{\partial y^i} P^\varepsilon(x,y) \asymp
-\frac{1}{2} \int_{-\infty}^0 d\alpha\: \alpha\: \partial^i h_{jk}\big|_{\alpha y + (1-\alpha)\, x}\: \xi^j \xi^k\:
\frac{\partial}{\partial y^i}  P^\varepsilon(x,y) \:.
\eeq
Using spherical symmetry and homogeneity, we can write the unregularized kernel of the vacuum as
\beq \label{Pepsstatic}
P^\varepsilon(x,y) = P^\varepsilon\big[ \xi^i \xi_i, \xi^0 \big] \:.
\eeq
Moreover, since~$h_{jk}$ is static, the index~$i$ in~\eqref{DelPeps} is not zero.
Therefore, the partial derivatives do not act on the second argument of the kernel in~\eqref{Pepsstatic}.
Consequently, the product rule gives again a factor~$\xi_i$, making it possible 
to integrate by parts exactly as explained after~\eqref{contract}. We conclude that
\[ \Delta \gamma^i|_x \: \frac{\partial}{\partial y^i}  P^\varepsilon\big[ \xi^i \xi_i, \xi^0 \big]
= -\frac{1}{2} \int_{-\infty}^0 d\alpha \;h^i_{\;k} 
\big|_{\alpha y + (1-\alpha)\, x}\: \xi^k\: \frac{\partial}{\partial y^i} P^\varepsilon\big[ \xi^i \xi_i, \xi^0 \big] \:. \]
Using this formula in~\eqref{DelP} gives~\eqref{Pformstatic}.

We finally analyze how the terms~\eqref{D12P} contribute to the surface layer integral~\eqref{osilin2}
describing the total mass. Since the jets in~\eqref{D12P} describe an infinitesimal diffeomorphism,
their contribution to~\eqref{osilin2} can be written in analogy to~\eqref{MR2} as
\beq \label{Mcontrib}
\Mass(\tilde{\mu}, \mu) \asymp \lim_{\Omega \nearrow N}
\int_\Omega d\mu(\x) \int_{\R \times (N\setminus \Omega)} \!\!\!\!\!\!\!\!\!\!\!\!\!\!\!\!d\rho(y)\: 
\bigg(v^j(x,y)\: \frac{\partial}{\partial x^j} - v^j(y,x)\: \frac{\partial}{\partial y^j} \bigg) \L_\kappa(\x,\y)
\eeq
with
\beq \label{vform}
v^j(x,y) := \frac{1}{4} \int_{-\infty}^\infty d\alpha\: \epsilon(\alpha)\:
h^j_k|_{\alpha y + (1-\alpha)\, x} \:\xi^k \:.
\eeq

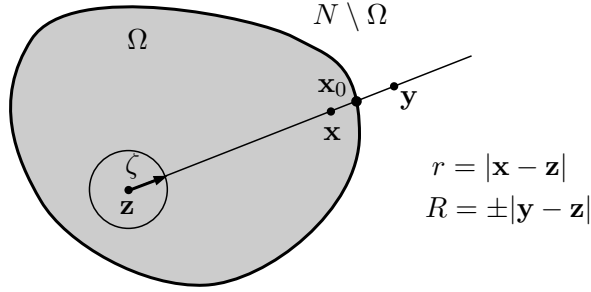
\begin{figure}
%
\psscalebox{1.0 1.0} 
{
\begin{pspicture}(-2.5,-1.898266)(10.909601,1.898266)
\definecolor{colour0}{rgb}{0.8,0.8,0.8}
\psbezier[linecolor=black, linewidth=0.04, fillstyle=solid,fillcolor=colour0](0.45893574,-0.71596164)(0.7942031,-1.2851541)(1.8190572,-1.8779153)(2.6472247,-1.8359616088867188)(3.4753923,-1.7940079)(4.6958613,-0.8269353)(4.694601,0.16403839)(4.693341,1.155012)(4.2754164,1.2150158)(3.8136125,1.4640384)(3.3518085,1.7130611)(0.8573324,2.2456412)(0.33460113,1.4340384)(-0.18813013,0.62243557)(0.12366837,-0.14676915)(0.45893574,-0.71596164)
\pscircle[linecolor=black, linewidth=0.02, fillstyle=solid,fillcolor=colour0, dimen=outer](1.6221012,-0.5634616){0.5275}
\rput[bl](1.5096011,-0.8909616){\normalsize{$\z$}}
\rput[bl](1.6046011,1.3040384){\normalsize{$\Omega$}}
\rput[bl](4.074601,1.5390384){\normalsize{$N \setminus \Omega$}}
\psline[linecolor=black, linewidth=0.02](1.6396011,-0.5709616)(6.1846013,1.2140384)
\pscircle[linecolor=black, linewidth=0.04, fillstyle=solid,fillcolor=black, dimen=outer](1.6246011,-0.5709616){0.055}
\pscircle[linecolor=black, linewidth=0.04, fillstyle=solid,fillcolor=black, dimen=outer](4.314601,0.4740384){0.055}
\pscircle[linecolor=black, linewidth=0.04, fillstyle=solid,fillcolor=black, dimen=outer](5.154601,0.8090384){0.055}
\rput[bl](4.214601,0.089038394){\normalsize{$\x$}}
\rput[bl](5.234601,0.5040384){\normalsize{$\y$}}
\psline[linecolor=black, linewidth=0.04, arrowsize=0.05291667cm 2.0,arrowlength=1.4,arrowinset=0.0]{->}(1.6546011,-0.5659616)(2.144601,-0.3809616)
\rput[bl](1.5796012,-0.42096162){\normalsize{$\zeta$}}
\rput[bl](5.654601,-0.4659616){\normalsize{$r=|\x-\z|$}}
\rput[bl](5.569601,-1.0209616){\normalsize{$R=\pm |\y-\z|$}}
\pscircle[linecolor=black, linewidth=0.04, fillstyle=solid,fillcolor=black, dimen=outer](4.654601,0.6090384){0.07}
\rput[bl](4.139601,0.6740384){\normalsize{$\x_0$}}
\end{pspicture}
}
\caption{The unbounded line integrals in the surface layer integral.}
\label{figlin}
\end{figure}

\begin{Prp} \label{prpdifflin} The contribution to the mass given by~\eqref{Mcontrib} and~\eqref{vform}
vanishes if the perturbation of the metric~$h_{ij}$ is compactly supported and trace-free.
\end{Prp}
\Proof It suffices to consider the summand involving the $x$-derivative in~\eqref{Mcontrib},
because the other summand can be treated in the same way. 
Thus our task is to analyze the integral expression
\[ I := \int_{\Omega} d^3\x \int_{N\setminus \Omega} \!\!\!\!\!\!\!d^3\y\:
\int_{-\infty}^\infty d\alpha\: \epsilon(\alpha) \:h^j_k \big|_{\alpha \y + (1-\alpha)\, \x} \int_{-\infty}^\infty
d\xi^0\:\xi^k \: \frac{\partial}{\partial x^j}\: \L_\kappa(x,y) \:. \]
The first step is to transform the integral over~$\y$ to an integral over~$\z := \alpha \y + (1-\alpha)\,\x$,
\begin{align*}
I &= \left\{ \begin{array}{c} \z-\x = \alpha\: (\y-\x) \:,\;\; d^3\z = \alpha^3\: d^3\y \end{array} \right\} \\
&= \int_{-\infty}^\infty d\alpha\: \epsilon(\alpha) 
\int_{\R^3} \frac{d^3\z}{\alpha^3}\:h^j_k(\z) \\
&\qquad \times
\int_{\Omega} d^3\x\: \chi_{N \setminus \Omega}(\y) 
\int_{-\infty}^\infty
d\xi^0\:\xi^k \: \frac{\partial}{\partial x^j}\: \L_\kappa(x,y) \Big|_{\y = \x + \frac{\z-\x}{\alpha}} \:.
\end{align*}
Next, for the $\x$-integration we choose polar coordinates around~$\z$, i.e.
\beq \label{xform}
\x = \z + r\:\zeta \qquad \text{with} \qquad r \in \R^+ \text{ and } \zeta \in S^2 \:.
\eeq
We thus obtain
\beq \label{Isphere}
I = \int_{\R^3} d^3\z \:h^j_k(\z) \int_{S^2} d^2 \zeta \; J^k_j(\z,\zeta)
\eeq
with
\begin{align*}
J^k_j(\z,\zeta) &:=  \int_{-\infty}^\infty d\alpha\: \frac{\epsilon(\alpha)}{\alpha^3}
\int_0^\infty r^2\: dr\;
\chi_\Omega(\x)\: \chi_{N \setminus \Omega}(\y)
\int_{-\infty}^\infty d\xi^0\:\xi^k \: \frac{\partial}{\partial x^j}\: \L_\kappa(x,y) \:,
\end{align*}
where~$\x$ and~$\y$ are given by~\eqref{xform} and
\[ \y = \x + \frac{\z-\x}{\alpha} 
= \z + \Big( 1 - \frac{1}{\alpha} \Big)\:(\x-\z) = \z + \Big( 1 - \frac{1}{\alpha} \Big)\:r \,\zeta  \:. \]
Finally, we transform from the integration variable~$\alpha$ to~$R$ defined by
\[ \y = \z + R\, \zeta \quad \text{and thus} \quad R = \Big( 1 - \frac{1}{\alpha} \Big)\:r \:. \]
Then
\begin{align*}
J^k_j(\z,\zeta)
&= \left\{ \begin{array}{c} \displaystyle \alpha = \frac{r}{r-R} \:,\;\; d\alpha = \frac{\alpha^2}{r}\: dR  \end{array} \right\} \\
&= \int_0^\infty r^2\: dr
\int_{-\infty}^\infty \frac{dR}{r\,\alpha}\: \epsilon(r-R)
\; \chi_\Omega(\x)\: \chi_{N \setminus \Omega}(\y)
\int_{-\infty}^\infty d\xi^0\:\xi^k \: \frac{\partial}{\partial x^j}\: \L_\kappa(x,y) \\
&= \int_0^\infty dr \int_{-\infty}^\infty dR\: |r-R|\; \chi_\Omega(\x)\: \chi_{N \setminus \Omega}(\y)
\int_{-\infty}^\infty d\xi^0\:\xi^k \: \frac{\partial}{\partial x^j}\: \L_\kappa(x,y) \:.
\end{align*}
The integrals over~$r$ and~$R$ are integrals along the straight line~$z + \zeta\, \R$,
as shown in Figure~\ref{figlin}.
Let us assume that the ray~$z + \zeta\, \R^+$ intersects
the boundary of~$\Omega$ only once at a point~$\x_0 = z + r_0\, \zeta$
(this is clearly the case if we chose~$\Omega$ as
a ball or a convex set). Moreover, assume that~$z$ lies inside~$\Omega$.
Then, introducing the new integration variables~$\tau = r-r_0$ and~$\tau' = R-r_0$,
we obtain
\[ J^k_j(\z,\zeta)
= \int_{-r_0}^0 d\tau \int_0^\infty d\tau'\: \big( \tau'-\tau \big)\; 
\int_{-\infty}^\infty d\xi^0\:\xi^k \: \frac{\partial}{\partial x^j}\: \L_\kappa(x,y) \Big|_{
\tiny{ \begin{array}{l} \x=\x_0 + \tau \zeta \\
\y = \x_0 + \tau' \zeta \end{array} } }\:. \]
If the set~$\Omega$ is chosen as a ball whose radius tends to infinity, this simplifies to
\[ J^k_j(\z,\zeta)
= \int_{-\infty}^0 d\tau \int_0^\infty d\tau'\: \big( \tau'-\tau \big)\; 
\int_{-\infty}^\infty d\xi^0\:\xi^k \: \frac{\partial}{\partial x^j}\: \L_\kappa(x,y) \Big|_{
\tiny{ \begin{array}{l} \x=\x_0 + \tau \zeta \\
\y = \x_0 + \tau' \zeta \end{array} } }\:. \]
Moreover, we can evaluate these integrals for the regularized Lagrangian
in Minkowski space. In particular, using that the Lagrangian depends only
on the difference vector~$\xi=y-x$, one of the line integrals can be carried out,
\begin{align*}
J^k_j(\z,\zeta)
&= -\int_{-\infty}^0 d\tau \int_0^\infty d\tau'\: \big( \tau'-\tau \big)\; 
\int_{-\infty}^\infty d\xi^0\:\xi^k \: \frac{\partial}{\partial \xi^j}\: \L_\kappa[\xi] \bigg|_{\xi=\big(\xi^0, (\tau'-\tau) \zeta \big)} \\
&= \left\{ \begin{array}{c} \beta = \tau'-\tau \\ d\beta = d\tau' \end{array} \right\} 
= -\int_{-\infty}^0 d\beta \int_{-\beta}^0 d\tau\: \beta
\int_{-\infty}^\infty d\xi^0\:\xi^k \: \frac{\partial}{\partial \xi^j}\: \L_\kappa[\xi] \bigg|_{\xi=(\xi^0, \beta \zeta)} \\
&= -\int_{-\infty}^0 d\beta \:\beta^2
\int_{-\infty}^\infty d\xi^0\:\xi^k \: \frac{\partial}{\partial \xi^j}\: \L_\kappa[\xi] \Big|_{\xi=(\xi^0, \beta \zeta)} \:.
\end{align*}

Let us consider the different cases for the tensor indices. If~$j=0$, we can integrate by parts in~$y^0$ to obtain
\[ J^k_0(\z,\zeta)
= \delta^k_0 \int_0^\infty d\beta\: \beta^2\; 
\int_{-\infty}^\infty d\xi^0 \:\L_\kappa[\xi] \bigg|_{\xi=(\xi^0, \beta \zeta)}\:. \]
We next compute the trace,
\begin{align*}
J^k_k(\z,\zeta) &= -\int_0^\infty d\beta\: \beta^2\; 
\int_{-\infty}^\infty d\sigma\:\beta\: \xi^k \: \frac{\partial}{\partial \xi^k}\: \L_\kappa[\xi] \Big|_{\xi=\beta\, (\sigma, \zeta)} \\
&= -\int_0^\infty d\beta\: \beta^2\; 
\int_{-\infty}^\infty d\sigma\:\beta^2\: \frac{\partial}{\partial \beta} \L_\kappa[\xi] \Big|_{\xi=\beta\, (\sigma, \zeta)} \\
&= -\int_0^\infty d\beta\: \beta^4\; \frac{\partial}{\partial \beta} 
\int_{-\infty}^\infty d\sigma^0\: \L_\kappa[\xi] \Big|_{\xi=\beta\, (\sigma, \zeta)}  \\
&= 4 \int_0^\infty d\beta\: \beta^3
\int_{-\infty}^\infty d\sigma^0\: \L_\kappa[\xi] \Big|_{\xi=\beta\, (\sigma, \zeta)} \:.
\end{align*}
Comparing the above formulas, we conclude that
\[ J^k_k(\z, \zeta) = 4\: J^0_0(\z, \zeta) \:. \]
Using spherical symmetry, it follows that
\[ J^k_j(\z,\zeta) = c(\z)\: \big( \delta^k_0\: \delta^0_j + 3\: \zeta^k \zeta_j \big) \]
with a scalar function~$c(\z)$.
As a consequence, integrating over~$\zeta$ in~\eqref{Isphere} gives
\[ I = 4 \pi \int_{\R^3} d^3\z \:h^j_k(\z) \:c(\z)\: \delta^k_j = 0 \:, \]
because~$h^j_k(\z)$ is assumed to be trace-free.
\QED

\Thanks{{{\em{Acknowledgments:}} We are grateful to Niky Kamran for helpful discussions.
We would like to thank Magdalena Lottner for useful comments on the manuscript.
We are grateful for support by the German Science Foundation (DFG) within the
research training group GRK 1692 ``Curvature, Cycles, and Cohomology.''

\providecommand{\bysame}{\leavevmode\hbox to3em{\hrulefill}\thinspace}
\providecommand{\MR}{\relax\ifhmode\unskip\space\fi MR }
\providecommand{\MRhref}[2]{%
  \href{http://www.ams.org/mathscinet-getitem?mr=#1}{#2}
}
\providecommand{\href}[2]{#2}

\end{document}